\documentclass[amsmath,amssymb,twocolumn,showpacs,superscriptaddress,aps,prd,10pt]{revtex4-1}  

\usepackage[T1]{fontenc}
\usepackage[utf8]{inputenc}

\usepackage[dvipsnames,svgnames]{xcolor}
\usepackage{amssymb}
\usepackage{amsmath}
\usepackage{amsfonts}
\usepackage{mathtools}
\usepackage{graphicx}
\usepackage{slashed}
\usepackage{wasysym}
\usepackage{ulem}

\usepackage{tabularx}
\usepackage{graphicx}

\usepackage{nicefrac}
\usepackage{booktabs}
\usepackage{tikz}

\usepackage{siunitx}
\usepackage{units}
\usepackage{textcomp}

\newlength{\figlen}
\newlength{\figlenFull}
\newlength{\figlenThree}

\newcommand{\Eqref}[1]{Eq.~\eqref{#1}}

\newcommand*{\I}{ {\rm i} }
\newcommand*{\E}{ {\rm e} }

\usepackage{etoolbox}
\newcommand{\dd}[2][]{\ifstrempty{#1}{\mathop{\mathrm{d}#2}}{\mathop{\mathrm{d}^{#1}#2}}}

\let\vv\vec

\definecolor{orange}{RGB}{255,102,0}
\definecolor{holger}{RGB}{0,120,0}

\newcommand{\omite}[1]{}

\begin{document}

\title{All-optical signatures of quantum vacuum nonlinearities in generic laser fields}

\author{Alexander Blinne}
\affiliation{Helmholtz-Institut Jena, Fr\"obelstieg 3, 07743 Jena, Germany}
\affiliation{Institut für Optik und Quantenelektronik, Abbe Center of Photonics, \\
Friedrich-Schiller-Universit\"at Jena, Max-Wien-Platz 1, 07743 Jena, Germany}

\author{Holger Gies}
\affiliation{Helmholtz-Institut Jena, Fr\"obelstieg 3, 07743 Jena, Germany}
\affiliation{Theoretisch-Physikalisches Institut, Abbe Center of Photonics, \\
Friedrich-Schiller-Universit\"at Jena, Max-Wien-Platz 1, 07743 Jena, Germany}

\author{Felix Karbstein}\email{Corresponding author: \url{felix.karbstein@uni-jena.de}}
\affiliation{Helmholtz-Institut Jena, Fr\"obelstieg 3, 07743 Jena, Germany}
\affiliation{Theoretisch-Physikalisches Institut, Abbe Center of Photonics, \\
Friedrich-Schiller-Universit\"at Jena, Max-Wien-Platz 1, 07743 Jena, Germany}

\author{Christian Kohlf\"urst}
\affiliation{Helmholtz-Institut Jena, Fr\"obelstieg 3, 07743 Jena, Germany}
\affiliation{Theoretisch-Physikalisches Institut, Abbe Center of Photonics, \\
Friedrich-Schiller-Universit\"at Jena, Max-Wien-Platz 1, 07743 Jena, Germany}

\author{Matt Zepf}
\affiliation{Helmholtz-Institut Jena, Fr\"obelstieg 3, 07743 Jena, Germany}
\affiliation{Institut für Optik und Quantenelektronik, Abbe Center of Photonics, \\
Friedrich-Schiller-Universit\"at Jena, Max-Wien-Platz 1, 07743 Jena, Germany}

\date{\today}

\begin{abstract}
All-optical experiments at the high-intensity frontier offer a
promising route to unprecedented precision tests of quantum
electrodynamics in strong macroscopic electromagnetic fields.
So far, most theoretical studies of all-optical signatures of quantum vacuum
nonlinearity are based on simplifying approximations of the beam
profiles and pulse shapes of the driving laser fields.
Since precision tests require accurate quantitative theoretical predictions, we
introduce an efficient numerical tool facilitating the quantitative
theoretical study of all-optical signatures of quantum vacuum
nonlinearity in generic laser fields.  Our approach is based on the
vacuum emission picture, and makes use of the fact that the dynamics
of the driving laser fields are to an excellent approximation governed
by classical Maxwell theory in vacuum.  In combination with a Maxwell
solver, which self-consistently propagates any given laser field
configuration, this allows for accurate theoretical predictions of
photonic signatures of vacuum nonlinearity in high-intensity laser
experiments from first principles.
We employ our method to simulate photonic signatures of quantum vacuum nonlinearity in laser pulse collisions involving a few-cycle pulse,
and show that the angular and spectral distributions of the emitted signal photons deviate from those of the driving laser beams.
\end{abstract}

\maketitle

\section{Introduction}

The omnipresence of quantum fluctuations makes the vacuum of quantum electrodynamics (QED) far from being trivial.
Quantum fluctuations of charged particles mediate effective couplings between electromagnetic fields, supplementing Maxwell's linear theory of classical electrodynamics in vacuo with nonlinear self-interactions of the electromagnetic field \cite{Euler:1935zz,Heisenberg:1935qt,Weisskopf:1936bu,Schwinger:1951nm}, as
reviewed, e.g., in Ref.~\cite{Dittrich:1985yb,Dittrich:2000zu,Marklund:2008gj,Heinzl:2008an,DiPiazza:2011tq,Dunne:2012vv,Battesti:2012hf,King:2015tba,Karbstein:2016hlj,Inada:2017lop}.
Such vacuum nonlinearities have no classical analogue and are rather elusive in experiments.

Prominent theoretical proposals of all-optical signatures of quantum vacuum nonlinearity include vacuum birefringence \cite{Kotkin:1996nf,Heinzl:2006xc,DiPiazza:2006pr,Dinu:2013gaa,Dinu:2014tsa,Karbstein:2015xra,Schlenvoigt:2016,Karbstein:2016lby,King:2016jnl,Bragin:2017yau,Ataman:2018ucl,Karbstein:2018omb}, photon-photon scattering in the form of laser-pulse collisions \cite{Lundstrom:2005za,Lundin:2006wu,Tommasini:2009nh,Tommasini:2010fb,King:2012aw,Gies:2017ezf,King:2018wtn}, quantum reflection \cite{Gies:2013yxa,Gies:2014wsa}, photon merging \cite{Yakovlev:1966,Fedotov:2006ii,DiPiazza:2007prw,DiPiazza:2009cq,King:2014vha,Gies:2014jia,Gies:2016czm}, photon splitting \cite{Adler:1970gg,BialynickaBirula:1970vy,Adler:1971wn,Papanyan:1971cv,DiPiazza:2007yx}, and QED vacuum-fluctuation-triggered interference effects in multiple beam configurations \cite{King:2013am,King:2013zz,Hatsagortsyan:2011}.
While QED vacuum nonlinearities in macroscopic electromagnetic fields so far have not been directly verified experimentally, laboratory searches of vacuum birefringence in macroscopic magnetic fields \cite{DellaValle:2015xxa,Cadene:2013bva,Fan:2017fnd} have already demonstrated the need for high field strengths and, at the same time, a high signal detection sensitivity.
Besides, vacuum birefringence might also be relevant for the optical polarimetry of neutron stars \cite{Mignani:2016fwz,Capparelli:2017mlv,Turolla:2017tqt}.

In a series of recent papers \cite{Karbstein:2014fva,Karbstein:2015xra,Karbstein:2015qwa,Karbstein:2016lby,Gies:2017ygp,Gies:2017ezf}, it has been argued and demonstrated that such signatures can conveniently be analyzed within the vacuum emission picture \cite{Karbstein:2014fva}.
In this picture, no distinction between pump and probe fields is made. Instead, one formally studies the process of signal photon emission from the quantum vacuum subject to the electromagnetic fields of all driving laser pulses.
Given that the diameters of the interaction/strong field-volume are much smaller than its distance to the detectors, as is typically the case in high-field experiments, the precise microscopic origin of the signal photons within the interaction volume cannot be resolved and the interaction volume appears point-like. The kinematics of all outgoing signal photons can then be characterized by common-origin wave vectors $\vec{k}$.
A non-vanishing signal photon amplitude constitutes a -- potentially measurable -- signal of quantum vacuum nonlinearity.
In the absence of vacuum fluctuations of particles with photon interactions, the signal photon amplitude is identical to zero.
Due to the fact that the effects of QED vacuum nonlinearities are very small in typical experimental scenarios based on present and near-future laser technology, it amounts to an excellent approximation to assume the dynamics of the driving laser pulses to be governed by classical Maxwell theory in vacuum. This corresponds to neglecting quantum vacuum nonlinearity induced modifications of the driving fields themselves, such as beam depletion and other forms of back-reaction. The latter requires numerical solvers for the full nonlinear effective theory \cite{Bohl:2015uba,Domenech:2016xx,Carneiro:2016qus}.

Up to now, theoretical studies of all-optical signatures of quantum vacuum nonlinearity have typically involved various simplifying approximations of the beam profiles and pulse shapes of the driving laser fields, such as plane-wave based models, constant crossed fields amended with a pulse shape envelope, or more realistic laser pulses in the paraxial approximation. However, any such approximation inherently limits the accuracy of predictions for experiments, and thus the precision with which theory can be tested.
In the present article, we substantially advance the theoretical modeling of all-optical signatures of QED vacuum nonlinearities in experimentally realistic field configurations by evaluating the effect in generic laser fields, manifestly fulfilling Maxwell's equations in vacuum. 

To this end, we rely on a few well-justified and parametrically controlled fundamental approximations detailed below, and employ a Maxwell solver, recently put forward in Ref.~\cite{Blinne:2018}, to numerically solve the dynamics of the macroscopic electromagnetic fields driving the vacuum emission effect self-consistently. This facilitates the study of all-optical signatures of QED vacuum nonlinearities in very generic, experimentally relevant electromagnetic fields in full 3+1 space-time dimensions.
Most importantly, our approach allows us to easily overcome the limitations of any approximation for the laser beams.
No ad-hoc ansatz for the temporal pulse profile is required, allowing for the study of all-optical signatures of QED vacuum nonlinearities in arbitrary, experimentally determined 'real world' laser fields. 

For electric and magnetic fields of strengths below the critical electric and magnetic fields, given by $E_\text{cr}=m_e^2c^3/(e\hbar)
\approx 1.3 \times 10^{18}\,\frac{\rm V}{\rm m}$ and $B_\text{cr}=E_\text{cr}/c \approx 4 \times 10^{9}\,{\rm T}$, respectively, the effective self-interactions are parametrically suppressed with powers of $E/E_{\rm cr}$ and $B/B_{\rm cr}$.
If these electromagnetic fields in addition vary on spatial (temporal) scales much larger than the Compton wavelength (time) $\lambdabar_{\rm C}=\hbar/m_e\approx3.86 \cdot 10^{-13}\,{\rm m}$ ($\tau_{\rm
  C}=\lambdabar_{\rm C}/c\approx1.29 \cdot 10^{-21}\,{\rm s}$) of the electron, they can be considered as locally constant, and the Heisenberg-Euler effective Lagrangian \cite{Heisenberg:1935qt} formally derived for infinitely extended constant fields can be used for their study.
  
State-of-the-art high-intensity lasers reach peak fields $E={\cal O}(10^{14})\frac{\rm V}{\rm m}$ and $B={\cal O}(10^6){\rm T}$, implying that $E\ll E_{\rm cr}$ and $B\ll B_{\rm cr}$.
Moreover, their typical scale of variation is given by the laser wavelength $\lambda={\cal O}(1)$\textmu${\rm m}$, clearly fulfilling $\lambda\gg\lambdabar_{\rm C}$.
Correspondingly, for all-optical studies of QED vacuum nonlinearities in high-intensity laser fields, we can limit ourselves to the perturbative weak-field limit of the Heisenberg-Euler effective Lagrangian.
In the Heaviside-Lorentz System with $c=\hbar=1$, adopted throughout this article, its leading contribution corresponding to a quantum vacuum fluctuation induced effective quartic self-interaction of the electromagnetic field is given by \cite{Euler:1935zz}
\begin{equation}
 {\cal L}_\text{HE}^{1\text{-loop}}\simeq\frac{m_e^4}{8\pi^2}\frac{1}{45}\Bigl(\frac{e}{m_e^2}\Bigr)^4(4{\cal F}^2+7{\cal G}^2\bigr) .
 \label{eq:HEpert}
\end{equation}
Here ${\cal F} = \frac{1}{4}F_{\mu\nu}F^{\mu\nu}=\frac{1}{2}(\vec{B}^2-\vec{E}^2)$ and ${\cal G} = \frac{1}{4} F_{\mu\nu} ({}^\star\!F)^{\mu\nu}=-\vec{B}\cdot\vec{E}$ are the scalar invariants of the electromagnetic field, and $({}^\star\!F)^{\mu\nu}=\frac{1}{2}\epsilon^{\mu\nu\alpha\beta}F_{\alpha\beta}$ is the dual field strength tensor; our metric convention is $g_{\mu \nu}=\mathrm{diag}(-1,+1,+1,+1)$.
Deviations from a full-fledged QED calculation in manifestly inhomogeneous electromagnetic fields are suppressed parametrically by a factor of $(\upsilon/m_e)^2$, where $\upsilon\sim1/\lambda$ denotes the typical frequency/momentum scale of variation of the electromagnetic fields \cite{Galtsov:1982,Karbstein:2015cpa}.
In addition, note that the expression given in \Eqref{eq:HEpert} amounts to the result of a one-loop computation.
Contributions from higher loops are parametrically suppressed with the fine-structure constant $\alpha=\frac{e^2}{4\pi}\simeq\frac{1}{137}$, such that we can expect \Eqref{eq:HEpert} to allow for the reliable study of all-optical signatures of QED vacuum nonlinearity driven by high-intensity lasers with an accuracy on the one percent level.
Equation~\eqref{eq:HEpert} constitutes the typical starting point of theoretical studies of all-optical signatures of quantum vacuum nonlinearity available in the literature, and also the present article is based thereon.

Our article is organized as follows:
After briefly summarizing the underlying formalism in Sec.~\ref{sec:formalism}, we present our novel vacuum emission solver in full detail in Sec.~\ref{sec:numerics}.
Here, we explain how the self-consistent propagation of the driving laser fields is implemented by means of a Maxwell solver. We put special emphasis on how to fix the spectral amplitudes characterizing the driving laser fields in order to accurately describe experimentally realistic beam profiles and pulse shapes.
The primary focus of Sec.~\ref{sec:results} is on benchmarking our numerical tools in situations where the paraxial approximation is expected to provide a reasonable description of the driving laser fields. After considering the effect of signal photon emission from a single high-intensity laser pulse, we focus on the collision of two laser pulses of the same parameters.
Section~\ref{sec:showcase} is devoted to the discussion of first results on the frequency and angular distribution of scattered photons for tightly focused femtosecond laser pulses.
One of the driving laser pulses is assumed to be a few-cycle pulse, challenging the applicability of the paraxial approximation.
We show that the spatio-temporal envelope of tightly focused femtosecond laser pulses significantly affects both the angular and frequency distribution of the emitted photons. 
This underlines the importance of an accurate description of the driving high-intensity laser pulses and demonstrates the great potential of our approach for the quantitative study of all-optical signatures of quantum vacuum nonlinearities from first principles.
Finally, we end with conclusions and an outlook in Sec.~\ref{sec:conclusions}.

\section{Formalism}
\label{sec:formalism}

\omite{Beyond the Standard Model (BSM) and the world-average of Higgs bounds obtained from Higgs-inflation models, there is a coupling constant, which can give rise to jet formation, via swampland-induced CP violation. It can increase the chemical potential at finite temperature by stimulating decay channels into the flat zero-modes of left-handed Hawking radiation with chirally symmetric string vaua. Moreover, the decay of string resonances and LHC data, can phenomenologically affect the matter content of extended gauge theories. The direct searches for anomalies and axions is highly relevant for Gauge-Higgs Unification Models. In this context, the interactions among quarks and gluons, may provide important indications of how to solve the dark energy puzzle with scalar-vector-tensor theories. Besides, the Higgs portal might provide helpful insights into the renormalization of the chiral magnetic effect in hot dense baryonic matter with holographic dijet systems mapped via the AdS/CFT duality for D brane N=6 supergravity on the bulk with boundary-like super multiplets mixed with scalar tensor theories. The strong weak duality might moreover be used to determine the temperature of a rotating black whole with gray boundary and stimulate invisible neutrino decay channels with adequate truncation for the associated superpartners in large N 't Hooft Witten index modulated coarse graining dumps. N=8 Supersymmetry might be solution.}

As detailed in Ref.~\cite{Karbstein:2014fva}, the zero-to-single signal photon transition amplitude to a state of wave vector $\vec{k}$ and transverse polarization $p\in\{1,2\}$ is given by
\begin{equation}
 S_{(p)}(\vec{k})=\frac{\epsilon_{(p)}^{*\mu}(\vec{k})}{\sqrt{2k^0}}\int{\rm d}^4x\,{\rm e}^{{\rm i}kx}\,j_{\mu}(x)\,\biggr|_{k^0=|\vec{k}|}\,, \label{eq:Sp}
\end{equation}
where $\epsilon_{(p)}^{*\mu}(\vec{k})$ is the polarization vector characterizing the polarization state $p$, and
\begin{equation}
 j_\mu(x)=2\partial^\alpha\frac{\partial{\cal L}_\text{HE}^{1\text{-loop}}(F)}{\partial F^{\alpha\mu}}
 \label{eq:j}
\end{equation}
is the single signal photon current induced by the macroscopic electromagnetic field $F^{\mu\nu}$, which amounts to the superposition of all prescribed fields driving the effect.
Using spherical coordinates $\vec{k}={\rm k}(\cos\varphi\sin\vartheta,\sin\varphi\sin\vartheta,\cos\vartheta)$, it is straightforward to show that the unit vectors perpendicular to $\vec{k}$ can be parameterized by a single angle $\beta$, as
\begin{equation}
 \vec{e}_\perp(\beta)=\vec{e}_1(\vec{k})\cos\beta + \vec{e}_2(\vec{k})\sin\beta\,,
 \label{eq:ebeta}
\end{equation}
where we introduced the orthonormal vectors
\begin{equation}
\vec{e}_1(\vec{k})=
\left(\begin{array}{c}
  \cos\varphi\cos\vartheta \\
  \sin\varphi\cos\vartheta \\
  -\sin\vartheta
 \end{array}\right),\quad
\vec{e}_2(\vec{k})=
\left(\begin{array}{c}
  -\sin\varphi \\
  \cos\varphi \\
  0
 \end{array}\right).
 \label{eq:e1&e2}
\end{equation}
An alternative representation of these vectors is
\begin{equation}
\vec{e}_1(\vec{k})=\frac{1}{{\rm k}{\rm k}_{\rm xy}}
\left(\begin{array}{c}
  k_{\rm x}k_{\rm z} \\
  k_{\rm y}k_{\rm z} \\
  -{\rm k}_{\rm xy}^2
 \end{array}\right),
 \quad
\vec{e}_2(\vec{k})=\frac{1}{{\rm k}_{\rm xy}}
\left(\begin{array}{c}
  -k_{\rm y} \\
  k_{\rm x} \\
  0
 \end{array}\right),
 \label{eq:e1&e2_alt}
\end{equation}
with ${\rm k}_{\rm xy}=\sqrt{k_{\rm x}^2+k_{\rm y}^2}$ and ${\rm k}=\sqrt{k_{\rm xy}^2+k_{\rm z}^2}$.
Equation~\eqref{eq:ebeta} can be employed to span the transverse polarizations of signal photons of wave vector $\vec{k}$.
Using a linear polarization basis, we define
\begin{equation}
 \epsilon^\mu_{(p)}(\vec{k})=\bigl(0,\vec{e}_\perp(\beta_p)\bigr)\,,\quad\text{with}\quad \beta_p=\beta_0+\frac{\pi}{2}(p-1)
\label{eq:polv1&2}
\end{equation}
and a suitably chosen $\beta_0$
The signal photon current~\eqref{eq:j} derived from the terms given explicitly in \Eqref{eq:HEpert} is cubic in $F^{\mu\nu}$, and results in the following expression for the signal photon transition amplitude \cite{Gies:2017ygp},
\begin{align}
{\cal S}_{(p)}(\vec{k}) = & \frac{1}{\rm i}\frac{1}{2\pi}\frac{m_e^2}{45}\sqrt{\frac{\alpha}{\pi}\frac{\rm k}{2}}\Bigl(\frac{e}{m_e^2}\Bigr)^3  \nonumber\\
 &\times \Bigl\{\cos\beta_p\bigl[{\cal I}_{11}(\vec{k})-{\cal I}_{22}(\vec{k})\bigr] \nonumber\\
 &\ \ +\sin\beta_p\bigl[{\cal I}_{12}(\vec{k})+{\cal I}_{21}(\vec{k})\bigr]\Bigr\} ,
 \label{eq:S1pert}
\end{align}
where we made use of the definition
\begin{equation}
 {\cal I}_{ij}(\vec{k})=\int{\rm d}^4 x\, {\rm e}^{{\rm i}(\vec{k}\cdot\vec{x}-{\rm k}t)}\,\vec{e}_i\cdot\vec{U}_j\,, \label{eq:Iij}
\end{equation}
with
\begin{align}
 \label{eqn:Uj}
 \begin{split}
 \vec{U}_1&=2\vec{E}(\vec{B}^2-\vec{E}^2)-7\vec{B}(\vec{B}\cdot\vec{E}) \,, \\
 \vec{U}_2&=2\vec{B}(\vec{B}^2-\vec{E}^2)+7\vec{E}(\vec{B}\cdot\vec{E}) \,,
 \end{split}
\end{align}
encoding the dependence on the electromagnetic fields driving the vacuum emission process.
In the present paper, we consider the driving electromagnetic fields to be delivered by lasers.

In summary, the determination of the vacuum emission amplitude boils down to performing the four-dimensional Fourier integrals in \Eqref{eq:Iij} for given spacetime dependent fields $\vec{E}(x)$, $\vec{B}(x)$.
The differential number of signal photons of polarization $p$ follows from the modulus squared of \Eqref{eq:Sp} and is given by \cite{Karbstein:2014fva}
\begin{equation}
{\rm d}^3N_{(p)}(\vec{k})=\,\frac{{\rm d}^3k}{(2\pi)^3}\bigl|{\cal
S}_{(p)}(\vec{k})\bigr|^2 \,. \label{eq:d3Np_polarcoords}
\end{equation}
Accordingly, the differential number of signal photons of arbitrary polarization is given by ${\rm d}^3N(\vec{k})=\sum_{p=1}^2{\rm d}^3N_{(p)}(\vec{k})$.
Upon summation over both transverse polarizations $p$, the dependence on the angle $\beta_0$ drops out completely as it should: the orientation of the polarization basis does not matter.

\section{Numerical Implementation}
\label{sec:numerics}

\subsection{Fourier integrals}

As we aim at evaluating signal photon numbers in generic laser fields, we perform the Fourier integrals~\eqref{eq:Iij} numerically.
To this end, we first note that \Eqref{eq:Iij} does not correspond to a 4D Fourier transform, but rather amounts to a 3D Fourier transform with an additional temporal integration, the reason for this being the fact that ${\rm k}=|\vec{k}|$ is not an independent variable for on-shell signal photons.

Various methods of discretization are available for performing the 3D Fourier transform and the 1D integral.
For simplicity, we adopt a uniform Cartesian grid for space and time, leading to
\begin{align}
   {\cal I}_{ij}(\vec{k}) &= {\int \dd{t} \E^{-\I {\rm k} t}}
 {\int \dd[3]{x} \E^{\I \vv{k}\cdot\vv{x}}}
    \vec{e}_i\cdot\vec{U}_j
 \nonumber\\
 &\approx  { \sum_n \Delta t \,\E^{-\I {\rm k} t_n}}
 \;{\mathrm{FFT_3}[} \vec{e}_i\cdot\vec{U}_j]\,,
\end{align}
where FFT$_3$ denotes a 3D discrete Fourier transform using a fast Fourier transform algorithm. For simplicity, we perform the time integration by a simple trapezoidal rule.

The requirements on the extent and resolution of the Cartesian grid depend on the details of the considered electromagnetic fields. Given that a field is characterized by a set of spatial and temporal frequencies $\upsilon$, the fact that the polynomials $\vec{U}_j$ defined in \Eqref{eqn:Uj} are cubic in the field immediately implies the need to resolve all frequencies in a range including $3\upsilon$.
The temporal extent needs to cover the whole interval in which the driving electromagnetic fields are strong enough as to induce significant contributions to the vacuum emission signal.
For collisions of two or more laser pulses, this typically corresponds to the time interval, where the pulses overlap.
Aiming at the study of the effect of self-emission from a single laser pulse, the relevant longitudinal scale is of the order of the maximum of the pulse duration and the Rayleigh range.
Analogously, the spatial extents need to be large enough to cover the entire interaction/strong-field volume. In case vastly different scales are introduced by the driving fields or specific long-time observables \cite{Briscese:2017htx,Briscese:2017wuh}, multiscale methods may be advantageous. 
An example study of the convergence properties of the algorithm is sketched in the appendix.

Depending on the specific scenario under consideration, these constraints lead to very different requirements on the computational resources.
In its current implementation, our code uses single node parallelization: Simple scenarios can be run in a few seconds with a few hundred megabytes of memory, while more complex scenarios require tens of hours and hundreds of gigabytes of memory on a high performance node.
An MPI enabled version of the code might be a desirable option in the future.

\subsection{Driving laser fields}

The most direct approach towards the realistic modeling of the driving laser fields, serving as input in the above routines, is to specify them as a set of initial data, and numerically implement their self-consistent propagation according to free Maxwell theory in vacuum. This procedure guarantees that -- up to discretization effects -- the driving laser fields exactly fulfill Maxwell's equations in vacuum. This is our approach of choice, which is implemented here for the first time.

Another option is to model the electric and magnetic fields by explicit analytic solutions of Maxwell's equations in vacuum that can be directly evaluated for any space-time coordinate.
As a main drawback, this approach is limited to a subset of fields. It does not fully extend to all relevant fields, in particular, tightly focused and ultrashort laser pulses.
A first step towards the description of a focused Gaussian high-intensity laser beam has been the paraxial approximation.
With regard to the vacuum emission picture, this approach has been put forward by Refs.~\cite{Gies:2017ygp,Gies:2017ezf}, modeling the driving laser pulses as leading (zeroth) order paraxial Gaussian beams supplemented with a finite pulse duration.
The paraxial approximation is valid for small diffraction angles $\theta\simeq\frac{w_0}{{\rm z}_R}$, where $w_0$ and ${\rm z}_R$ are the beam radius in the focus and the Rayleigh range, respectively.
For a beam of wavelength $\lambda$, the latter is given by ${\rm z}_R=\frac{\pi w_0^2}{\lambda}$.
Explicit analytical expressions for the electric and magnetic fields up to order $\theta^{11}$ have been worked out \cite{Davis:1979zz,Barton:1989,Salamin:2002dd,Salamin:2006ff,Salamin:2006}.
The leading order paraxial approximation only accounts for terms of order $\theta^0$.

To benchmark our results obtained from the self-consistent numerical Maxwell solver, we adopt the paraxial approximation up to order $\theta^5$ \cite{Salamin:2006ff}, supplemented with a finite pulse duration $\tau$ by an overall factor of $\exp\{-(\hat{\vec{\kappa}}\cdot\vec{x}-t)^2/(\tau/2)^2\}$, where the unit vector $\hat{\vec{\kappa}}$ points along the laser's beam axis; cf., e.g., Ref.~\cite{Karbstein:2015cpa}.
Since we mainly aim at benchmarking our novel numerical approach with results based on multi-cycle paraxial fields, the dominant signal photon emission channels are independent of the phase of the beam; cf., e.g., Refs.~\cite{Karbstein:2015xra,Karbstein:2016lby}. In the remainder, we ignore possible constant phase shifts.

Our focus is on exact numerical solutions of Maxwell's equations in vacuum closely resembling experimentally realistic field profiles.
Particularly in the benchmark scenarios, these solutions are modeled after the corresponding analytical approximations.
Let us already emphasize here that these comparisons are by no means unique: generically, various exact numerical solutions can be invoked to closely resemble a single analytical approximation in the region of interest.

\subsection{Field decomposition}
\label{sect:fieldsolver}

As detailed in Ref.~\cite{Blinne:2018}, our numerical solver of Maxwell's equations in vacuum makes use of complex representations for all electromagnetic potentials, fields and associated amplitudes.
Linearity of Maxwell's equations guarantees that real and imaginary parts of a complex solution satisfy these equations individually.
By contrast, a manifestly real representation of the fields is needed for the determination of the vacuum emission amplitude.
Correspondingly, we switch to manifestly real-valued fields by means of the prescription $\vec{E}(x)\to\Re\{\vec{E}(x)\} \equiv \vec{\mathcal{E}}(x)$ and $\vec{B}(x)\to\Re\{\vec{B}(x)\}\equiv \vec{\mathcal{B}}(x)$ for the formalism detailed in Sec.~\ref{sec:formalism}.

Our aim is to ensure that the driving laser fields which can be considered as ensembles of propagating on-shell photons (${\rm k}=|\vec{k}|)$ fulfill Maxwell's equations in vacuum exactly.
To this end, we rely on a complete representation of the electromagnetic potentials in radiation gauge, $A^\mu\equiv(0,\vec{A})$ and $\vec{\nabla}\cdot\vec{A}=0$, with
\begin{align}
 \vec{A} (x) = \int\mathop{\frac{\dd[3]{k}}{\left( 2\pi \right)^3}} {\rm e}^{\I \vec{k} \cdot \vec{x}}\, \tilde{\vec{A}} (t, \vec{k}) \,,
 \label{equ:A}
\end{align}
where we have defined
\begin{equation}
 \tilde{\vec{A}} (t, \vec{k}) = {\rm e}^{-\I {\rm k}t }\,\sum_{p=1}^2 \vec{e}_p (\vec{k})\,a_{p}(\vec{k})\,.
 \label{equ:A_components}
\end{equation}
We span the two linear polarizations transverse to $\vec{k}$ by the unit vectors $\vec{e}_p(\vec{k})$ defined in Eqs.~\eqref{eq:e1&e2} and \eqref{eq:e1&e2_alt}.
A specific field configuration is realized by a suitable choice of the complex spectral amplitudes $a_{p}(\vec{k})$.
Any possible choice for the spectral amplitudes constitutes a viable solution of Maxwell's equations.

Defining the spatial Fourier transforms of the associated electric $\vec{E}(x)=-\partial_t\vec{A}(x)$ and magnetic $\vec{B}(x)=\vec{\nabla}\times\vec{A}(x)$ fields analogous to \Eqref{equ:A}, we obtain
\begin{align}
 \tilde{\vec{E}} (t, \vec{k}) &= {\rm e}^{-\I{\rm k}t}\, \I {\rm k} \bigl[ \vec{e}_1(\vec{k})\, a_1(\vec{k}) + \vec{e}_2(\vec{k})\, a_2(\vec{k}) \bigr], \label{eqn:amplitude_E} \\
\tilde{\vec{B}} (t, \vec{k}) &= {\rm e}^{-\I{\rm k}t}\, \I {\rm k} \bigl[ \vec{e}_2(\vec{k})\, a_1(\vec{k}) - \vec{e}_1(\vec{k})\, a_2(\vec{k}) \bigr] . \label{eqn:amplitude_B}
\end{align}
It is straightforward to verify that these expressions fulfill the transversality conditions
\begin{equation}
 \vec{k}\cdot\tilde{\vec{E}}(t,\vec{k})=\vec{0}\,,\quad \vec{k}\cdot\tilde{\vec{B}}(t,\vec{k})=\vec{0}\,, \label{eq:transversality}
\end{equation}
as well as the spectral analogue of the Maxwell-Ampere equation,
\begin{align}
\label{eqn:spectral_ampere}
\tilde{\vec{E}} (t, \vec{k}) +  \hat{\vec{k}}\times \tilde{\vec{B}} (t, \vec{k}) = \vec{0}\,,
\end{align}
where $\hat{\vec{k}}=\vec{k}/{\rm k}$.

The advantage of the complex representation of the fields is that it allows for a clear distinction of Fourier modes with wave vectors of opposite sign.
Contrarily, purely real-valued fields inevitably mix Fourier modes with wave-vectors of opposite sign.
This is obvious from the fact that real-valued fields can be expressed as
\begin{equation}
 {\vec{\mathcal{E}}} (x) =\frac{1}{2} \int\mathop{\frac{\dd[3]{k}}{\left( 2\pi \right)^3}} {\rm e}^{\I \vec{k} \cdot \vec{x}}\, \tilde{\vec{E}} (t,\vec{k})\ +\ \text{c.c.}\,, \label{eq:ReE} 
\end{equation}
and analogously for $\vec{\cal B}(x)$; c.c. denotes the complex conjugate.

\subsection{Spectral amplitudes}
\label{sec:specam}

The remaining task is to fix the spectral amplitudes $a_p(\vec{k})$ such as to reproduce the macroscopic electromagnetic fields of experimentally realistic high-intensity laser pulses in position space.

\subsubsection{Extraction from model fields}
\label{sec:extmod}

Starting point of this endeavor often is a model field configuration exhibiting the desired properties. 
The latter is specified by the complex fields $\vec{E}_{\rm m}(t_0,\vec{x})$ and $\vec{B}_{\rm m}(t_0,\vec{x})$ at all spatial coordinates $\vec{x}$ and a fixed time $t_0$. 

The fact that the fields $\vec{E}_\mathrm{m}(t_0,\vec{x})$ and $\vec{B}_\mathrm{m}(t_0,\vec{x})$ typically do not correspond to exact solutions of Maxwell's equations in vacuum comes along with violations of the transversality conditions~\eqref{eq:transversality}, such that in general
\begin{equation}
 \vec{k}\cdot\tilde{\vec{E}}_{\rm m}(t_0,\vec{k})\neq\vec{0}\,,\quad \vec{k}\cdot\tilde{\vec{B}}_{\rm m}(t_0,\vec{k})\neq\vec{0}\,.
\end{equation}
Here, we have employed the inverse of \Eqref{equ:A} to transform the fields to momentum space.
This implies that, contrarily to Eqs.~\eqref{eqn:amplitude_E} and \eqref{eqn:amplitude_B}, $\tilde{\vec{E}}_{\rm m}(t_0,\vec{k})$ and $\tilde{\vec{B}}_{\rm m}(t_0,\vec{k})$ cannot be spanned by the vectors $\vec{e}_p(\vec{k})$ alone, inhibiting a simple inversion for the associated spectral amplitudes.

In response, we do not aim at extracting the spectral amplitudes from the model fields, but instead define manifestly transverse fields based on the model input.
The electromagnetic fields constructed along these lines exactly fulfill Maxwell's equations in vacuum.
More specifically, our strategy is to assume these fields to be given by Eqs.~\eqref{eqn:amplitude_E} and \eqref{eqn:amplitude_B} and define the spectral amplitudes $a_p(\vec{k})$ in terms of $\tilde{\vec{E}}_{\rm m}(t_0,\vec{k})$ and $\tilde{\vec{B}}_{\rm m}(t_0,\vec{k})$ based on a suitable projection.
Such an {\it ad hoc} prescription is of course not unique and various choices are possible.
However, all the prescriptions discussed below should result in field configurations closely resembling the model fields at $t=t_0$, if the latter violate the transversality conditions only mildly.

Having manifestly complex model fields at our disposal, a rather generic prescription is to define the spectral amplitudes motivated by the structure of \Eqref{eqn:amplitude_E} as
\begin{align}
 a_p(\vec{k}) &:= {\rm e}^{\I {\rm k}t_0}\frac{1}{\I{\rm k}} \vec{e}_p(\vec{k})\cdot\tilde{\vec{E}}_\mathrm{m}(t_0,\vec{k})\,. \label{eq:a_p}
\end{align}
Inserting \Eqref{eq:a_p} into \Eqref{eqn:amplitude_E}, we obtain
\begin{equation}
 \tilde{\vec{E}} (t, \vec{k}) = {\rm e}^{-\I{\rm k}(t-t_0)} \bigl[ \tilde{\vec{E}}_\mathrm{m}(t_0,\vec{k})-\hat{\vec{k}} \bigl(\hat{\vec{k}}\cdot\tilde{\vec{E}}_\mathrm{m}(t_0,\vec{k})\bigr) \bigr], \label{eq:Etk}
\end{equation}
where we have made use of the fact that the unit vectors $\vec{e}_1$, $\vec{e}_2$ and $\hat{\vec{k}}$ form a complete orthonormalized basis of $\mathbb{R}^3$.

For the special case of laser pulses polarized in z direction in the focus and beam axes in the xy-plane, one might want to match the z component of the electric field, leaving all other components of the fields to be determined self-consistently by Maxwell's equations in vacuum.
Taking into account that $\vec{e}_2\cdot\vec{e}_{\rm z}=0$, where $\vec{e}_{\rm z}$ is the unit vector in z direction, \Eqref{eqn:amplitude_E} prompts us to define
\begin{align}
 a_{1}(\vec{k}) &:= {\rm e}^{\I {\rm k}t_0}\frac{1}{\I{\rm k}}\frac{1}{\vec{e}_{\rm z}\cdot\vec{e}_1} \vec{e}_{\rm z}\cdot\tilde{\vec{E}}_\mathrm{m}(t_0,\vec{k})\,. \label{eq:a_1}
\end{align}
For completeness, also note that $\vec{e}_1|_{k_{\rm z}=0}\sim\vec{e}_{\rm z}$.
The spectral amplitude $a_2(\vec{k})$ is not at all constrained by this condition and can be chosen freely.
For convenience, in the benchmark calculations employing this prescription in Sec.~\ref{sec:results} below, we use $a_{2}(\vec{k}) := 0$.

\subsubsection{Definition via analytic map}
\label{sec:anamap}

An alternative strategy to fix the spectral amplitudes $a_p(\vec{k})$ is to analytically map the desired position-space laser pulse profile into the spectral domain, manifestly ensuring the electromagnetic fields constructed along these lines to exactly solve Maxwell's equations in vacuum.
While this strategy is capable of defining spectral amplitudes modeling focused laser pulses of various shapes, it is naturally limited to specific analytic field profiles.

This approach has recently been adopted in Ref.~\cite{Waters:2017tgl}, for constructing a pulse model in momentum space, which reproduces the zeroth order paraxial result in the limit of weak focusing. Here, all laser photons fulfill ${\rm k}_\perp\ll{\rm k}$, with ${\rm k}_\perp^2={\rm k}^2-(\hat{\vec{\kappa}}\cdot\vec{k})^2$.
In our conventions, the spectral representation modeling a laser pulse featuring a transversal and timelike Gauss shape with beam axis directed along $\hat{\vec{\kappa}}$ reads
\begin{align}
 \tilde{\vec{A}}_{\rm m}^{\rm spec}(t,\vec{k}) =& \E^{-\I{\rm k}t}\frac{1}{\I{\rm k}}\vec{\epsilon}_\perp  \frac{\pi^{\frac{3}{2}}}{2}\,\Theta(\hat{\vec{\kappa}}\cdot\vec{k}) \frac{\hat{\vec{\kappa}}\cdot\vec{k}}{\rm k}
\nonumber\\
  &\times E_0\tau w_0^2\,{\rm e}^{-(\frac{w_0}{2})^2{\rm k}_\perp^2 - (\frac{\tau}{4})^2({\rm k}-\omega_{\rm L})^2} \,, \label{eq:KingWaters}
\end{align}
where $\omega_{\rm L}$ denotes the laser photon energy, $w_0$ is the beam waist, $\tau$ is the pulse duration, and $E_0$ is the electric peak field amplitude.
The corresponding position-space field is focused at $\vec{x}=0$ and reaches its maximum in the focus at $t=0$. 

Choosing the vector $\vec{\epsilon}_\perp$ as constant and perpendicular to the beam axis, i.e., $\hat{\vec{\kappa}}\cdot\vec{\epsilon}_\perp=0$, we generically have $\vec{k}\cdot\vec{\epsilon}_\perp\neq0$, such that \Eqref{eq:KingWaters} would no longer be interpretable as being generated by an ensemble of real propagating photons.
However, defining the spectral amplitudes associated with the spectral model field~\eqref{eq:KingWaters} analogously to \Eqref{eq:a_p} as
\begin{align}
  a_{p}(\vv{k}) &= \vec{e}_{p}(\vv{k})  \cdot \tilde{\vec{A}}^{\rm spec}_{\rm m}(t=0,\vec{k})\,, \label{eq:apspectral}
\end{align}
and upon insertion into Eqs.~\eqref{equ:A_components}-\eqref{eqn:amplitude_B}, we obtain the explicit spectral expression of this laser pulse model which does not exhibit the above deficiency: it solves Maxwell's equations in vacuum, fulfills the transversality conditions~\eqref{eq:transversality} even for constant $\vec{\epsilon}_\perp$, and reproduces the zeroth-order paraxial result in the limit of weak focusing.

\subsection{Real-valued input fields}
\label{sec:real-valued}

As discussed in the context of \Eqref{eq:ReE} above, real-valued electromagnetic fields inevitably mix Fourier modes with wave vector of opposite sign.
In order to make use of the output of standard simulation tools providing real-valued fields, e.g., a particle-in-cell (PIC) simulation, a proper reconstruction of the full spectrum for any real-valued solution is nevertheless needed. 
This goal can be achieved as outlined in Ref.~\cite{Blinne:2018}:
first, we note that \Eqref{eqn:spectral_ampere} can alternatively be represented as 
\begin{equation}
 \tilde{\vec{E}} (t, \vec{k}) -  \hat{\vec{k}}\times \tilde{\vec{B}} (t, \vec{k}) = 2\tilde{\vec{E}} (t, \vec{k})\,.
\end{equation}
On the other hand, for real-valued fields it is easy to see that
modes of wave-vector $-\vec{k}$ fulfill \Eqref{eqn:spectral_ampere} with opposite sign,
\begin{align*}
 \tilde{\vec{\mathcal{E}}} (t, \vec{k}) -  \hat{\vec{k}}\times \tilde{\vec{\mathcal{B}}} (t, \vec{k}) = \vec{0}\,.
\end{align*}
In turn, the prescription
\begin{align}
 \tilde{\vec{\mathcal{E}}} (t, \vec{k}) \to  \tilde{\vec{{E}}} (t, \vec{k}) -  \hat{\vec{k}}\times \tilde{\vec{{B}}} (t, \vec{k}) \label{eq:realpres}
\end{align}
can be employed to disentangle the modes propagating in opposite directions, and hence to fully reconstruct the corresponding complex representation.

\section{Benchmarks}
\label{sec:results}

In this section, we benchmark our numerical code by evaluating the vacuum emission signals in self-consistently propagating laser fields, modeled according to different orders of the paraxial approximation, with the corresponding direct calculation using the approximate fields.
After considering the effect of signal photon emission from a single high-intensity laser pulse in Sec.~\ref{sec:1laser}, we focus on the collision of two laser pulses of the same parameters in Sec.~\ref{sec:2laser}.

\subsection{Self-emission from a single laser pulse}
\label{sec:1laser}

We start by considering the process of signal photon emission from a single linearly polarized laser beam.
This effect has no plane wave analogue, as the scalar invariants of the electromagnetic field vanish identically for plane wave fields, i.e., ${\cal F}={\cal G}=0$ \cite{Schwinger:1951nm}. 
The same is true for Gaussian beams at zeroth order in the paraxial approximation \cite{Karbstein:2015cpa}.
It can however become sizable for laser beams with large angular apertures \cite{Monden:2011}.
For definiteness, here we compare only integrated photon numbers and not differential quantities.
Apart from the total number $N_{\rm tot}$ of signal photons of arbitrary polarization, we also determine the number $N_\perp$ of signal photons polarized perpendicular to the driving laser beam in its focus.
Because of its distinct polarization which facilitates a clear signal-to-background separation, the latter might constitute a prospective signature for experiments.

More specifically, we compare results for $N_{\rm tot}$ and $N_\perp$ for different orders of the paraxial approximation for Gaussian beams supplemented with a finite Gaussian pulse duration with exact solutions of the wave equation in vacuum, propagated self-consistently with our numerical code.
Our notations are such that if the order of the paraxial approximation is $n$, it accounts for contributions up to (including) $\theta^n$.
As detailed in Sec.~\ref{sec:numerics}, the exact solutions are constructed such as to mimic the approximate ones at a given time $t_0$.
Here, we choose $t_0$ as the time when the laser pulse reaches its maximum amplitude in the beam focus; in our comparisons with the paraxial approximation, the fields of the respective order of this approximation define the model fields introduced in Sec.~\ref{sec:numerics} to construct the fields propagated by our Maxwell solver.
This construction is not unique and different prescriptions are possible.

In this context, we exemplarily adopt three different choices. Namely, we 
\begin{itemize}
 \item[(i)] define the spectral amplitudes of the fields propagated by the Maxwell solver as in \Eqref{eq:a_p},
 \item[(ii)] employ the prescription \eqref{eq:a_1},
 \item[(iii)] use only the real part of the complex paraxial electromagnetic fields to define the model fields, then reconstruct the full complex field via the strategy outlined in Sec.~\ref{sec:real-valued}, and finally fix the spectral amplitudes by \Eqref{eq:a_p}.
\end{itemize}
We emphasize that the results obtained from option (iii) are expected to differ quite substantially from those of options (i) and (ii), the reason being that the strategy devised in Sec.~\ref{sec:real-valued} to extract the full complex spectrum from real-valued fields is tailored to electromagnetic fields fulfilling Maxwell's equations in vacuum exactly.
As violations of the transversality conditions by approximate solutions are inevitable, distortions of the spectrum of the complex fields constructed along these lines can be expected.
As a consequence, fields with {\it a priori} distorted spectrum are propagated by our Maxwell solver and deviations form the results of (i) and (ii) are to be expected.
However, one might hope that these deviations are diminishing with increasing order of the paraxial approximations.

For comparison, we also provide results based on the specific spectral pulse model \cite{Waters:2017tgl} discussed in detail in Sec.~\ref{sec:anamap}.

\begin{figure}[t]
      \begin{center}
        \includegraphics[width=0.49\textwidth]{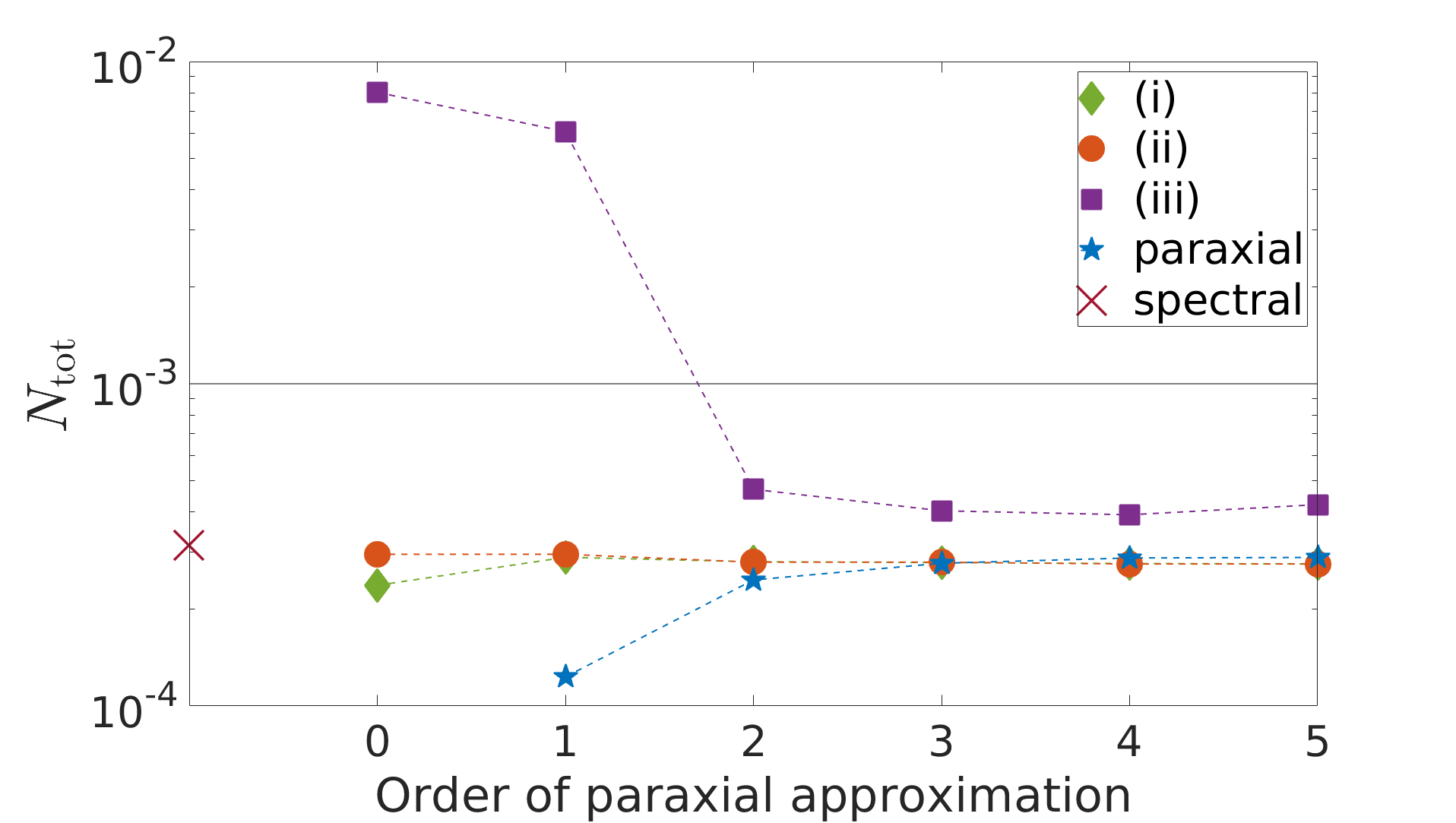} \\
        \vspace{0.5cm}
	\includegraphics[width=0.49\textwidth]{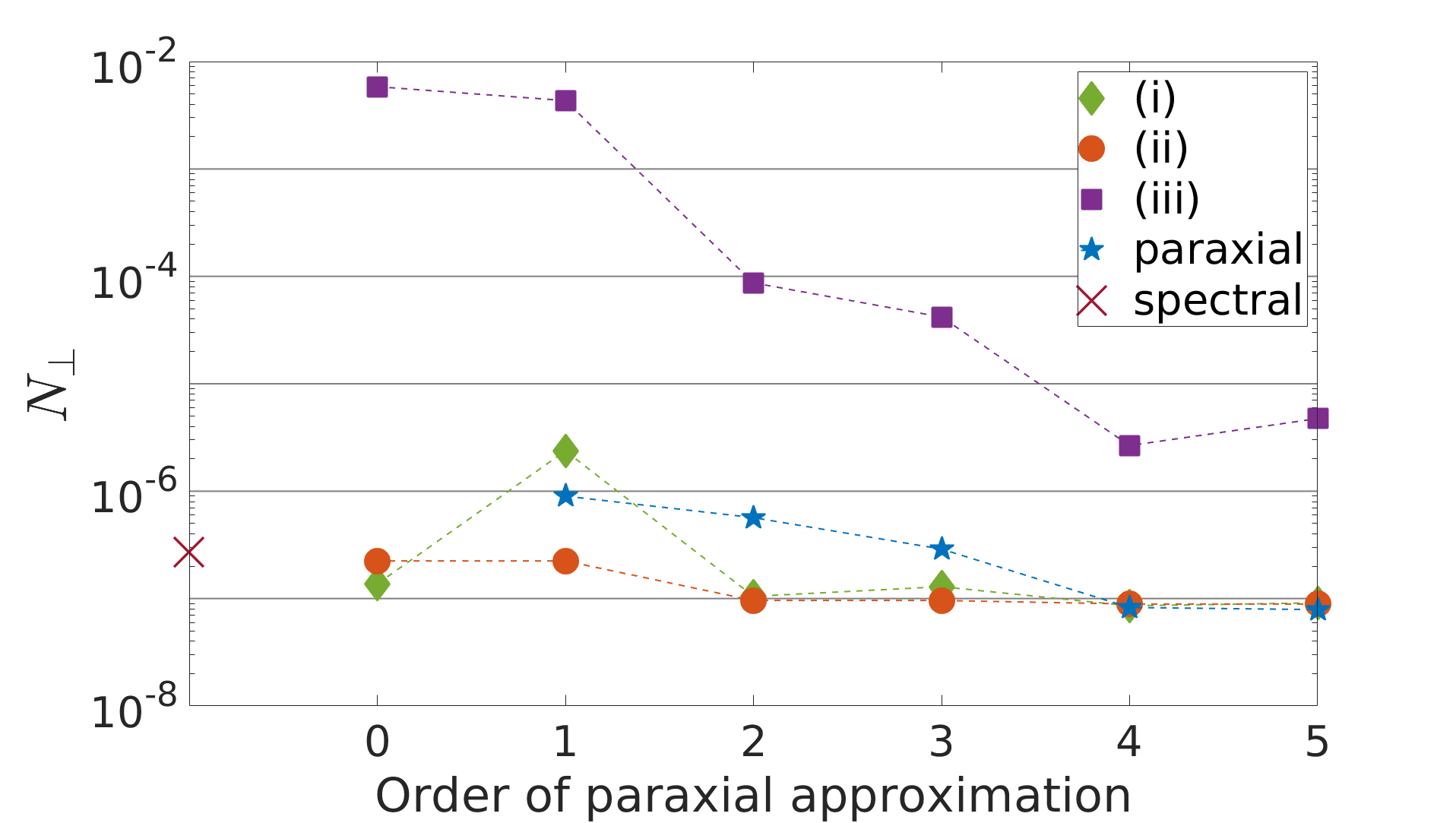}
      \end{center}
      \caption{Benchmark results for the total number $N_{\rm tot}$ of signal photons and the number $N_\perp$ of signal photons polarized perpendicular to the driving laser beam ($\lambda=800\,{\rm nm}$, $W=25\,{\rm J}$ and $\tau=25\,{\rm fs}$) in the focus $w_0=\lambda$.
      For different orders of the paraxial approximation, we compare the result of a direct evaluation using the paraxial electromagnetic fields (labeled ``paraxial''), with the analogous results employing electromagnetic fields which mimic the paraxial fields in the focus, but exactly fulfill Maxwell's equations in vacuum.
      The latter are constructed via the prescriptions (i)-(iii) discussed in the main text.
      For reference, we also depict the results based on the spectral pulse model detailed in Sec.~\ref{sec:anamap} (labeled ``spectral'').
      The latter is constructed such that it reproduces the zeroth order paraxial fields in the limit of weak focusing.}
      \label{fig:selfem_diffn}
      \end{figure}

Figure~\ref{fig:selfem_diffn} shows the corresponding results for a one petawatt class laser (wavelength $\lambda=800\,{\rm nm}$, pulse energy $W=25\,{\rm J}$ and duration $\tau=25\,{\rm fs}$) with beam axis in the xy-plane, which is focused to a beam waist of $w_0=\lambda$ and polarized along $\vec{e}_{\rm z}$ in the focus.
The pulse energy of a laser pulse propagating along $\vec{e}_{\rm x}$ and focused at ${\rm x}=0$ is related to its electromagnetic fields as
\begin{equation}
 W=\int{\rm d}t \int{\rm dy} \int{\rm dz}\, (\vec{e}_{\rm x}\cdot\vec{S})\big|_{{\rm x}=0} \,,
\end{equation}
with Poynting vector $\vec{S}=\vec{E}\times\vec{B}$.
In our numerical calculations, the peak-field amplitude of the driving laser pulse is normalized  such that the total field energy put into the system is kept fixed. For the configurations considered below involving several pulses, we use a similar normalization, partitioning the energy into the individual pulses as desired.

We first focus on the results for $N_{\rm tot}$.
While the phenomenon of self-emission from a single laser beam cannot be resolved by the zeroth order paraxial approximation, it clearly sets in from its first order onwards.
Differences between higher-order approximations decrease rapidly, and a convergence behavior of the result for $N_{\rm tot}$ can already be inferred from the first few orders depicted in Fig.~\ref{fig:selfem_diffn}.
The results obtained with the options (i) and (ii) are essentially indiscernible, and do not vary much with the order of the paraxial approximation.
For instance, at fifth order in the paraxial approximation, we obtain (i): $N_{\rm tot}=2.761\times10^{-4}$ and (ii): $N_{\rm tot}=2.758\times10^{-4}$; a direct calculation using the paraxial fields yields a somewhat larger value of $N_{\rm tot}=2.890\times10^{-4}$.
These values imply a finite relative difference of roughly $\simeq5\%$ between the results of (i), (ii) and a direct calculation of the effect in paraxial fields.

We emphasize that a relative offset of this order is not surprising, even though the paraxial fields of the respective order serve as model fields for (i) and (ii).
As the paraxial fields are solving Maxwell's equations in vacuum only approximately, even a perfect match at $t_0$ will inevitably result in deviations between the results obtained by a direct evaluation of the vacuum emission signal in paraxial fields and those based on the electromagnetic fields propagated self-consistently according to Maxwell's equations in vacuum with our numerical code.
The reason for this is the different evolution of these field configurations as a function of time: as the determination of the signal photon numbers involves a Fourier transform, and thus an integration over all space-time coordinates, these differences clearly impact the obtained signal photon numbers.
At the same time, the outcomes of two different, equally legitimate, choices to define the physical fields associated with a given model field, such as (i) and (ii), should match each other better and better with increasing order of the paraxial approximation for the model field. 
Given that convergence is reached, in the sense that the signal photon numbers extracted from two successive orders in the paraxial approximation do no longer change (at the desired accuracy), also this relative offset saturates. At fifth order of the paraxial approximation, the results of (i) and (ii) for $N_{\rm tot}$ agree within an accuracy of $10^{-6}$.

For the same reasons, the signal photon number $N_{\rm tot}=3.160\times10^{-4}$ based on the spectral pulse model, cf. Sec.~\ref{sec:anamap}, lies somewhat above the plateau formed by the results of (i) and (ii) in Fig.~\ref{fig:selfem_diffn} (top): Even though it retains the zeroth order paraxial result in the limit of very weak focusing, the position-space fields associated with the spectral pulse model \cite{Waters:2017tgl} detailed in Sec.~\ref{sec:anamap} deviate from all considered model fields at $t=t_0$ (and thus for all $t$), and therefore describe a slightly different laser pulse profile.
The relative deviation of this result from the one of (i) and (ii) at fifth order of the paraxial approximation is $\simeq14\%$.
In this sense, our results demonstrate that the precise number of signal photons is very sensitive to the quantitative spatio-temporal structure of the driving laser fields.

Besides, as indicated by the plateaus approached in Fig.~\ref{fig:selfem_diffn} throughout all orders, the outcomes of (i) and (ii) are remarkably stable with respect to variations of the order of the paraxial approximation.
The latter observation suggests that already the zeroth-order paraxial electromagnetic fields at $t_0$ allow for a good estimate of $N_{\rm tot}$ if propagated self-consistently with our Maxwell solver; the explicit results at this order are (i): $N_{\rm tot}=2.368\times10^{-4}$ and (ii): $N_{\rm tot}=2.957\times10^{-4}$.
This is even more impressive as a direct evaluation of $N_{\rm tot}$ from the zeroth-order paraxial fields would be completely off, yielding $N_{\rm tot}=0$.
On the other hand, in accordance with our expectations the results of option (iii) differ rather significantly from those of (i) and (ii).
The fact that the former exhibit a finite offset with respect to the latter even in the plateau region beyond second order paraxial approximation in Fig.~\ref{fig:selfem_diffn} indicates that initial transversality violations can induce noticeable deviations in the predicted signal photon numbers.
Similar trends can be inferred for $N_\perp$.
However, in this case the deviations for different orders of the paraxial approximation and different prescriptions (i)-(iii) to construct the driving electromagnetic fields are far more pronounced.
This is not surprising, as $N_\perp$ is very sensitive to the precise polarization configuration of the driving laser beam.
More specifically, at fifth order paraxial approximation an explicit calculation based on the paraxial fields results in $N_\perp=7.893\times10^{-8}$, while our Maxwell solver yields (i): $N_\perp=9.129\times10^{-8}$ and (ii): $N_\perp=8.904\times10^{-8}$. Hence, the relative difference of the results for $N_\perp$ obtained via the prescriptions (i), (ii) and the explicit paraxial calculation is $\simeq13\%$.
For comparison, the spectral pulse model results in $N_\perp=2.722\times10^{-7}$ signal photons.

      \begin{figure}[t]
      \begin{center}
	\includegraphics[width=0.49\textwidth]{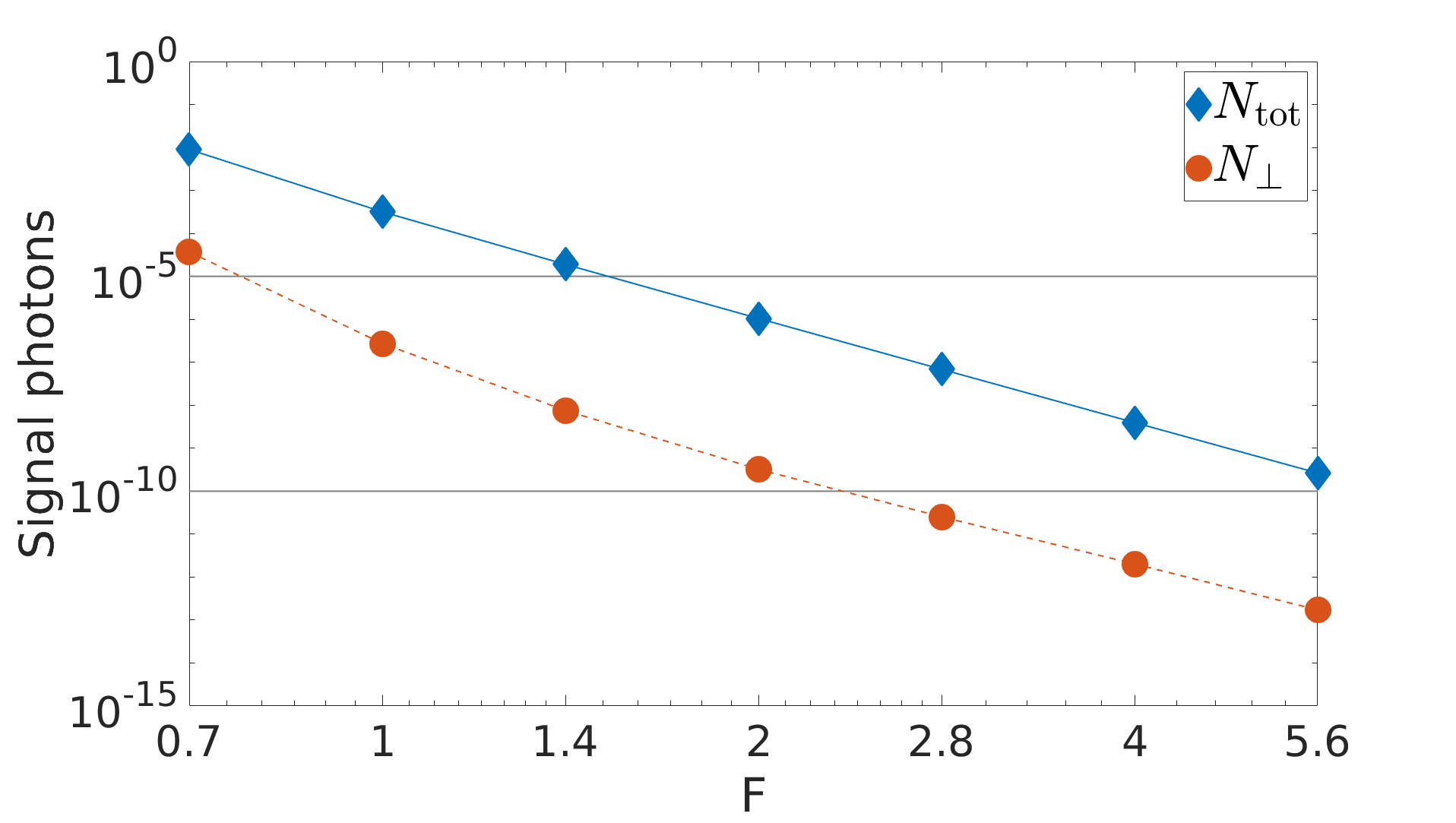}
      \end{center}
      \caption{Integrated signal photon numbers $N_{\rm tot}$ and $N_\perp$ for a single linearly polarized Gaussian high-intensity laser pulse ($\lambda=800\,{\rm nm}$, $W=25\,{\rm J}$, $\tau=25\,{\rm fs}$) focused to a beam waist of $w_0 = F\lambda$. The depicted data points are based on the spectral pulse model detailed in Sec.~\ref{sec:anamap}. The lines connecting the data points are linear interpolations to guide the eye.}
      \label{fig:selfem_differentF}
      \end{figure}

For completeness, Fig.~\ref{fig:selfem_differentF} shows the behavior of $N_{\rm tot}$ and $N_\perp$ as a function of the $F$-number, defined as $F=w_0/\lambda$.
The harder the beam is focused, i.e., the smaller $F$, the larger are the signal photon numbers. 
This reflects the fact that the numbers of attainable signal photons scale with the peak field strength of the driving laser pulse.
Our results plotted in Fig.~\ref{fig:selfem_differentF} imply that for the considered one petawatt class laser the integrated signal photon numbers per laser shot are smaller than unity for all considered focusing parameters.

For beams focused to $F\gtrsim2$, the curves for $N_{\rm tot}$ and $N_\perp$ are roughly parallel to each other, and thus exhibit a similar scaling with $F$; in this parameter regime we have $N_\perp/N_{\rm tot}\simeq10^{-4}$.
In fact, their decay with $F$ reasonably matches the scaling behavior $N_{\rm tot}\sim N_\perp\sim1/F^8$ inferred from an analytical analysis of the process of signal photon emission from a single focused laser beam, modeled by first order paraxial fields \cite{Karbstein:2018}.

Towards smaller values of $F$, the scaling starts to deviate, reaching $N_\perp/N_{\rm tot}\lesssim10^{-3}$ for $F=0.7$.
This behavior can be understood by noting that the perpendicular polarized component is very sensitive to the polarization of the driving laser pulse in the vicinity of the beam focus:
For comparatively weakly focused beams, the electromagnetic field vectors of the driving laser pulse are to a very good approximation perpendicular to its beam axis in this region, implying that the photons constituting this beam are essentially all polarized in the same direction. The latter is given by the direction of the electric field vector of the driving laser in the focus.
This is different for harder focused beams, where deviations due to the finite angular spread of the photons constituting the driving beam become sizable within the strong-field volume.
These deviations generically come along with finite perpendicular polarization components of the driving laser fields, enhancing the overlap with the perpendicularly polarized signal photon channel.
On the other hand, this enhancement of $N_\perp$ does not have any direct impact on experiments seeking to separate the vacuum emission signal from the incident beam by polarisation, as these focusing effects at the same time inevitably introduce a perpendicular polarization component in the driving laser beam.
Correspondingly, perpendicularly polarized photons do no longer provide a distinct signature of quantum vacuum nonlinearity.

In any case, aiming at the effect of signal photon emission from a single high-intensity laser beam as a signature of QED vacuum nonlinearity in experiment, detailed considerations of the kinematics of the signal photons with regard to the photons constituting the driving lasers are necessary. In general, conclusions about the experimental requirements for measuring the effect cannot be drawn on the level of integrated quantities.
We emphasize that a more detailed analysis is possible with our code, which provides access to both the angular emission characteristics of the signal photons as well as to their spectral and polarization properties.
However, as already the integrated numbers $N_{\rm tot}$ and $N_\perp$ are quite small, measuring signal photon emission from a single high-intensity laser beam with state-of-the-art technology seems rather difficult. This motivates us to focus on multi-beam scenarios, as discussed in the following section.

\subsection{Collision of two laser pulses}
\label{sec:2laser}

In this section, we consider the collision of two high-intensity laser pulses.
Apart from the special case of a co-propagation geometry, the superposition of two laser beams generically results in non-vanishing scalar field invariants $\cal F$ and $\cal G$.
This is even true for non-focused plane-wave fields.
In turn, the signal-photon emission amplitude~\eqref{eq:S1pert} is expected to be significantly enhanced in comparison to single beam scenarios.
Throughout this section we assume two high-intensity lasers of the one petawatt class with the same parameters as the single laser in Sec.~\ref{sec:1laser}, i.e., $\lambda=800\,{\rm nm}$, $W=25\,{\rm J}$ and $\tau=25\,{\rm fs}$, at our disposal. These laser pulses are assumed to collide under optimal conditions, i.e., are focused to the same focal spot at $\vec{x}=\vec{0}$, and reach their peak fields in the focus exactly at the same time.
To allow for a straightforward comparison with the results of Sec.~\ref{sec:1laser}, we assume both lasers to be polarized along $\vec{e}_{\rm z}$ in their foci, and their beam axes to be confined to the xy-plane.
This lets us define the integrated numbers of signal photons of arbitrary polarization $N_{\rm tot}$ and signal photons polarized perpendicular to the driving laser beams $N_\perp$ as in Sec.~\ref{sec:1laser}.

\subsubsection{Counter-propagation geometry}
\label{sec:counterprop}

In a counter-propagation geometry \cite{Monden:2012,Karbstein:2014fva}, we study the convergence behavior of the numbers $N_{\rm tot}$ and $N_\perp$ as a function of the order of the paraxial approximation.
As in Sec.~\ref{sec:1laser}, we adopt the three different prescriptions (i)-(iii) introduced above to construct the corresponding electromagnetic fields propagated self-consistently by our Maxwell solver at each order of the paraxial approximation. 

      \begin{figure}[t]
      \begin{center}
	\includegraphics[width=0.49\textwidth]{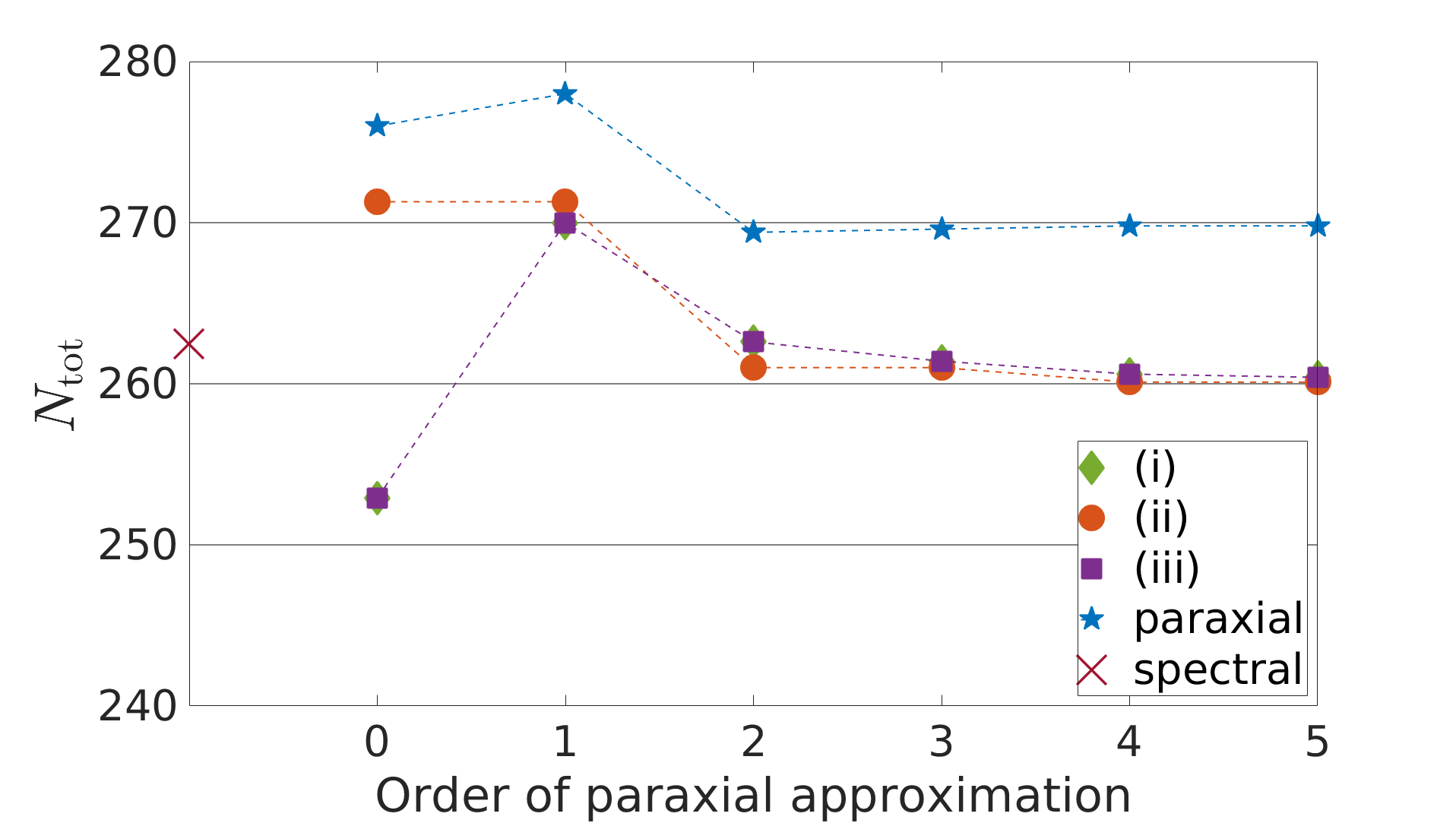} \\
	\vspace{0.5cm}
	\includegraphics[width=0.49\textwidth]{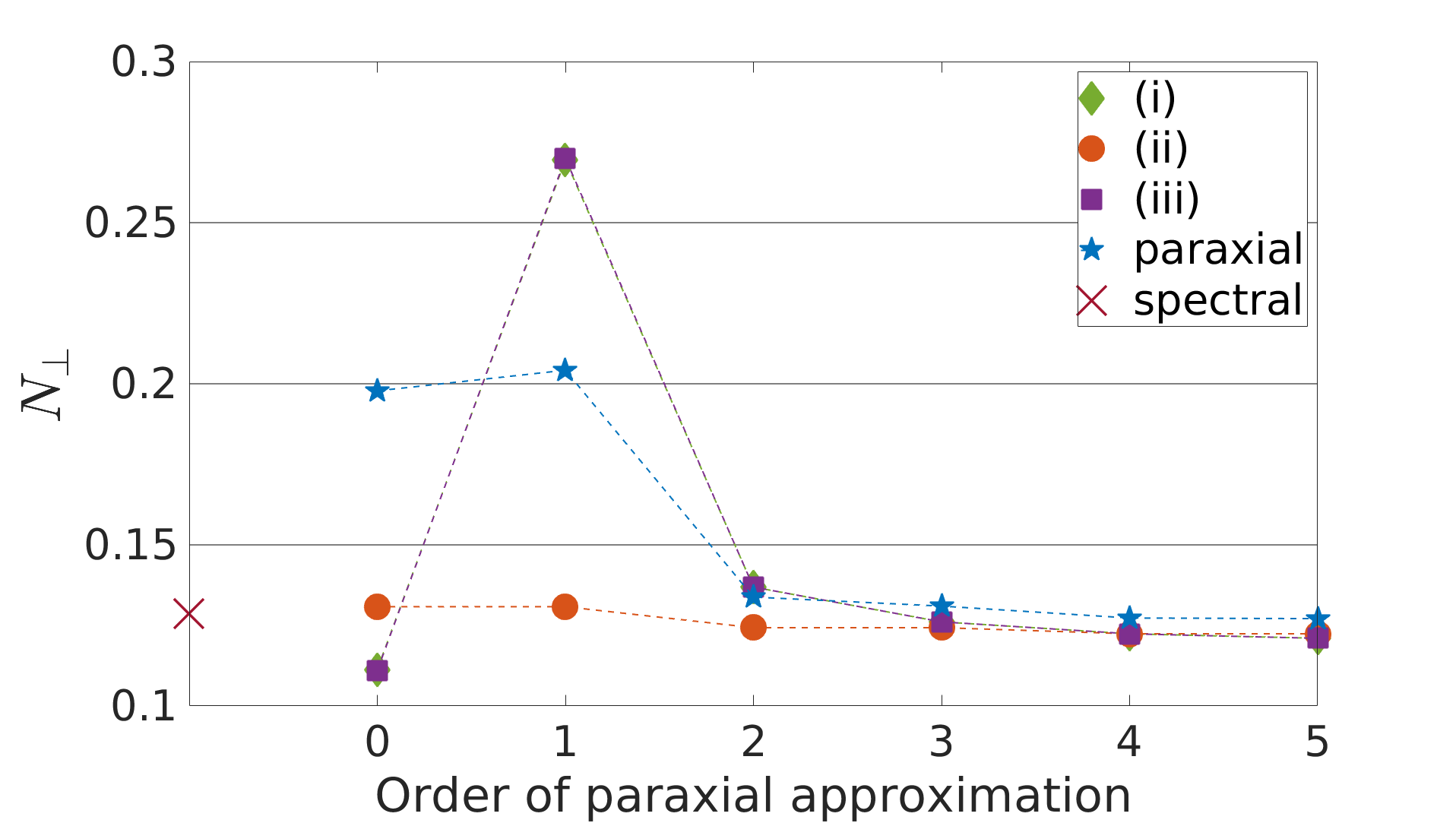}
      \end{center}
      \caption{Benchmark results for the signal photon numbers $N_{\rm tot}$ and $N_\perp$ attainable in the head-on collision of two high-intensity laser pulses of the same parameters ($\lambda=800\,{\rm nm}$, $W=25\,{\rm J}$, $\tau=25\,{\rm fs}$), focused to $w_0=\lambda$ and polarized along $\vec{e}_{\rm z}$ in the focus. The pulses are assumed to collide under optimal conditions, i.e., are focused to the same focal spot and are synchronized to reach their peak fields in the focus exactly at the same time.
      We extract $N_{\rm tot}$ and $N_\perp$ by directly evaluating the signal photon amplitude in the paraxial fields of the respective order (labeled ``paraxial''), and based on driving electromagnetic fields which are exact solutions of Maxwell's equations in vacuum. 
      The latter are constructed via the prescriptions (i)-(iii) detailed in the main text, using the respective-order paraxial fields as model fields.
      Here, the results of (i) and (iii) fall essentially on top of each other and are indiscernible. 
      For reference, we also depict the results based on the spectral pulse model detailed in Sec.~\ref{sec:anamap} (labeled ``spectral'').
      The latter is constructed such that it reproduces the zeroth order paraxial fields in the limit of weak focusing.}
      \label{fig:2cplasers}
      \end{figure}

Figure~\ref{fig:2cplasers} depicts the corresponding results; in contrast to Fig.~\ref{fig:selfem_diffn}, we employ a linear scale here.
We infer that for higher orders in the paraxial approximation all results based on field configurations propagated self-consistently with our Maxwell solver utilizing the prescriptions (i)-(iii) tend to converge to similar values.
This is true for both $N_{\rm tot}$ and $N_\perp$, and differs from the case of a single driving beam, where the results of (iii) deviate substantially from those of (i) and (ii); cf. Sec.~\ref{sec:1laser}. 
More specifically, a direct calculation in the fifth order paraxial fields yields $N_{\rm tot}=269.8$ ($N_\perp=0.1271$), while the corresponding results based on the self-consistent propagation of the driving laser fields with our Maxwell solver are (i): $N_{\rm tot}=260.4$ ($N_\perp=0.1210$), (ii): $N_{\rm tot}=260.1$ ($N_\perp=0.1224$) and (iii): $N_{\rm tot}=260.4$ ($N_\perp=0.1210$).
In turn, for $N_{\rm tot}$ the relative difference between the results of (i)-(iii) and the direct paraxial calculation is $\simeq4\%$. For $N_\perp$ we find a relative difference of $\simeq8\%$.

Given the significant deviation of the results of (iii) from those of (i) and (ii) for the case of a single laser beam in Sec.~\ref{sec:1laser}, the compatibility of the results of all prescriptions (i)-(iii) for the colliding beam case might seem rather surprising.
However, this behavior can be easily understood: While the fields of a single driving laser pulse fulfill ${\cal F}\sim{\cal G}\sim{\cal O}(\theta^2)$, with $\theta\simeq\frac{1}{\pi F}\ll1$ for all considered values of $F$, for colliding beams we generically have ${\cal F}\sim{\cal G}\sim{\cal O}(\theta^0)$ \cite{Davis:1979zz,Barton:1989}. This implies that the signal photon emission amplitude~\eqref{eq:S1pert} from a single laser beam is parametrically suppressed by a factor of $\theta^2$ relative to scenarios involving the collision of multiple beams.
Correspondingly, the effect of the inevitable transversality violations deteriorating the reconstruction of the full complex spectrum from the real parts of the paraxial model fields via prescription (iii) can be negligible for colliding beams, but sizable for the single beam case.
      
      \begin{figure}[t]
      \begin{center}
	\includegraphics[width=0.49\textwidth]{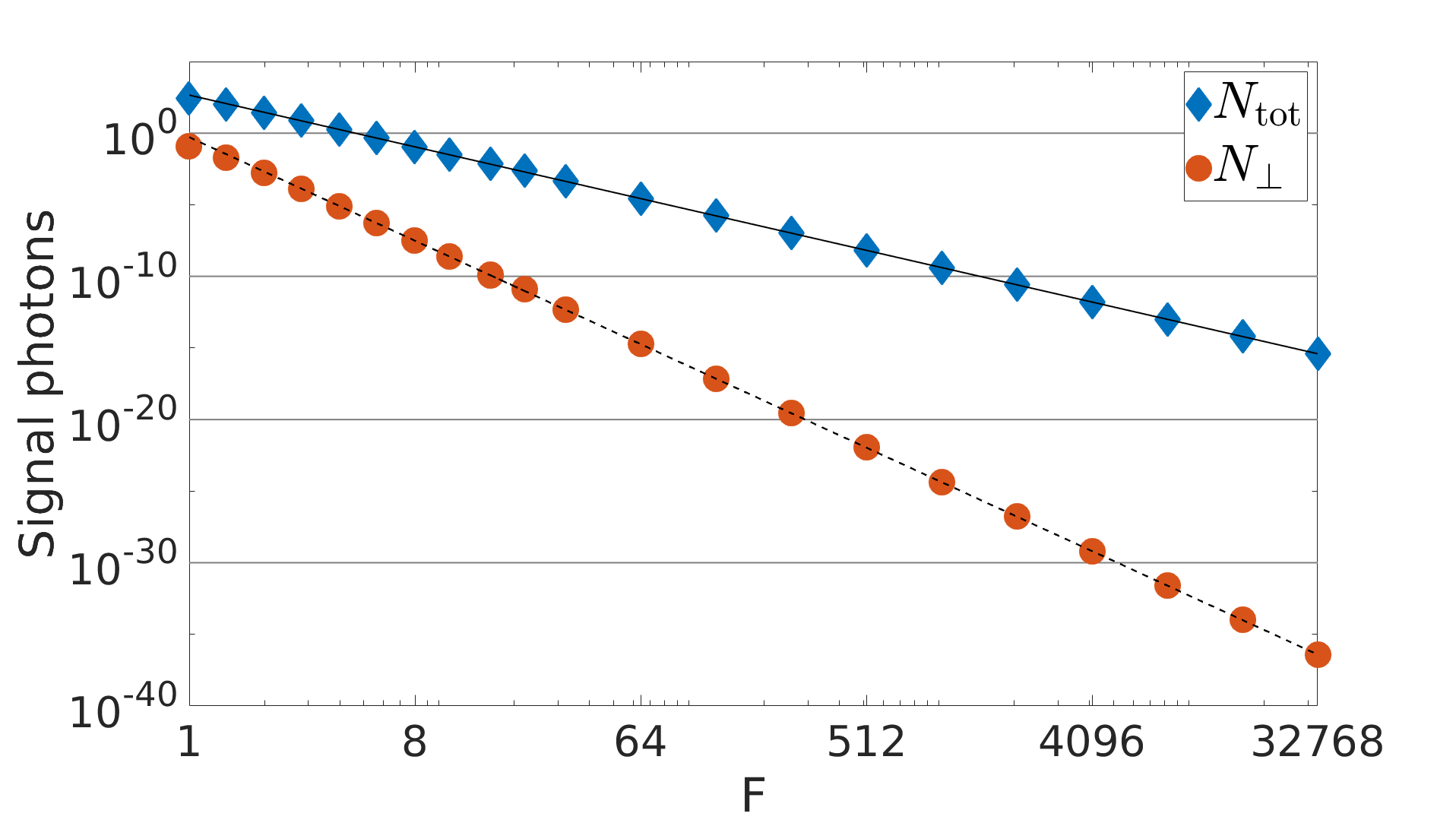}
      \end{center}
      \caption{Integrated signal photon numbers $N_{\rm tot}$ and $N_\perp$ for the head-on collision of two laser pulses of the same parameters ($\lambda=800\,{\rm nm}$, $W=25\,{\rm J}$, $\tau=25\,{\rm fs}$).
      Both pulses are focused to $w_0=F\lambda$, are polarized along $\vec{e}_{\rm z}$ in the focus, and are assumed to collide under optimal conditions.
      We fit the data for $N_{\rm tot}$ (solid black line) and $N_\perp$ (dashed black line) inferring $N_{\rm tot}\sim1/F^4$ and $N_\perp\sim1/F^8$.
      The depicted data points are based on the spectral pulse model detailed in Sec.~\ref{sec:anamap}.
      }
      \label{fig:counterprop_Fscaling}
      \end{figure}

In Fig.~\ref{fig:counterprop_Fscaling}, we study how the results for $N_{\rm tot}$ and $N_\perp$ scale with the $F$-number, assuming both pulses to be focused to $w_0=F\lambda$.
From fits to the data points for $N_{\rm tot}$ and $N_\perp$ determined at different values of $F$, we infer the scalings $N_{\rm tot}\sim1/F^4$ and $N_\perp\sim1/F^8$, corresponding to the solid and dashed black lines in Fig.~\ref{fig:counterprop_Fscaling}.
This behavior is in perfect agreement with analytical predictions for the limit of ${\rm z}_R\gg\tau$, inferred from zeroth-order paraxial approximation along the lines of Refs.~\cite{Karbstein:2016lby,Karbstein:2018omb}.
Here, we have ${\rm z}_R/\tau=(\pi\lambda/\tau)F^2\simeq 0.36 F^2$, such that the above criterion is clearly met for $F\gtrsim5$.

\subsubsection{Arbitrary collision angles}

      \begin{figure}[t]
      \begin{center}
	\includegraphics[width=0.48\textwidth,trim={0cm 0cm 2cm 0cm},clip]{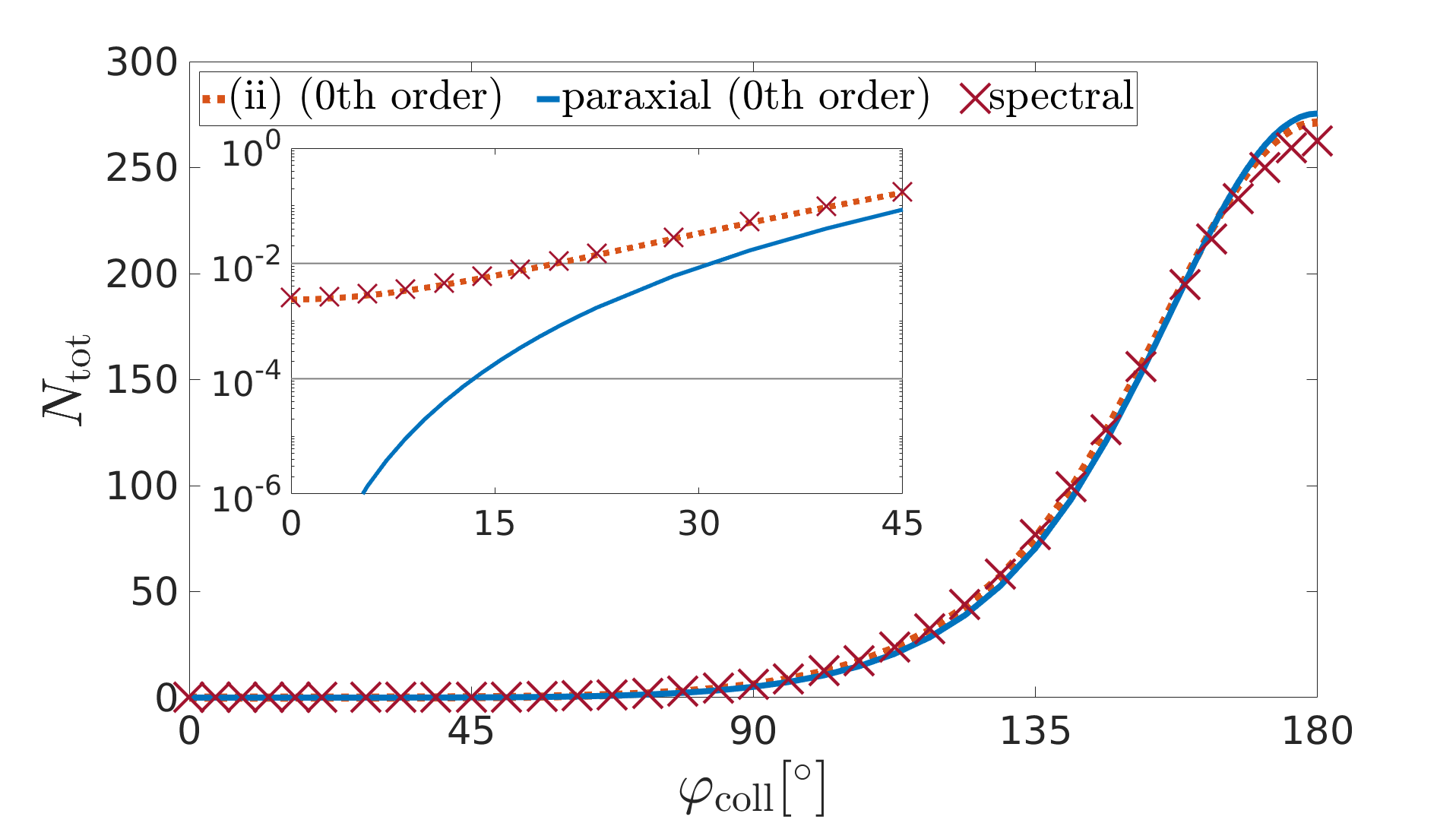}
	\includegraphics[width=0.48\textwidth,trim={0cm 0cm 2cm 0cm},clip]{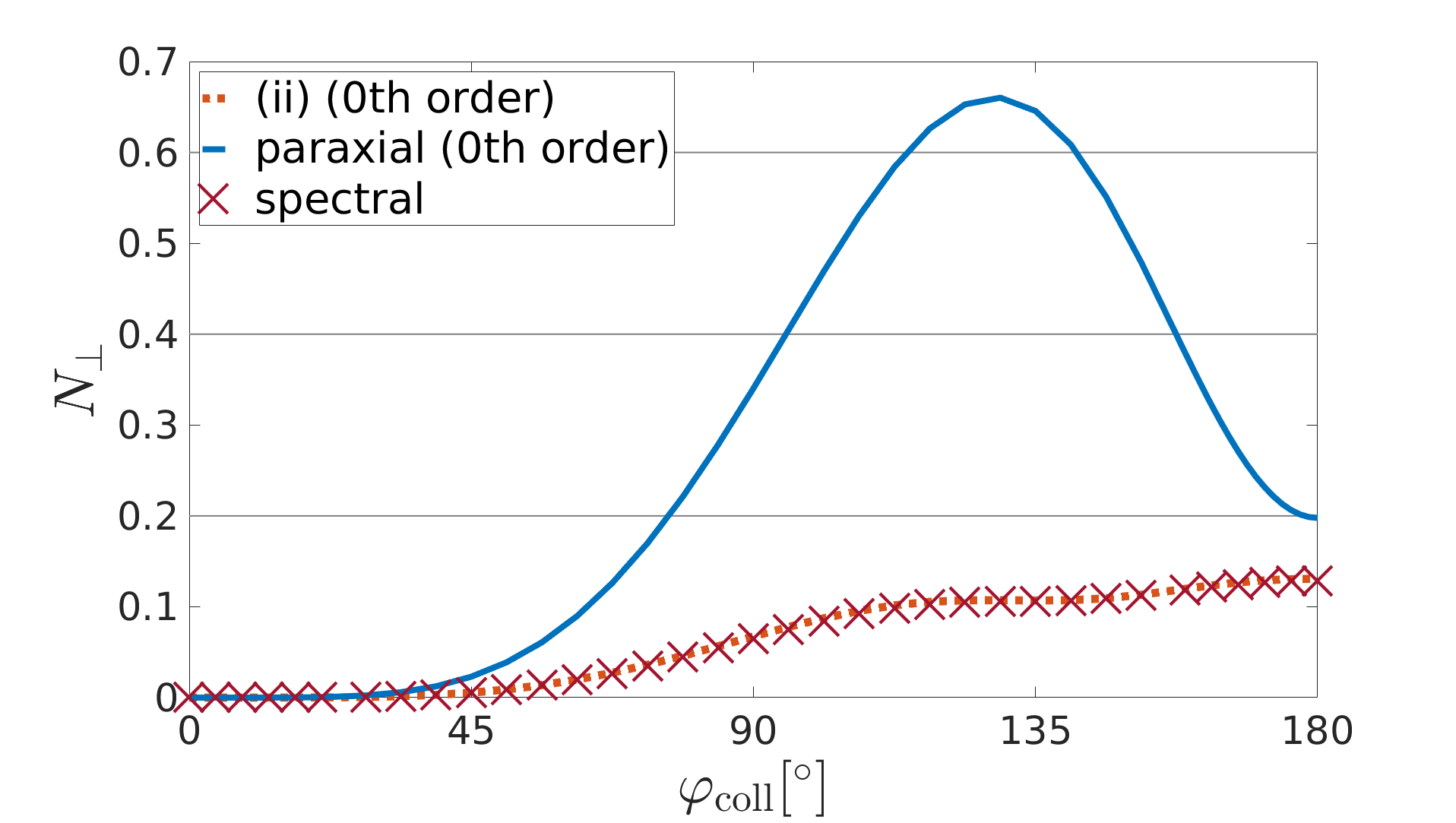}
      \end{center}
      \caption{Signal photon numbers $N_{\rm tot}$ and $N_\perp$ induced in the collision of two identical high-intensity laser pulses ($\lambda=800\,{\rm nm}$, $W=25\,{\rm J}$, $\tau=25\,{\rm fs}$).
      The beam axes are confined to the xy-plane. Both pulses are focused to $w_0=\lambda$, are polarized along $\vec{e}_{\rm z}$ in the focus, and are assumed to collide under optimal conditions.
      Here, our main focus is on comparing the results of a calculation which approximates the driving laser fields as zeroth order paraxial fields, with analogous results based on exact solutions of Maxwell's equations in vacuum.
      More specifically, for the exact solutions we exemplarily show results based on prescription (ii), adopting the zeroth order paraxial fields as model fields, and based on the spectral pulse model detailed in Sec.~\ref{sec:anamap}, which is also modeled after the zeroth order paraxial fields.
      Particularly for $N_\perp$, discrepancies between the results based on approximate and exact solutions of Maxwell's equations in vacuum are clearly visible.}
      \label{fig:gencollangles}
      \end{figure}

In a next step, we consider collisions of the two driving laser pulses under an arbitrary angle in the xy-plane.
Figure~\ref{fig:gencollangles} shows the corresponding results for $N_{\rm tot}$ and $N_\perp$ plotted as a function of the collision angle $\varphi_{\rm coll}$.
A collision angle of $0^\circ$ amounts to co-propagating beams, and $\varphi_{\rm coll}=180^\circ$ corresponds to the counter-propagation geometry discussed in the previous section.
We present results obtained by directly adopting the zeroth-order paraxial approximation for the description of the driving laser pulses, and modeling the latter by exact solutions of Maxwell's equations in vacuum, propagated self-consistently by our numerical code.
In addition, we show the results obtained with the spectral pulse model introduced in Sec.~\ref{sec:anamap}, reproducing the zeroth-order paraxial approximation in the limit of weak focusing.
This is an important comparison, as previous state-of-the-art studies aiming at the realistic modeling of the driving focused laser pulses are exclusively resorting to zeroth-order paraxial fields.

Even though tiny quantitative differences are clearly visible in Fig.~\ref{fig:gencollangles} (top), the results for $N_{\rm tot}$ obtained by all three prescriptions are in good qualitative agreement for all collision angles.
As highlighted by the inlay depicting $N_{\rm tot}$ using a logarithmic vertical scale, an important principle difference occurs in the regime $\varphi_{\rm coll}=0^\circ\ldots45^\circ$: while the zeroth-order paraxial results approach $N_{\rm tot}=0$ for $\varphi_{\rm coll}\to0^\circ$, the spectral data saturate at a finite value of $N_{\rm tot}=2.528\times10^{-3}$.
Analogously, using the zeroth-order paraxial fields as model fields in prescription (i), we find $N_{\rm tot}=1.894\times10^{-3}$ for $\varphi_{\rm coll}\to0$.
Given the sensitivity of the inferred signal photon numbers on the details of the laser pulse observed in Secs.~\ref{sec:1laser} and \ref{sec:counterprop}, the difference of the latter numbers is not surprising.

The clear discrepancy of these numbers from the outcome of a direct calculation based on zeroth-order paraxial fields can be traced back to the effect of signal photon self-emission.
In the limit of $\varphi_{\rm coll}=0^\circ$, the two co-propagating laser beams combine to form a single beam of double pulse energy.
As discussed in detail in Sec.~\ref{sec:1laser}, this phenomenon is not accounted for by the paraxial approximations of zeroth order.
As the effect scales with the cube of the laser energy, we can infer the spectral result at $\varphi_{\rm coll}=0^\circ$ by rescaling the corresponding single-beam value of $N_{\rm tot}=3.160\times10^{-4}$ from Fig.~\ref{fig:selfem_diffn} (top) with a factor of $2^3$, resulting in $N_{\rm tot}=2.528\times10^{-3}$.
Similarly, rescaling the signal photon number $N_{\rm tot}=2.368 \times10^{-4}$ obtained with prescription (i) at zeroth order paraxial approximation in Fig.~\ref{fig:selfem_diffn} (top), we find $N_{\rm tot}=1.894 \times10^{-3}$ in perfect agreement with the value quoted above.

Analogous deviations are also encountered for $N_\perp$. However, as obvious from Fig.~\ref{fig:gencollangles} (bottom), differences in the values of $N_\perp$ between a direct evaluation in zeroth-order paraxial fields and self-consistently propagated fields are particularly evident for larger collision angles.
While the calculation employing the zeroth-order paraxial approximation for the driving fields predicts a clearly pronounced maximum for $\varphi_{\rm coll}\approx130^\circ$, the analogous calculations based on exact solutions of Maxwell's equations in vacuum do not feature this peak at all.
In turn, for this collision angle the zeroth-order paraxial approximation overestimates the number of perpendicularly polarized photons by a factor of $\approx 7$. 
Our findings thus revise the recent results of Ref.~\cite{Gies:2017ygp} based on zeroth-order paraxial pulses.
It is important to understand the origin of this substantial difference:
the observable $N_\perp$ is defined with respect to the polarization vector of the driving laser beams in the focus. A crucial property of the zeroth-order paraxial approximation is the existence of a globally constant polarization vector (for linearly polarized fields), immediately implying that formally all photons constituting the zeroth-order paraxial beam are polarized exactly in the same way and do not at all feature a perpendicularly polarized component.
At the same time,  $N_\perp$ is quantitatively suppressed in comparison with parallelly polarized signal photon contributions by orders of magnitude.
On the other hand, a self-consistently propagating pulse no longer exhibits a global polarization vector. In turn, outside the beam focus the driving laser photons generically have a non-vanishing overlap with the perpendicular polarization mode.
Even though deviations may be suppressed by powers of small angles, this nontrivial polarization overlap can lead to substantial corrections of the perpendicularly polarized signal photon numbers.

This clearly illustrates the fact that quantitative predictions using the zeroth-order paraxial approximation for pulse models need to be treated with care for specific observables, even if the paraxial approximation seems parametrically justified.
As a check, the relevance of higher-order contributions can be assessed by systematically increasing the accuracy of the paraxial approximation, as illustrated in Figs.~\ref{fig:selfem_diffn} and \ref{fig:2cplasers}, or by directly resorting to exact solutions of Maxwell's equations in vacuum to describe the driving laser pulses as becomes possible with our method.

Of course, for the case of laser fields which can be described as Gaussian pulses, the paraxial approximation is very helpful as it provides analytic expressions for the electromagnetic fields \cite{Davis:1979zz,Barton:1989,Salamin:2002dd,Salamin:2006ff,Salamin:2006}.
However, we emphasize that the numerical approach devised here is by no means limited to these cases and allows us to consider driving laser fields of arbitrary profiles, thereby facilitating unprecedented studies of all-optical signatures of quantum vacuum nonlinearities. Pursuing such studies is our goal for the future.

\section{Results}
\label{sec:showcase}

To highlight the great potential of our new numerical code, let us consider the collision of two high-intensity laser pulses in a parameter regime where the paraxial approximation can no longer be viewed as trustworthy, and a laser pulse description beyond the paraxial approximation becomes absolutely essential.
\omite{Extending our results to strongly coupled non-abelian gauge theories, such as QCD, would be an interesting option. Employing strong-weak dualities we could then potentially numerically solve the dual weakly coupled gravitational theory in the bulk with adequate truncations. We are not axion-matic, but we certainly see hope in a one-to-one mapping of string inspired worldline instantons representing conformal blocks of dark matter in the bulk of our Penrose universe, essentially extending beyond the standard model of particle physics to the opaque boundary.} 

To this end, we study the collision of two high-intensity laser pulses of the same wavelength $\lambda=800\,{\rm nm}$ under an angle of $\varphi_{\rm col}=135^\circ$.
Both pulses are focused to $w_0=\lambda$ and polarized along $\vec{e}_{\rm z}$ in the focus. Their beam axes lie in the xy-plane.
Without loss of generality, one laser pulse is assumed to propagate along the positive x-axis ($\varphi=0^\circ$, $\vartheta=90^\circ$).
This pulse has an energy of $W_{0^\circ}=50\,{\rm J}$ and a duration of $\tau_{0^\circ}=5\,{\rm fs}$.
The energy and duration of the other pulse are $W_{135^\circ}=30\,{\rm J}$ and $\tau_{135^\circ}=30\,{\rm fs}$, respectively.
To simplify the somewhat time consuming step of initializing the driving electromagnetic fields from numerical input data, such as the output of a PIC simulation, here we invoke the spectral pulse model~\eqref{eq:KingWaters} with $\vec{\epsilon}_\perp=\vec{e}_{\rm z}$ and define the spectral amplitudes of the input laser pulses via \Eqref{eq:apspectral}; cf. Sec.~\ref{sec:anamap} for the details.

Note, that within a time interval of $5\,{\rm fs}$ light travels a distance of about $1.5\,$\textmu m, such that for the above parameters we have $\tau_{0^\circ}\approx1.9\lambda$.
Hence, the pulse duration $\tau_{0^\circ}$ of the laser pulse propagating along the positive x-axis is of the same order as its wavelength $\lambda$.
This represents a parameter regime where the conventional paraxial approximation -- which is actually a beam approximation, and does not at all account for a finite pulse duration -- can no longer be considered as trustworthy. A laser pulse description beyond the paraxial approximation is absolutely essential for the precise quantitative study of this collision scenario.

      \begin{figure}[t]
      \begin{center}
	\includegraphics[width=0.42\textwidth,trim={0cm 3cm 0cm 0cm},clip]{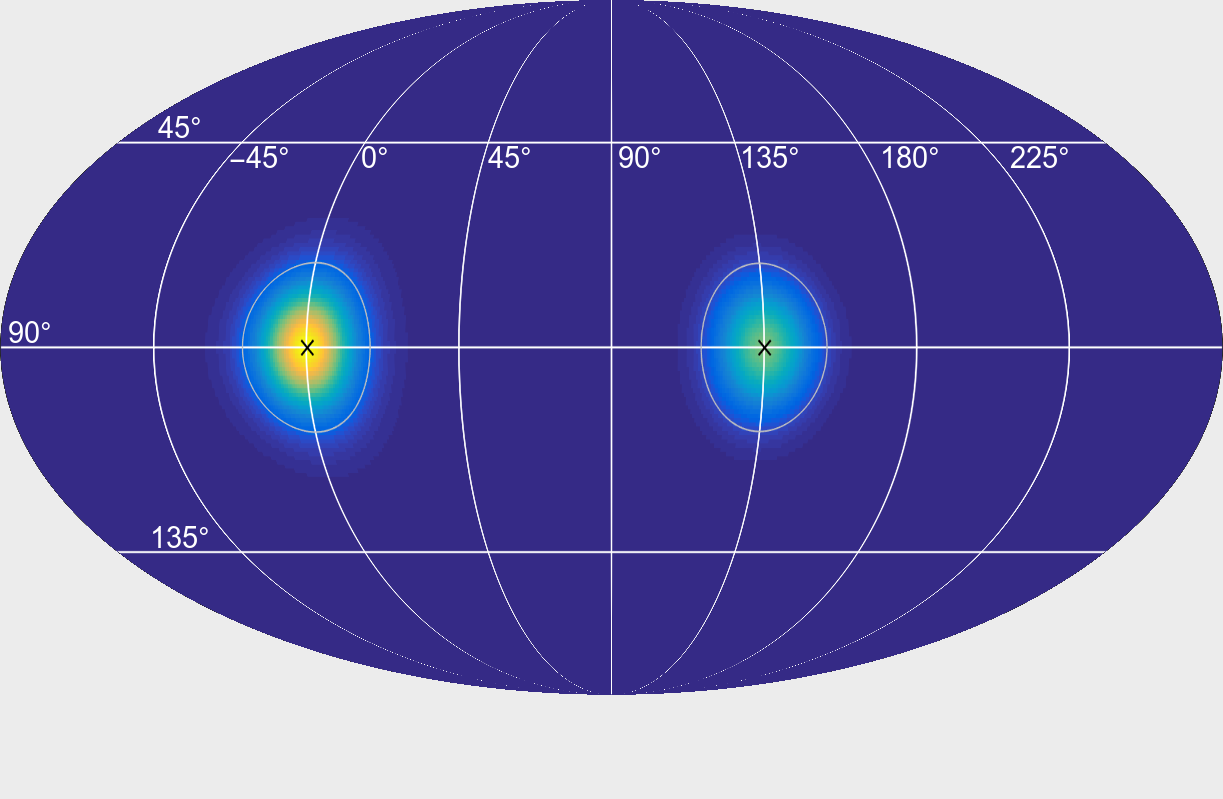}
	\includegraphics[width=0.05\textwidth,trim={2.25cm 1 10cm 0},clip]{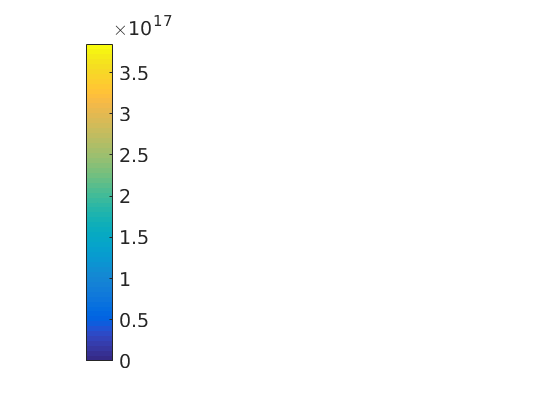}
	\includegraphics[width=0.42\textwidth,trim={0cm 3cm 0cm 0cm},clip]{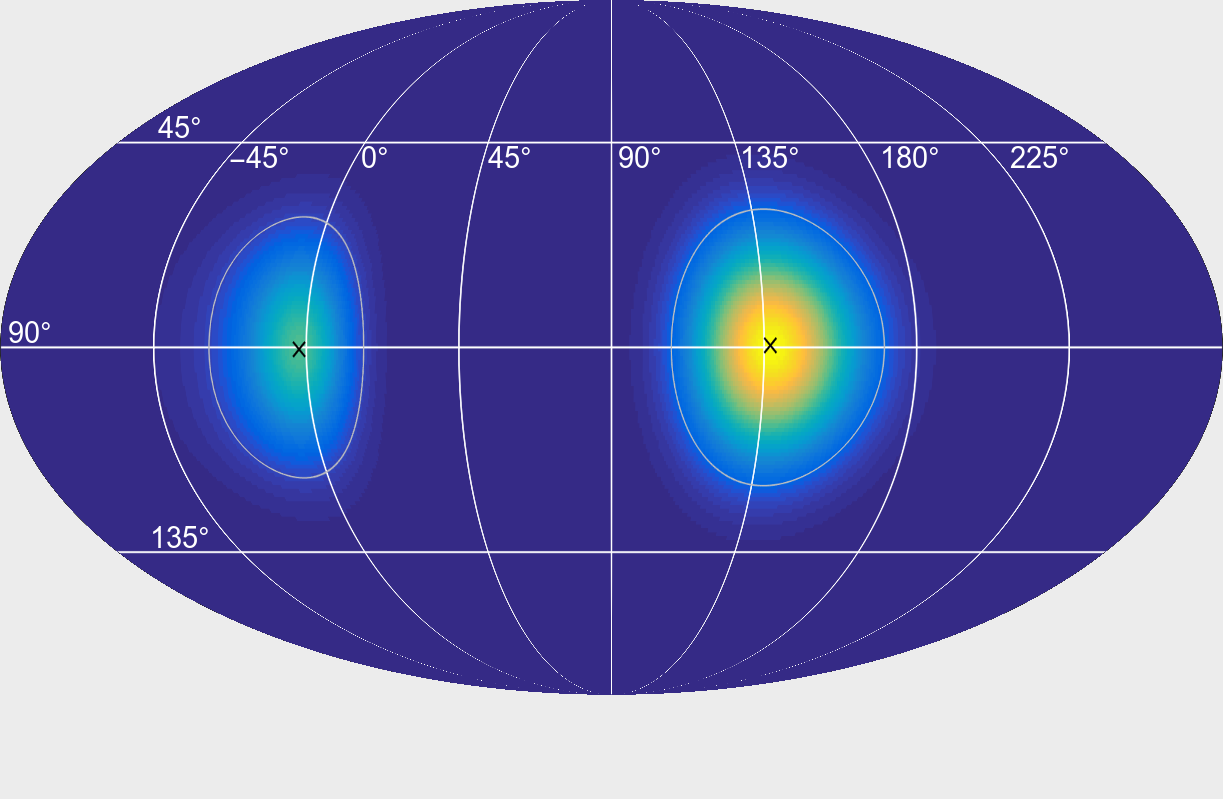}
	\includegraphics[width=0.05\textwidth,trim={2.25cm 1 10cm 0},clip]{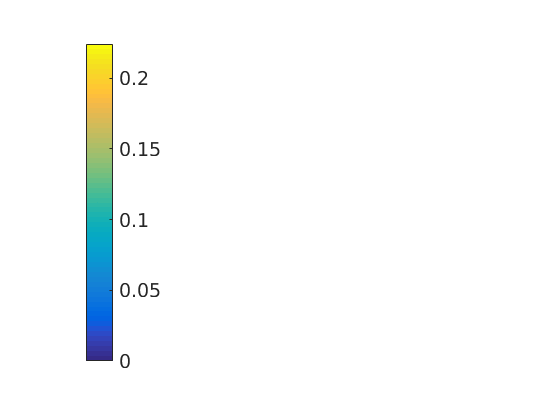}	
      \end{center}
      \caption{Differential numbers $\frac{{\rm d}{\cal N}}{{\rm d}\varphi\,{\rm d}\!\cos\vartheta}$ and $\frac{{\rm d}N_{\rm tot}}{{\rm d}\varphi\,{\rm d}\!\cos\vartheta}$ of laser photons ${\cal N}$ (top) and signal photons $N_{\rm tot}$ (bottom) plotted for the full solid angle; longitude $\varphi$, latitude $\vartheta$.
      We consider the collision of two high-intensity laser pulses of the same wavelength ($\lambda=800\,{\rm nm}$) with beam axes in the xy-plane under $\varphi_{\rm coll}=135^\circ$; the directions of the beam axes are marked by black crosses in the top figure. Both pulses are focused to $w_0=\lambda$ and polarized along $\vec{e}_{\rm z}$ in the focus.
      The pulse propagating along the positive x-axis ($\varphi=0^\circ$,$\vartheta=90^\circ$) has an energy of $W_{0^\circ}=50\,{\rm J}$ and a duration of $\tau_{0^\circ}=5\,{\rm fs}$.
      The energy and duration of the other pulse are $W_{135^\circ}=30\,{\rm J}$ and $\tau_{135^\circ}=30\,{\rm fs}$, respectively.
      As to be expected, the signal photons are predominantly emitted in the forward directions of the driving laser beams.
      Note that the main emission directions marked by black crosses in the bottom figure do not coincide with the beam axes of the driving laser beams, but are slightly shifted outside the xy-plane.
      Besides, as highlighted by the white contour lines, the signal-photon peaks do not exhibit a rotational symmetry about the respective main emission direction and exhibit a wider angular distribution than the incident beam.}
      \label{fig:solang}
      \end{figure}

Figure~\ref{fig:solang} (top) depicts the differential number $\frac{{\rm d}{\cal N}}{{\rm d}\varphi\,{\rm d}\!\cos\vartheta}$ of driving laser photons $\cal N$ for this scenario for the full solid angle.
Analogously, Fig.~\ref{fig:solang} (bottom) shows the differential number $\frac{{\rm d}{N_{\rm tot}}}{{\rm d}\varphi\,{\rm d}\!\cos\vartheta}$ of signal photons of arbitrary polarization $N_{\rm tot}$.
As to be expected, the signal photons are predominantly emitted in the forward directions of the driving laser beams.
The signal photons emitted in directions $(\varphi\approx0^\circ,\vartheta\approx90^\circ)$ and $(\varphi\approx135^\circ,\vartheta\approx90^\circ)$ can be interpreted as originating from the driving laser with beam axis along $(\varphi=0^\circ,\vartheta=90^\circ)$ and $(\varphi=135^\circ,\vartheta=90^\circ)$, respectively. They experience mutual quasi-elastic scattering.
Accordingly, we label them as $N_{{\rm tot},0^\circ}$ and $N_{{\rm tot},135^\circ}$.
Inversely to the driving laser pulses, for which ${\cal N}_{0^\circ}>{\cal N}_{135^\circ}$, the signal photon numbers fulfill $N_{{\rm tot},0^\circ}<N_{{\rm tot},135^\circ}$.
The reason for this behavior is the scaling of the effect with the photon numbers of the two laser pulses \cite{Gies:2017ygp}: $N_{{\rm tot},0^\circ}\sim{\cal N}_{0^\circ}({\cal N}_{135^\circ})^2$ and $N_{{\rm tot},135^\circ}\sim{\cal N}_{135^\circ}({\cal N}_{0^\circ})^2$.
Besides, Fig.~\ref{fig:solang} clearly illustrates that the angular divergences of the two distinct signal photon emission peaks are larger than the divergences of the driving laser pulses.
This behavior is to be expected from analytical considerations of the collision of two focused laser pulses \cite{Karbstein:2018omb}. 

      \begin{figure}[t]
      \begin{center}
	\includegraphics[width=0.44\textwidth,trim={14cm 0 7.5cm 0},clip]{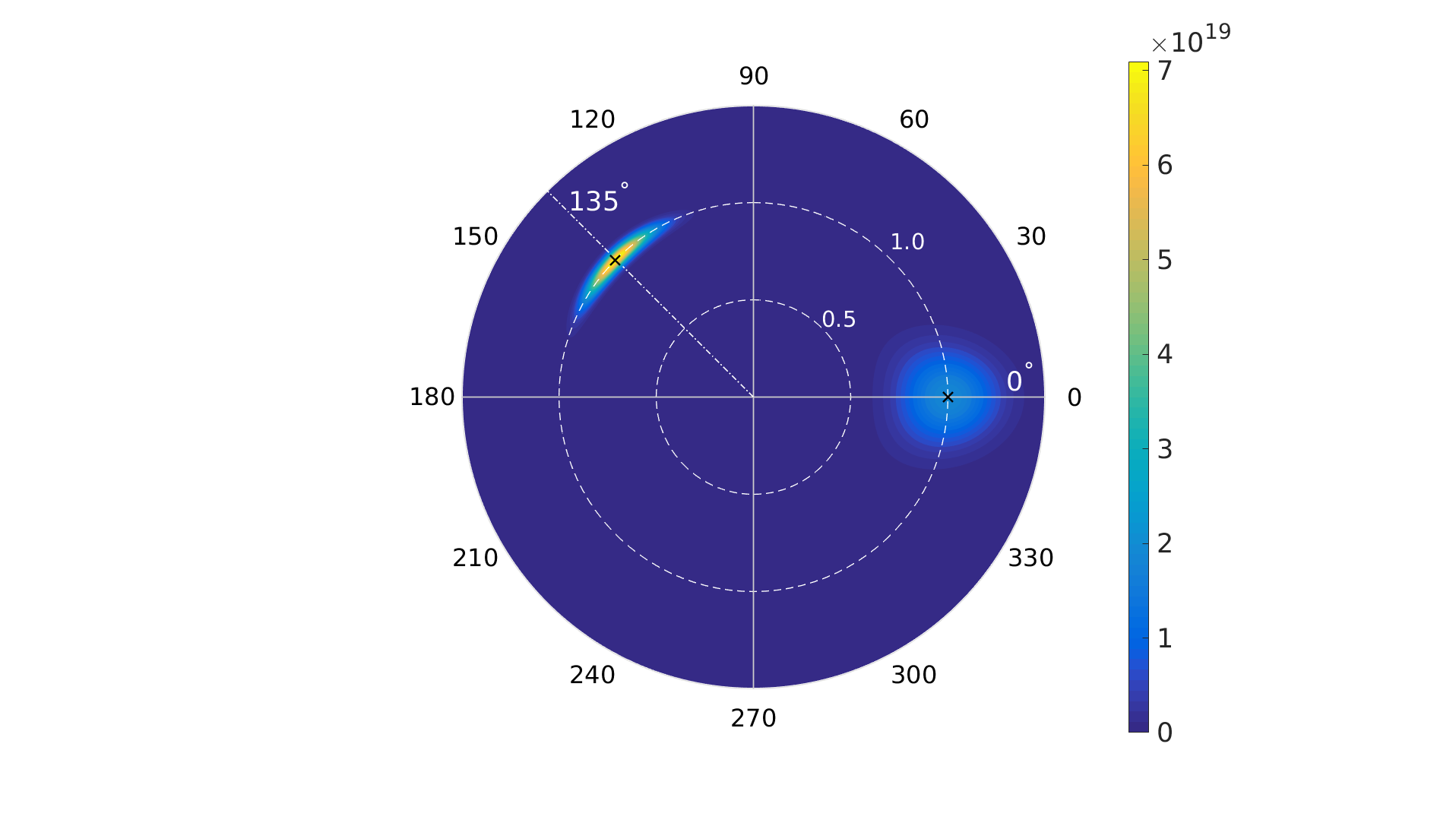}
	\includegraphics[width=0.44\textwidth,trim={14cm 0 7.5cm 0},clip]{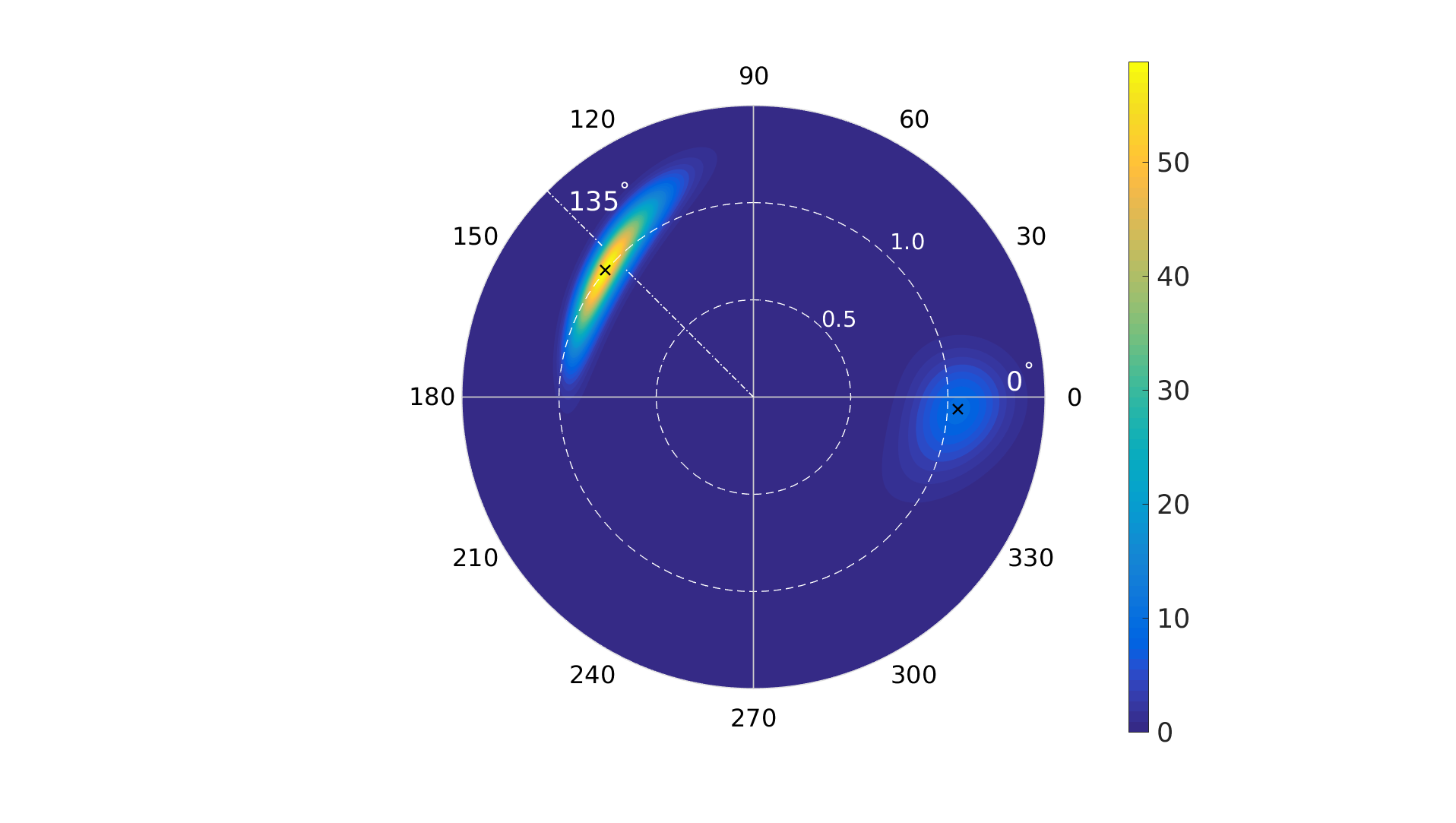}
	\vspace*{-5mm}
      \end{center}
      \caption{Differential numbers $\frac{{\rm d}^2{\cal N}}{{\rm d}\omega{\rm d}\varphi}$ and $\frac{{\rm d}^2N_{\rm tot}}{{\rm d}\omega{\rm d}\varphi}$ of laser photons $\cal N$ (top) and signal photons $N_{\rm tot}$ (bottom).
      We consider the collision of two laser pulses with beam axes in the xy-plane under an angle of $\varphi_{\rm coll}=135^\circ$.
      The two pulses of different energy ($W_{0^\circ}=50\,{\rm J}$, $W_{135^\circ}=30\,{\rm J}$) and duration ($\tau_{0^\circ}=5\,{\rm fs}$, $\tau_{135^\circ}=30\,{\rm fs}$) have the same wavelength and beam waist ($\lambda=w_0=800\,{\rm nm}$) and are polarized along $\vec{e}_{\rm z}$ in the focus.
      Here, the radial coordinate measures the signal photon energy $\omega$ in units of the laser frequency $\omega_{\rm L}$.        
      Note, that the two maxima (black crosses) in $\frac{{\rm d}^2N_{\rm tot}}{{\rm d}\omega{\rm d}\varphi}$ are notably shifted from the dashed circle marking the condition $\omega=\omega_{\rm L}$ and that the angular distribution is also significantly wider.}
      \label{fig:dNdkdphi}
      \end{figure}    
      
\begin{figure}[t]
      \begin{center}
	\includegraphics[width=0.49\textwidth]{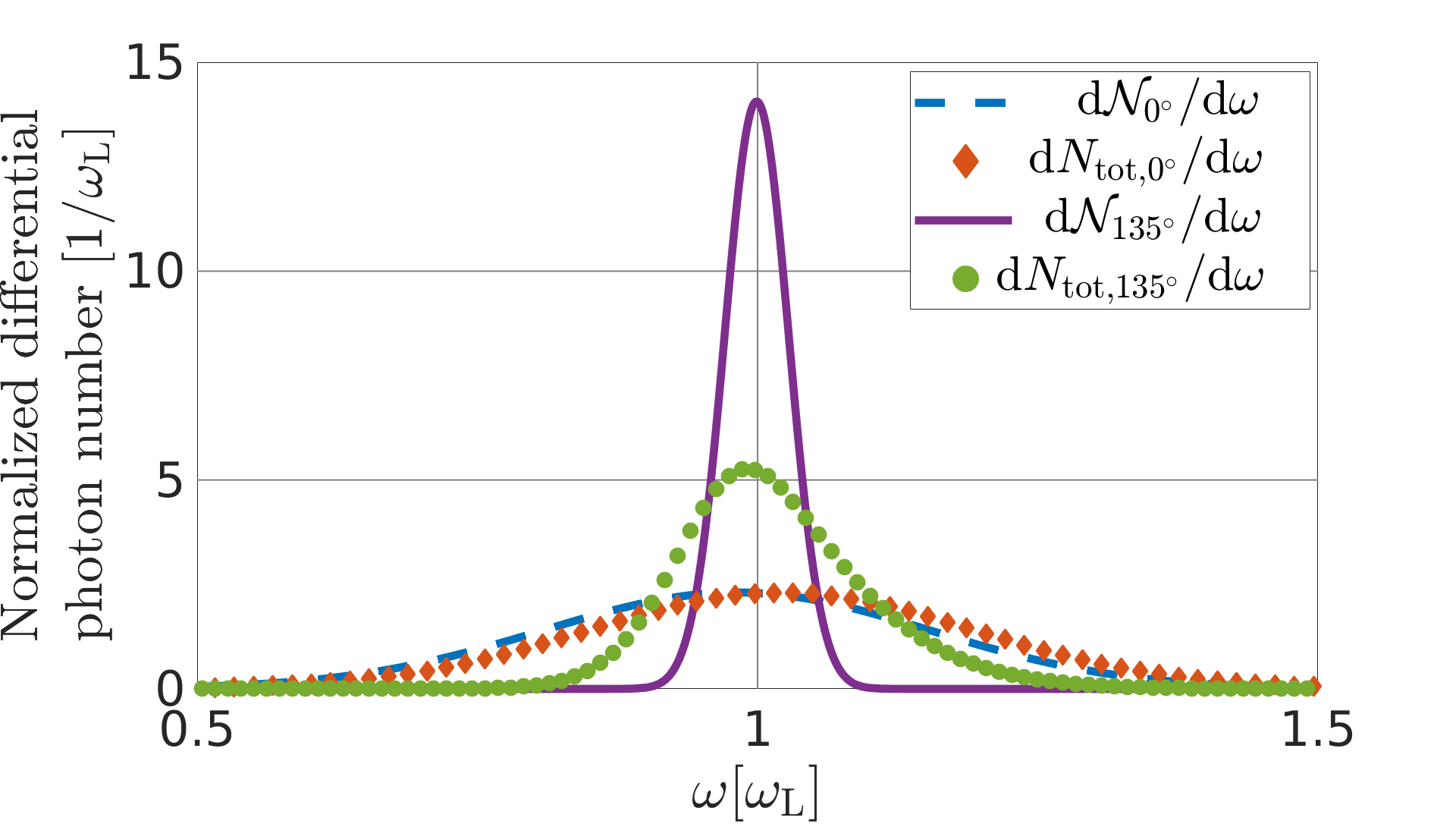}
      \end{center}
      \caption{Comparison of the spectral distributions of laser photons $\cal N$ (lines) and signal photons $N_{\rm tot}$ (symbols) for the collision of two laser pulses with beam axes in the xy-plane under an angle of $\varphi_{\rm coll}=135^\circ$, highlighting the different decay behaviors of the laser and signal photon spectra, and the emergence of an asymmetry in the signal photon spectrum.
      The two pulses of different energy ($W_{0^\circ}=50\,{\rm J}$, $W_{135^\circ}=30\,{\rm J}$) and duration ($\tau_{0^\circ}=5\,{\rm fs}$, $\tau_{135^\circ}=30\,{\rm fs}$) have the same wavelength and beam waist ($\lambda=w_0=800\,{\rm nm}$) and are polarized along $\vec{e}_{\rm z}$ in the focus.
      Here, we distinguish between components propagating in directions about $\varphi=0^\circ$ and $\varphi=135^\circ$.
      The signal photons emitted at an angle of $\varphi\simeq0^\circ$ ($135^\circ$) can be considered as originating from the driving laser pulse propagating in $\varphi=0^\circ$ ($135^\circ$) direction, and being scattered quasi-elastically off the other laser pulse.
      To allow for a straightforward comparison of their decay properties, the curves have been normalized such that the areas below them agree.}
      \label{fig:normspec}
      \end{figure}

Figure~\ref{fig:dNdkdphi} shows the energy spectra of the laser (top) and signal (bottom) photons as a function of the longitude $\varphi$; to this end the latitude $\vartheta$ has been integrated out.
More specifically, we depict the differential numbers $\frac{{\rm d}^2{\cal N}}{{\rm d}\omega{\rm d}\varphi}$ (top) and $\frac{{\rm d}^2N_{\rm tot}}{{\rm d}\omega{\rm d}\varphi}$ (bottom), where we have made use of the fact that the modulus of the signal-photon wave vector equals the signal photon energy, i.e., $\omega={\rm k}$.
Apart from the clear differences in the absolute numbers of driving laser photons and induced signal photons, these spectra unveil additional distinct features.
The spectra of the driving laser pulses in Fig.~\ref{fig:dNdkdphi} (top) are symmetric with respect to their beam axes, are peaked at the laser frequency $\omega$, and reach their maxima in strict forward direction, i.e., at $\vartheta=0^\circ$ and $\vartheta=135^\circ$, respectively.
For the signal photons, this symmetry is broken and both emission channels exhibit obvious asymmetries. The peak values in Fig.~\ref{fig:dNdkdphi} (bottom) are shifted by a few degrees from the forward directions of the driving beams, such that the angle between these peaks is somewhat larger than the collision angle of the driving beams $\varphi_{\rm coll}=135^\circ$ in Fig.~\ref{fig:dNdkdphi} (top). This behavior is also visible in Fig.~\ref{fig:solang}.
In particular for the signal photons originating from the shorter laser pulse propagating along $\varphi=0^\circ$, the peak value is shifted to higher energies $\omega>\omega_{\rm L}$.
On the other hand, especially the signal photons emitted about $\varphi=135^\circ$ are shifted to higher energies $\omega\gtrsim\omega_{\rm L}$ for $\varphi<135^\circ$, and lower energies $\omega\lesssim\omega_{\rm L}$ for $\varphi>135^\circ$.
The combined effect on the signal photon spectrum after integration over $\varphi$ can be inferred from Fig.~\ref{fig:normspec} depicting results for the spectral distributions $\frac{{\rm d}{\cal N}}{{\rm d}\omega}$ and $\frac{{\rm d}N_{\rm tot}}{{\rm d}\omega}$.
While the photons of the driving pulses are symmetrically distributed  about $\omega=\omega_{\text{L}}$, the signal photons show a clear spectral distortion with respect to this symmetry axis.
As both the laser photons and the associated signal photons are highly directional (cf. Figs. \ref{fig:solang} and \ref{fig:dNdkdphi}), allowing for their clear spatial separation, we distinguish between components propagating in directions about $\varphi=0^\circ$ and $\varphi=135^\circ$.
All the curves depicted in Fig.~\ref{fig:normspec} have been normalized such that upon integration over $\omega$ the same number is obtained.  

The spectral distributions of the driving laser photons ${\cal N}_{0^\circ}$ and ${\cal N}_{135^\circ}$ are peaked at the laser photon energy $\omega=\omega_{\rm L}$, and are -- to a good accuracy -- symmetric with respect to $\omega_{\rm L}$; rather small shifts arise from considering the differential with respect to the frequency.
By contrast, the maxima of the spectral distributions of the signal photons $N_{{\rm tot},0^\circ}$ and $N_{{\rm tot},135^\circ}$ are prominently displaced from the laser photon energy:
The former (latter) is shifted to a somewhat higher (lower) energy, indicating a spectral asymmetry.  

In a next step, we analyze the widths and decay behaviors of the various depicted curves in Fig.~\ref{fig:normspec}.
While the curve associated with $N_{{\rm tot},0^\circ}$ essentially parallels the one for ${\cal N}_{0^\circ}$ in its width and decay behavior, the curve for $N_{{\rm tot},135^\circ}$ is much wider and decays much slower than the one for ${\cal N}_{135^\circ}$.
To understand this behavior, recall that the signal photons $N_{{\rm tot},0^\circ}$ ($N_{{\rm tot},135^\circ}$) can be interpreted as originating from the driving laser pulse propagating in $\varphi=0^\circ$ ($135^\circ$) direction, and being scattered quasi-elastically at the other laser pulse; cf. above.
Obviously, the interaction time of the laser pulses is determined by their temporal overlap which critically depends on the shorter pulse duration.
In turn, the signal photon channels can be characterized by pulse durations $\tau\sim\tau_{0^\circ}<\tau_{135^\circ}$.
A shorter pulse duration translates to a larger peak width in the energy spectrum and vice versa.
This clearly explains why the spectral distributions for $N_{{\rm tot},135^\circ}$, $N_{{\rm tot},0^\circ}$ and ${\cal N}_{0^\circ}$ are wider than that for ${\cal N}_{135^\circ}$.
As an interesting feature, the signal photon spectra exhibit kinematical and spectral asymmetries in comparison with the spectrally symmetric driving pulses. These effects can be explained by the manifestly asymmetric collision scenario, involving different pulse shapes, and by the fact that the signal photons emitted in a given direction are effectively induced by the scattering process off a moving intensity profile: 
the signal photons are predominantly induced at times and positions where the intensity of the other field acting as scatterer are maximal.
From the perspective of the photons of a given laser pulse traversing the other one, these intensity maxima are reached at different times at the front and rear side of the scatterer, thereby naturally sourcing a spectral asymmetry.

The different spectral widths and decays of the signal photons $N_{{\rm tot},135^\circ}$ and the laser photons ${\cal N}_{135^\circ}$ might potentially be employed to distinguish the signal photons from the photons constituting the driving laser pulses.
We plan to investigate the potential of this idea in a dedicated follow up study.

Finally, note that this effect can be considered as the time domain analogue of the effect of signal photon scattering out of the forward cone of a weakly focused probe laser pulse in the collision with a more tightly focused pump pulse in the spatial domain.
It has recently been argued that the latter phenomenon may constitute an important means to enhance the signal-to-background separation in all-optical vacuum birefringence experiments \cite{Karbstein:2015xra,Karbstein:2016lby,Karbstein:2018omb}.

\section{Conclusions and Outlook}
\label{sec:conclusions}

\omite{We haven't discussed so far as to whether the set-up would be suitable for axion searches. Photon-axion oscillations are one of the powerful tool to search for axions, all the way down to the QCD-inspired axion models, possibly superseeding astrophysical bounds on axion parameters from direct and indirect searches. Axions could be emitted from the collision center and undergo axion-photon coversion including a momentum transfer to the outgoing axion. The axion spectrum is not known so far, and the axion momentum, the axion directions as well as the axion angular distribution would be interesting to know for axion-discovery studies. Provided axions exist. Axion-like particles could be in reach even though they do not solve the strong-CP problem. But they could be like axions or like axion-like particles which axion searchers like very much. In any case, the photon-axion interaction deserves to be included in the numerical solvers as the classical equations of motion are modified by the axion terms.} 
In this article, we have significantly advanced beyond previous theoretical studies of all-optical quantum vacuum nonlinearities at the high-intensity frontier. 
The new numerical tool devised here is tailored to study all-optical signatures of QED vacuum nonlinearities in generic laser fields.
As no simplifying approximations of the beam profiles and pulse shapes of the driving laser fields are needed, it allows us to overcome previous simplifying assumptions on the beam profiles and pulse shapes, and to obtain quantitative predictions in realistic field configurations available in experiment.
To achieve this goal, we have combined the vacuum emission picture with an efficient numerical Maxwell solver, which self-consistently propagates the driving laser fields from any given initial field configuration, manifestly ensuring the latter to fulfill Maxwell's equations in vacuum at all space-time coordinates. 

In turn, our approach resolves the obvious mismatch of analytical approaches studying signatures of quantum vacuum nonlinearities in electromagnetic fields which solve the wave equation only approximately, such as, e.g., plane-wave based models, crossed-field models with pulse-shape envelopes or pulsed paraxial beams.
This is of particular importance as the deviations of results derived from approximate vs. exact solutions of the wave-equation are typically hard to assess quantitatively.
They, however, amount to a critical unknown, limiting the achievable accuracy of predictions for experiments based on approximations of the driving laser fields.
In the present article, we have carefully assessed their importance in various benchmark scenarios, and demonstrated that depending on the specific scenario under consideration, such deviations can be sizable.
For definiteness and a more straightforward comparison of the obtained results, throughout all benchmark scenarios we have focused on integrated photon numbers.

Moreover, to illustrate the substantial potential of our code for the quantitative study of optical signatures of QED vacuum nonlinearity,
we have investigated a scenario where a moderately long tightly focused high-intensity laser pulse is brought into collision with a substantially stronger few-cycle pulse.
As the pulse duration is of the order of the wavelength, conventional approximations are no longer applicable in this case.
Here, we put special attention on the spectra of the driving laser photons and the attainable signal photons.
We note that the spectral and angular distribution of the emitted photons deviates from that of the input beams, e.g., giving rise to kinematical and spectral asymmetries. This highlights the need for a full description of the laser field for precision experiments.

Given the high flexibility and efficiency of our numerical code to self-consistently propagate initial data characterizing any possible driving laser field configuration conceivable in experiment, we are confident that it constitutes an important tool for the accurate theoretical study of all-optical signatures of QED vacuum nonlinearities in generic high-intensity laser pulse collisions.
In particular, it facilitates precise quantitative first-principles predictions for dedicated all-optical discovery experiments of quantum vacuum nonlinearities at the various high-intensity laser facilities coming online just now, such as CILEX \cite{CILEX}, CoReLS \cite{CoReLS}, ELI \cite{ELI} and SG-II \cite{SG-II}.
Besides, aiming at exploring more visionary parameter regimes, such as ultrastrong high-intensity laser pulses reaching field strengths of the order of the Schwinger critical field, our approach can be readily extended to account for higher order self-couplings of the external electromagnetic fields \cite{Heisenberg:1935qt}, higher orders in the derivative expansion \cite{Gusynin:1995bc,Gusynin:1998bt}, and higher loop orders \cite{Ritus:1975,Gies:2016yaa} of the Heisenberg-Euler effective Lagrangian.

\begin{acknowledgments}
  The work of C.K. is funded by the Helmholtz Association through the
  Helmholtz Postdoc Programme (PD-316).  We acknowledge support by the
  BMBF under grant No. 05P15SJFAA (FAIR-APPA-SPARC).  Computations
  were performed on the ``Supermicro Server 1028TR-TF'' in Jena, which
  was funded by the Helmholtz Postdoc Programme (PD-316).
  F.K. thanks the Baktschar Institute of Theoretical Physics for hospitality.
\end{acknowledgments}

\appendix

\section{Convergence test}
\label{app:conv}

\begin{figure}[t]
  \begin{center}
    \includegraphics[width=0.49\textwidth]{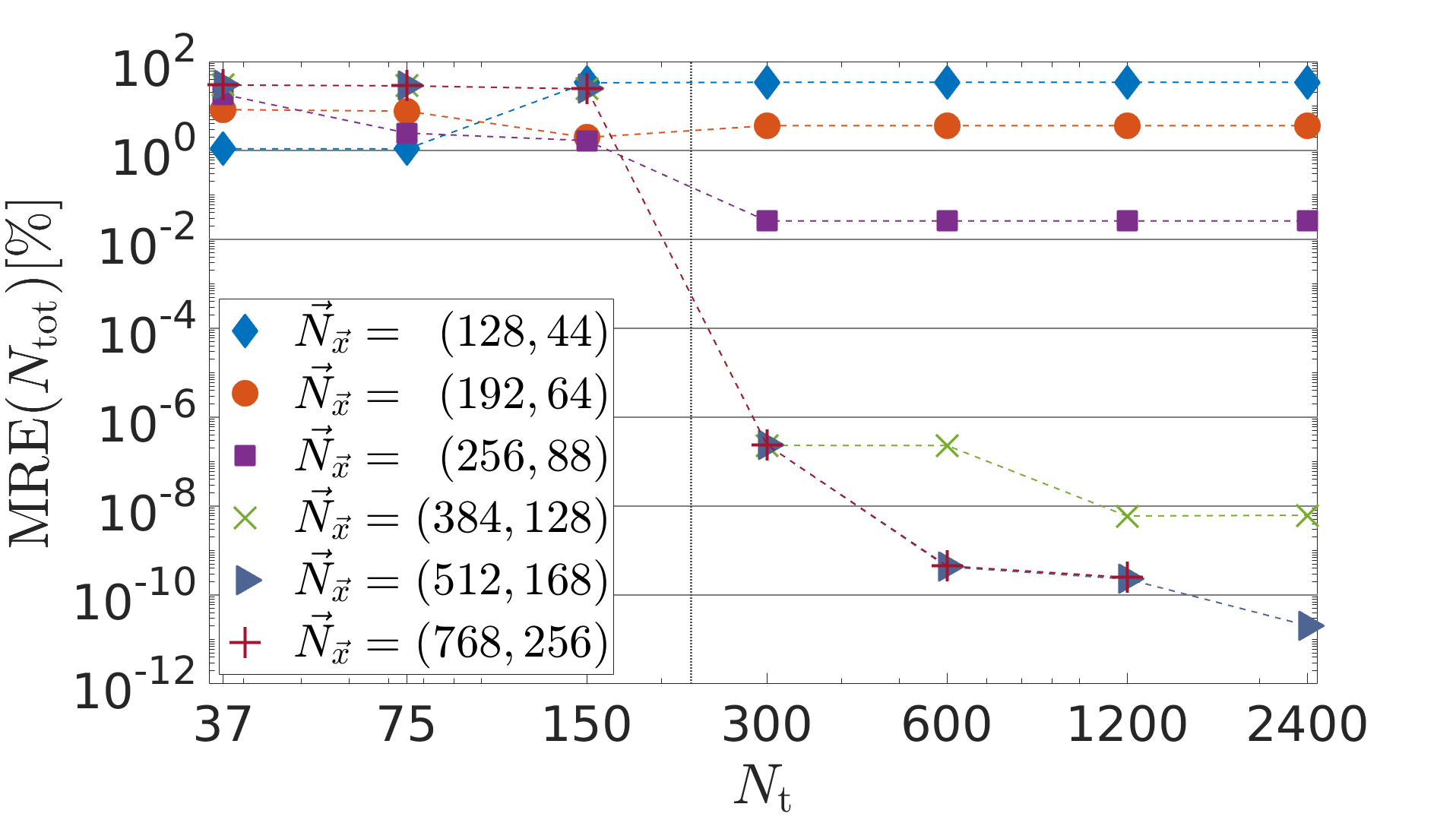}
  \end{center}
  \caption{Log-log plot showing convergence of the total number of signal photons $N_{\rm tot}$
    as a function of the number of time steps $N_{\rm t}$ for various grid sizes $\vec{N}_{\vec{x}}$ (specified by $(N_{\rm xy},N_{\rm z})$).
    The two  pulses (pulse energy $W=25\,{\rm J}$, duration $\tau=25\,{\rm fs}$, wavelength $\lambda=800\,{\rm nm}$) 
    focused to $w_0=\lambda$ and polarized along $\vec{e}_{\rm z}$ collide under an angle of $\varphi_2 = 135^\circ$.
    The results are presented in terms of the mean relative error MRE with respect to the 
    numerical result determined with the highest resolution ($N_{\rm tot} = 78.36$).}
  \label{fig:convergence}
\end{figure}

In addition to the benchmark tests described in the main text, we now
detail the convergence behavior of our numerical method. The code of
the Maxwell solver itself has been verified by quantitative comparison
of raw data as also produced by alternative solvers
\cite{Blinne:2018}. The present vacuum emission solver has also been
benchmarked with the code developed in \cite{Gies:2017ygp} in the
overlapping regime of validity. For a convergence test of the new code
in a regime which has remained unexplored so far, we  use the total
number of signal photons $N_{\rm tot}$ attainable in the collision of two high-intensity laser pulses  as
an observable example.  In absence of a quantitative reference
result in this regime, we perform a self-consistency test of the data
under variations of the numerical control parameters.

Within our Maxwell solver we can tune $8$ different, purely numerical
parameters. Four of these are $N_{\rm t}$ specifying the number of
time steps involved in a computation and $\vec{N}_{\vec{x}}$
determining the number of grid points per direction. In addition, we
have the four length parameters determining the spacetime volume of
the sampling region, $L_{\rm t} \times L_{\rm x} \times L_{\rm y}
\times L_{\rm z}$. The choice of the latter is essentially dictated by
the spacetime volume of the interaction region of the pulse foci. A
choice of $\mathcal{O}(1\dots 10)$ times the typical length or time scales
(pulse width, Rayleigh range, pulse duration) guarantees in our case
that the error is exponentially small in this length ratio.  Hence,
we focus here on the total number of grid points and study the
dependence of the observable on these control parameters.

As in the main part of this work, we focus on the collision
of two identical laser pulses (pulse energy $W=25\,{\rm J}$, duration $\tau=25\,{\rm fs}$ and wavelength $\lambda=800\,{\rm nm}$) under an angle of $\varphi = 135^\circ$ in the xy-plane, focused to $w_0=\lambda$ and polarized along $\vec{e}_{\rm z}$ in the focus. These laser pulses are assumed to collide under optimal conditions, i.e., are focused to the same focal spot at $\vec{x}=\vec{0}$, and reach their peak fields in the focus exactly at the same time $t_0=0$.
We begin our calculations at $t_i = -50\,{\rm fs}$, before the two pulses collide, and terminate at $t_f=50\,{\rm fs}$, when there is no
significant beam overlap any more ($L_{\rm t}=100\,{\rm fs}$). For the spatial sampling region, we choose $L_{\rm xy}\equiv L_{\rm x}=L_{\rm y} = 50\,$\textmu m and
$L_{\rm z} = 30\,$\textmu m; thus the field strengths at the boundaries reach maximally $\sim 7 \%$ of the peak field strengths, i.e., $\sim 0.5\%$ of the peak intensity.
Further out, the amplitudes fall off exponentially reducing the error correspondingly. 
A full calculation using a lattice of size $N_{\rm t} = 2400$, $N_{\rm xy} = 768$ and $N_{\rm z} = 256$ to discretize a 
box of size $L_{\rm t} = 37.2 \lambda$, $L_{\rm xy} = 62.5 \lambda$ and $L_{\rm z} = 37.2 w_0$ yields $N_{\rm tot}=78.36$ signal photons.
In the direction perpendicular to the collision plane, i.e., the z direction in our case, the width $w_0$ is the relevant scale rather than the wave length $\lambda$. In the present case, however, we have $w_0=\lambda$.
We use this result for $N_{\rm tot}$ as a reference value for the following discussion.

In Fig.~\ref{fig:convergence}, the mean relative error for the total signal photon number MRE($N_{\rm tot}$) compared with the reference value is displayed for various
coarser grid resolutions. As expected, a large number of grid points in the spatial coordinates $\vec x$ cannot compensate a poor resolution in time $t$
and vice versa. However, there seems to be a significant increase in accuracy as soon as $N_{\rm t}$ and $N_{\rm xy}$ reach $\sim 300$. This behavior
is directly connected to the fact, that we have to properly resolve the substructure of the pulses.
As the vacuum emission amplitude scales cubically with the field strength, every frequency component $\upsilon$ present in the input fields can translate to frequency components $3\upsilon$ present in the integrand of the zero-to-single photon transition amplitude~\eqref{eq:Sp}.
In order to properly resolve these frequency components in the spatial Fourier transform as well as in the temporal integration, six grid points per shortest length scale along the $\mathrm{x, y}$ or $\mathrm{z}$ axis as well as per shortest time scale occurring in the input fields are required as an absolute minimum.

The box size in  x and y directions is kept fixed at $L_{\rm xy}=62.5 \lambda$ in accordance with the reference calculation.
At $6$ points per cycle, a total of $N_{\rm xy}=375$
grid points is needed for a proper resolution of the integrand. The same argument holds also for the time component: the length of the time interval is
$L_{\rm t}=37.2 \lambda$; hence, having a total of $N_{\rm t}=224$ grid points in time is a prerequisite for resolving the relevant frequency scale. As a result, the minimal configuration in Fig. \ref{fig:convergence} which fulfills the requirements, $N_{\rm xy}=384$ and $N_{\rm t}=300$,
yields a mean relative error of $2.3 \times 10^{-7} \%$.

Analogous analyses have been performed for all configurations considered in this work, yielding similar outcomes in terms of 
precision and accuracy. In order to obtain reliable results with an error well below the per mille level, such considerations serve to adjust the numerical control parameters governing the size and the number of sampling points of the discretized spacetime volume.


\bibliography{libraryMendeley,libraryMore}

\begin{thebibliography}{84}%
\makeatletter
\providecommand \@ifxundefined [1]{%
 \@ifx{#1\undefined}
}%
\providecommand \@ifnum [1]{%
 \ifnum #1\expandafter \@firstoftwo
 \else \expandafter \@secondoftwo
 \fi
}%
\providecommand \@ifx [1]{%
 \ifx #1\expandafter \@firstoftwo
 \else \expandafter \@secondoftwo
 \fi
}%
\providecommand \natexlab [1]{#1}%
\providecommand \enquote  [1]{``#1''}%
\providecommand \bibnamefont  [1]{#1}%
\providecommand \bibfnamefont [1]{#1}%
\providecommand \citenamefont [1]{#1}%
\providecommand \href@noop [0]{\@secondoftwo}%
\providecommand \href [0]{\begingroup \@sanitize@url \@href}%
\providecommand \@href[1]{\@@startlink{#1}\@@href}%
\providecommand \@@href[1]{\endgroup#1\@@endlink}%
\providecommand \@sanitize@url [0]{\catcode `\\12\catcode `\$12\catcode
  `\&12\catcode `\#12\catcode `\^12\catcode `\_12\catcode `\%12\relax}%
\providecommand \@@startlink[1]{}%
\providecommand \@@endlink[0]{}%
\providecommand \url  [0]{\begingroup\@sanitize@url \@url }%
\providecommand \@url [1]{\endgroup\@href {#1}{\urlprefix }}%
\providecommand \urlprefix  [0]{URL }%
\providecommand \Eprint [0]{\href }%
\providecommand \doibase [0]{http://dx.doi.org/}%
\providecommand \selectlanguage [0]{\@gobble}%
\providecommand \bibinfo  [0]{\@secondoftwo}%
\providecommand \bibfield  [0]{\@secondoftwo}%
\providecommand \translation [1]{[#1]}%
\providecommand \BibitemOpen [0]{}%
\providecommand \bibitemStop [0]{}%
\providecommand \bibitemNoStop [0]{.\EOS\space}%
\providecommand \EOS [0]{\spacefactor3000\relax}%
\providecommand \BibitemShut  [1]{\csname bibitem#1\endcsname}%
\let\auto@bib@innerbib\@empty
\bibitem [{\citenamefont {Euler}\ and\ \citenamefont
  {Kockel}(1935)}]{Euler:1935zz}%
  \BibitemOpen
  \bibfield  {author} {\bibinfo {author} {\bibfnamefont {H.}~\bibnamefont
  {Euler}}\ and\ \bibinfo {author} {\bibfnamefont {B.}~\bibnamefont {Kockel}},\
  }\href {\doibase 10.1007/BF01493898} {\bibfield  {journal} {\bibinfo
  {journal} {Naturwissenschaften}\ }\textbf {\bibinfo {volume} {23}},\ \bibinfo
  {pages} {246} (\bibinfo {year} {1935})}\BibitemShut {NoStop}%
\bibitem [{\citenamefont {Heisenberg}\ and\ \citenamefont
  {Euler}(1936)}]{Heisenberg:1935qt}%
  \BibitemOpen
  \bibfield  {author} {\bibinfo {author} {\bibfnamefont {W.}~\bibnamefont
  {Heisenberg}}\ and\ \bibinfo {author} {\bibfnamefont {H.}~\bibnamefont
  {Euler}},\ }\href {\doibase 10.1007/BF01343663} {\bibfield  {journal}
  {\bibinfo  {journal} {Zeitschrift f{\"{u}}r Phys.}\ }\textbf {\bibinfo
  {volume} {98}},\ \bibinfo {pages} {714} (\bibinfo {year} {1936})}\BibitemShut
  {NoStop}%
\bibitem [{\citenamefont {Weisskopf}(1936)}]{Weisskopf:1936bu}%
  \BibitemOpen
  \bibfield  {author} {\bibinfo {author} {\bibfnamefont {V.~S.}\ \bibnamefont
  {Weisskopf}},\ }\href
  {http://users.physik.fu-berlin.de/{~}kleinert/kleinert/?p=histpapers}
  {\bibfield  {journal} {\bibinfo  {journal} {K. Danske Vidensk. Selsk. Mat.
  Meddelelser}\ }\textbf {\bibinfo {volume} {XIV}},\ \bibinfo {pages} {3}
  (\bibinfo {year} {1936})}\BibitemShut {NoStop}%
\bibitem [{\citenamefont {Schwinger}(1951)}]{Schwinger:1951nm}%
  \BibitemOpen
  \bibfield  {author} {\bibinfo {author} {\bibfnamefont {J.}~\bibnamefont
  {Schwinger}},\ }\href {\doibase 10.1103/PhysRev.82.664} {\bibfield  {journal}
  {\bibinfo  {journal} {Phys. Rev.}\ }\textbf {\bibinfo {volume} {82}},\
  \bibinfo {pages} {664} (\bibinfo {year} {1951})}\BibitemShut {NoStop}%
\bibitem [{\citenamefont {Dittrich}\ and\ \citenamefont
  {Reuter}(1985)}]{Dittrich:1985yb}%
  \BibitemOpen
  \bibfield  {author} {\bibinfo {author} {\bibfnamefont {W.}~\bibnamefont
  {Dittrich}}\ and\ \bibinfo {author} {\bibfnamefont {M.}~\bibnamefont
  {Reuter}},\ }\href {\doibase 10.1007/3-540-15182-6} {\enquote {\bibinfo
  {title} {{Effective Lagrangians in Quantum Electrodynamics}},}\ } (\bibinfo
  {year} {1985})\BibitemShut {NoStop}%
\bibitem [{\citenamefont {Dittrich}\ and\ \citenamefont
  {Gies}(2000)}]{Dittrich:2000zu}%
  \BibitemOpen
  \bibfield  {author} {\bibinfo {author} {\bibfnamefont {W.}~\bibnamefont
  {Dittrich}}\ and\ \bibinfo {author} {\bibfnamefont {H.}~\bibnamefont
  {Gies}},\ }\href {\doibase 10.1007/3-540-45585-X} {\emph {\bibinfo {title}
  {{Probing the quantum vacuum}}}},\ \bibinfo {series} {Springer Tracts in
  Modern Physics}, Vol.\ \bibinfo {volume} {166}\ (\bibinfo  {publisher}
  {Springer},\ \bibinfo {address} {Berlin, Heidelberg},\ \bibinfo {year}
  {2000})\ p.\ \bibinfo {pages} {241},\ \Eprint {http://arxiv.org/abs/9808028}
  {arXiv:9808028 [arXiv:physics]} \BibitemShut {NoStop}%
\bibitem [{\citenamefont {Marklund}\ and\ \citenamefont
  {Lundin}(2009)}]{Marklund:2008gj}%
  \BibitemOpen
  \bibfield  {author} {\bibinfo {author} {\bibfnamefont {M.}~\bibnamefont
  {Marklund}}\ and\ \bibinfo {author} {\bibfnamefont {J.}~\bibnamefont
  {Lundin}},\ }\href {\doibase 10.1140/epjd/e2009-00169-6} {\bibfield
  {journal} {\bibinfo  {journal} {Eur. Phys. J. D}\ }\textbf {\bibinfo {volume}
  {55}},\ \bibinfo {pages} {319} (\bibinfo {year} {2009})}\BibitemShut
  {NoStop}%
\bibitem [{\citenamefont {Heinzl}\ and\ \citenamefont
  {Ilderton}(2009)}]{Heinzl:2008an}%
  \BibitemOpen
  \bibfield  {author} {\bibinfo {author} {\bibfnamefont {T.}~\bibnamefont
  {Heinzl}}\ and\ \bibinfo {author} {\bibfnamefont {A.}~\bibnamefont
  {Ilderton}},\ }\href {\doibase 10.1140/epjd/e2009-00113-x} {\bibfield
  {journal} {\bibinfo  {journal} {Eur. Phys. J. D}\ }\textbf {\bibinfo {volume}
  {55}},\ \bibinfo {pages} {359} (\bibinfo {year} {2009})}\BibitemShut
  {NoStop}%
\bibitem [{\citenamefont {{Di Piazza}}\ \emph {et~al.}(2012)\citenamefont {{Di
  Piazza}}, \citenamefont {M{\"{u}}ller}, \citenamefont {Hatsagortsyan},\ and\
  \citenamefont {Keitel}}]{DiPiazza:2011tq}%
  \BibitemOpen
  \bibfield  {author} {\bibinfo {author} {\bibfnamefont {A.}~\bibnamefont {{Di
  Piazza}}}, \bibinfo {author} {\bibfnamefont {C.}~\bibnamefont
  {M{\"{u}}ller}}, \bibinfo {author} {\bibfnamefont {K.~Z.}\ \bibnamefont
  {Hatsagortsyan}}, \ and\ \bibinfo {author} {\bibfnamefont {C.~H.}\
  \bibnamefont {Keitel}},\ }\href {\doibase 10.1103/RevModPhys.84.1177}
  {\bibfield  {journal} {\bibinfo  {journal} {Rev. Mod. Phys.}\ }\textbf
  {\bibinfo {volume} {84}},\ \bibinfo {pages} {1177} (\bibinfo {year}
  {2012})},\ \Eprint {http://arxiv.org/abs/1111.3886} {arXiv:1111.3886}
  \BibitemShut {NoStop}%
\bibitem [{\citenamefont {Dunne}(2012)}]{Dunne:2012vv}%
  \BibitemOpen
  \bibfield  {author} {\bibinfo {author} {\bibfnamefont {G.~V.}\ \bibnamefont
  {Dunne}},\ }in\ \href {http://arxiv.org/abs/1202.1557} {\emph {\bibinfo
  {booktitle} {QFEXT11}}}\ (\bibinfo {year} {2012})\ \Eprint
  {http://arxiv.org/abs/1202.1557} {arXiv:1202.1557} \BibitemShut {NoStop}%
\bibitem [{\citenamefont {Battesti}\ and\ \citenamefont
  {Rizzo}(2013)}]{Battesti:2012hf}%
  \BibitemOpen
  \bibfield  {author} {\bibinfo {author} {\bibfnamefont {R.}~\bibnamefont
  {Battesti}}\ and\ \bibinfo {author} {\bibfnamefont {C.}~\bibnamefont
  {Rizzo}},\ }\href {\doibase 10.1088/0034-4885/76/1/016401} {\bibfield
  {journal} {\bibinfo  {journal} {Reports Prog. Phys.}\ }\textbf {\bibinfo
  {volume} {76}},\ \bibinfo {pages} {016401} (\bibinfo {year} {2013})},\
  \Eprint {http://arxiv.org/abs/1211.1933} {arXiv:1211.1933} \BibitemShut
  {NoStop}%
\bibitem [{\citenamefont {King}\ and\ \citenamefont
  {Heinzl}(2016)}]{King:2015tba}%
  \BibitemOpen
  \bibfield  {author} {\bibinfo {author} {\bibfnamefont {B.}~\bibnamefont
  {King}}\ and\ \bibinfo {author} {\bibfnamefont {T.}~\bibnamefont {Heinzl}},\
  }\href {\doibase 10.1017/hpl.2016.1} {\bibfield  {journal} {\bibinfo
  {journal} {High Power Laser Sci. Eng.}\ }\textbf {\bibinfo {volume} {4}},\
  \bibinfo {pages} {e5} (\bibinfo {year} {2016})},\ \Eprint
  {http://arxiv.org/abs/1510.08456} {arXiv:1510.08456} \BibitemShut {NoStop}%
\bibitem [{\citenamefont {Karbstein}(2016)}]{Karbstein:2016hlj}%
  \BibitemOpen
  \bibfield  {author} {\bibinfo {author} {\bibfnamefont {F.}~\bibnamefont
  {Karbstein}},\ }\href {\doibase 10.3204/DESY-PROC-2016-04} {\  (\bibinfo
  {year} {2016}),\ 10.3204/DESY-PROC-2016-04},\ \Eprint
  {http://arxiv.org/abs/1611.09883} {arXiv:1611.09883} \BibitemShut {NoStop}%
\bibitem [{\citenamefont {Inada}\ \emph {et~al.}(2017)\citenamefont {Inada},
  \citenamefont {Yamazaki}, \citenamefont {Yamaji}, \citenamefont {Seino},
  \citenamefont {Fan}, \citenamefont {Kamioka}, \citenamefont {Namba},\ and\
  \citenamefont {Asai}}]{Inada:2017lop}%
  \BibitemOpen
  \bibfield  {author} {\bibinfo {author} {\bibfnamefont {T.}~\bibnamefont
  {Inada}}, \bibinfo {author} {\bibfnamefont {T.}~\bibnamefont {Yamazaki}},
  \bibinfo {author} {\bibfnamefont {T.}~\bibnamefont {Yamaji}}, \bibinfo
  {author} {\bibfnamefont {Y.}~\bibnamefont {Seino}}, \bibinfo {author}
  {\bibfnamefont {X.}~\bibnamefont {Fan}}, \bibinfo {author} {\bibfnamefont
  {S.}~\bibnamefont {Kamioka}}, \bibinfo {author} {\bibfnamefont
  {T.}~\bibnamefont {Namba}}, \ and\ \bibinfo {author} {\bibfnamefont
  {S.}~\bibnamefont {Asai}},\ }\href {\doibase 10.3390/app7070671} {\bibfield
  {journal} {\bibinfo  {journal} {Appl. Sci.}\ }\textbf {\bibinfo {volume}
  {7}},\ \bibinfo {pages} {671} (\bibinfo {year} {2017})}\BibitemShut {NoStop}%
\bibitem [{\citenamefont {Kotkin}\ and\ \citenamefont
  {Serbo}(1997)}]{Kotkin:1996nf}%
  \BibitemOpen
  \bibfield  {author} {\bibinfo {author} {\bibfnamefont {G.}~\bibnamefont
  {Kotkin}}\ and\ \bibinfo {author} {\bibfnamefont {V.}~\bibnamefont {Serbo}},\
  }\href {\doibase 10.1016/S0370-2693(97)01086-1} {\bibfield  {journal}
  {\bibinfo  {journal} {Phys. Lett. B}\ }\textbf {\bibinfo {volume} {413}},\
  \bibinfo {pages} {122} (\bibinfo {year} {1997})}\BibitemShut {NoStop}%
\bibitem [{\citenamefont {Heinzl}\ \emph {et~al.}(2006)\citenamefont {Heinzl},
  \citenamefont {Liesfeld}, \citenamefont {Amthor}, \citenamefont {Schwoerer},
  \citenamefont {Sauerbrey},\ and\ \citenamefont {Wipf}}]{Heinzl:2006xc}%
  \BibitemOpen
  \bibfield  {author} {\bibinfo {author} {\bibfnamefont {T.}~\bibnamefont
  {Heinzl}}, \bibinfo {author} {\bibfnamefont {B.}~\bibnamefont {Liesfeld}},
  \bibinfo {author} {\bibfnamefont {K.-U.}\ \bibnamefont {Amthor}}, \bibinfo
  {author} {\bibfnamefont {H.}~\bibnamefont {Schwoerer}}, \bibinfo {author}
  {\bibfnamefont {R.}~\bibnamefont {Sauerbrey}}, \ and\ \bibinfo {author}
  {\bibfnamefont {A.}~\bibnamefont {Wipf}},\ }\href {\doibase
  10.1016/j.optcom.2006.06.053} {\bibfield  {journal} {\bibinfo  {journal}
  {Opt. Commun.}\ }\textbf {\bibinfo {volume} {267}},\ \bibinfo {pages} {318}
  (\bibinfo {year} {2006})},\ \Eprint {http://arxiv.org/abs/0601076}
  {arXiv:0601076 [hep-ph]} \BibitemShut {NoStop}%
\bibitem [{\citenamefont {{Di Piazza}}\ \emph {et~al.}(2006)\citenamefont {{Di
  Piazza}}, \citenamefont {Hatsagortsyan},\ and\ \citenamefont
  {Keitel}}]{DiPiazza:2006pr}%
  \BibitemOpen
  \bibfield  {author} {\bibinfo {author} {\bibfnamefont {A.}~\bibnamefont {{Di
  Piazza}}}, \bibinfo {author} {\bibfnamefont {K.~Z.}\ \bibnamefont
  {Hatsagortsyan}}, \ and\ \bibinfo {author} {\bibfnamefont {C.~H.}\
  \bibnamefont {Keitel}},\ }\href {\doibase 10.1103/PhysRevLett.97.083603}
  {\bibfield  {journal} {\bibinfo  {journal} {Phys. Rev. Lett.}\ }\textbf
  {\bibinfo {volume} {97}},\ \bibinfo {pages} {083603} (\bibinfo {year}
  {2006})},\ \Eprint {http://arxiv.org/abs/0602039v2} {arXiv:0602039v2
  [hep-ph]} \BibitemShut {NoStop}%
\bibitem [{\citenamefont {Dinu}\ \emph
  {et~al.}(2014{\natexlab{a}})\citenamefont {Dinu}, \citenamefont {Heinzl},
  \citenamefont {Ilderton}, \citenamefont {Marklund},\ and\ \citenamefont
  {Torgrimsson}}]{Dinu:2013gaa}%
  \BibitemOpen
  \bibfield  {author} {\bibinfo {author} {\bibfnamefont {V.}~\bibnamefont
  {Dinu}}, \bibinfo {author} {\bibfnamefont {T.}~\bibnamefont {Heinzl}},
  \bibinfo {author} {\bibfnamefont {A.}~\bibnamefont {Ilderton}}, \bibinfo
  {author} {\bibfnamefont {M.}~\bibnamefont {Marklund}}, \ and\ \bibinfo
  {author} {\bibfnamefont {G.}~\bibnamefont {Torgrimsson}},\ }\href {\doibase
  10.1103/PhysRevD.89.125003} {\bibfield  {journal} {\bibinfo  {journal} {Phys.
  Rev. D}\ }\textbf {\bibinfo {volume} {89}},\ \bibinfo {pages} {125003}
  (\bibinfo {year} {2014}{\natexlab{a}})},\ \Eprint
  {http://arxiv.org/abs/1312.6419} {arXiv:1312.6419} \BibitemShut {NoStop}%
\bibitem [{\citenamefont {Dinu}\ \emph
  {et~al.}(2014{\natexlab{b}})\citenamefont {Dinu}, \citenamefont {Heinzl},
  \citenamefont {Ilderton}, \citenamefont {Marklund},\ and\ \citenamefont
  {Torgrimsson}}]{Dinu:2014tsa}%
  \BibitemOpen
  \bibfield  {author} {\bibinfo {author} {\bibfnamefont {V.}~\bibnamefont
  {Dinu}}, \bibinfo {author} {\bibfnamefont {T.}~\bibnamefont {Heinzl}},
  \bibinfo {author} {\bibfnamefont {A.}~\bibnamefont {Ilderton}}, \bibinfo
  {author} {\bibfnamefont {M.}~\bibnamefont {Marklund}}, \ and\ \bibinfo
  {author} {\bibfnamefont {G.}~\bibnamefont {Torgrimsson}},\ }\href {\doibase
  10.1103/PhysRevD.90.045025} {\bibfield  {journal} {\bibinfo  {journal} {Phys.
  Rev. D}\ }\textbf {\bibinfo {volume} {90}},\ \bibinfo {pages} {045025}
  (\bibinfo {year} {2014}{\natexlab{b}})},\ \Eprint
  {http://arxiv.org/abs/1405.7291} {arXiv:1405.7291} \BibitemShut {NoStop}%
\bibitem [{\citenamefont {Karbstein}\ \emph {et~al.}(2015)\citenamefont
  {Karbstein}, \citenamefont {Gies}, \citenamefont {Reuter},\ and\
  \citenamefont {Zepf}}]{Karbstein:2015xra}%
  \BibitemOpen
  \bibfield  {author} {\bibinfo {author} {\bibfnamefont {F.}~\bibnamefont
  {Karbstein}}, \bibinfo {author} {\bibfnamefont {H.}~\bibnamefont {Gies}},
  \bibinfo {author} {\bibfnamefont {M.}~\bibnamefont {Reuter}}, \ and\ \bibinfo
  {author} {\bibfnamefont {M.}~\bibnamefont {Zepf}},\ }\href {\doibase
  10.1103/PhysRevD.92.071301} {\bibfield  {journal} {\bibinfo  {journal} {Phys.
  Rev. D}\ }\textbf {\bibinfo {volume} {92}},\ \bibinfo {pages} {071301}
  (\bibinfo {year} {2015})},\ \Eprint {http://arxiv.org/abs/1507.01084}
  {arXiv:1507.01084} \BibitemShut {NoStop}%
\bibitem [{\citenamefont {Schlenvoigt}\ \emph {et~al.}(2016)\citenamefont
  {Schlenvoigt}, \citenamefont {Heinzl}, \citenamefont {Schramm}, \citenamefont
  {Cowan},\ and\ \citenamefont {Sauerbrey}}]{Schlenvoigt:2016}%
  \BibitemOpen
  \bibfield  {author} {\bibinfo {author} {\bibfnamefont {H.-P.}\ \bibnamefont
  {Schlenvoigt}}, \bibinfo {author} {\bibfnamefont {T.}~\bibnamefont {Heinzl}},
  \bibinfo {author} {\bibfnamefont {U.}~\bibnamefont {Schramm}}, \bibinfo
  {author} {\bibfnamefont {T.~E.}\ \bibnamefont {Cowan}}, \ and\ \bibinfo
  {author} {\bibfnamefont {R.}~\bibnamefont {Sauerbrey}},\ }\href {\doibase
  10.1088/0031-8949/91/2/023010} {\bibfield  {journal} {\bibinfo  {journal}
  {Phys. Scr.}\ }\textbf {\bibinfo {volume} {91}},\ \bibinfo {pages} {023010}
  (\bibinfo {year} {2016})}\BibitemShut {NoStop}%
\bibitem [{\citenamefont {Karbstein}\ and\ \citenamefont
  {Sundqvist}(2016)}]{Karbstein:2016lby}%
  \BibitemOpen
  \bibfield  {author} {\bibinfo {author} {\bibfnamefont {F.}~\bibnamefont
  {Karbstein}}\ and\ \bibinfo {author} {\bibfnamefont {C.}~\bibnamefont
  {Sundqvist}},\ }\href {\doibase 10.1103/PhysRevD.94.013004} {\bibfield
  {journal} {\bibinfo  {journal} {Phys. Rev. D}\ }\textbf {\bibinfo {volume}
  {94}},\ \bibinfo {pages} {1} (\bibinfo {year} {2016})},\ \Eprint
  {http://arxiv.org/abs/1605.09294} {arXiv:1605.09294} \BibitemShut {NoStop}%
\bibitem [{\citenamefont {King}\ and\ \citenamefont
  {Elkina}(2016)}]{King:2016jnl}%
  \BibitemOpen
  \bibfield  {author} {\bibinfo {author} {\bibfnamefont {B.}~\bibnamefont
  {King}}\ and\ \bibinfo {author} {\bibfnamefont {N.}~\bibnamefont {Elkina}},\
  }\href {\doibase 10.1103/PhysRevA.94.062102} {\bibfield  {journal} {\bibinfo
  {journal} {Phys. Rev. A}\ }\textbf {\bibinfo {volume} {94}},\ \bibinfo
  {pages} {062102} (\bibinfo {year} {2016})}\BibitemShut {NoStop}%
\bibitem [{\citenamefont {Bragin}\ \emph {et~al.}(2017)\citenamefont {Bragin},
  \citenamefont {Meuren}, \citenamefont {Keitel},\ and\ \citenamefont {{Di
  Piazza}}}]{Bragin:2017yau}%
  \BibitemOpen
  \bibfield  {author} {\bibinfo {author} {\bibfnamefont {S.}~\bibnamefont
  {Bragin}}, \bibinfo {author} {\bibfnamefont {S.}~\bibnamefont {Meuren}},
  \bibinfo {author} {\bibfnamefont {C.~H.}\ \bibnamefont {Keitel}}, \ and\
  \bibinfo {author} {\bibfnamefont {A.}~\bibnamefont {{Di Piazza}}},\ }\href
  {\doibase 10.1103/PhysRevLett.119.250403} {\bibfield  {journal} {\bibinfo
  {journal} {Phys. Rev. Lett.}\ }\textbf {\bibinfo {volume} {119}},\ \bibinfo
  {pages} {250403} (\bibinfo {year} {2017})},\ \Eprint
  {http://arxiv.org/abs/1704.05234} {arXiv:1704.05234} \BibitemShut {NoStop}%
\bibitem [{\citenamefont {Ataman}(2018)}]{Ataman:2018ucl}%
  \BibitemOpen
  \bibfield  {author} {\bibinfo {author} {\bibfnamefont {S.}~\bibnamefont
  {Ataman}},\ }\href {\doibase 10.1103/PhysRevA.97.063811} {\bibfield
  {journal} {\bibinfo  {journal} {Phys. Rev. A}\ }\textbf {\bibinfo {volume}
  {97}},\ \bibinfo {pages} {063811} (\bibinfo {year} {2018})}\BibitemShut
  {NoStop}%
\bibitem [{\citenamefont {Karbstein}(2018)}]{Karbstein:2018omb}%
  \BibitemOpen
  \bibfield  {author} {\bibinfo {author} {\bibfnamefont {F.}~\bibnamefont
  {Karbstein}},\ }\href {\doibase 10.1103/PhysRevD.98.056010} {\bibfield
  {journal} {\bibinfo  {journal} {Phys. Rev. D}\ }\textbf {\bibinfo {volume}
  {98}},\ \bibinfo {pages} {056010} (\bibinfo {year} {2018})},\ \Eprint
  {http://arxiv.org/abs/1807.03302} {arXiv:1807.03302} \BibitemShut {NoStop}%
\bibitem [{\citenamefont {Lundstrom}\ \emph {et~al.}(2005)\citenamefont
  {Lundstrom}, \citenamefont {Brodin}, \citenamefont {Lundin}, \citenamefont
  {Marklund}, \citenamefont {Bingham}, \citenamefont {Collier}, \citenamefont
  {Mendonca},\ and\ \citenamefont {Norreys}}]{Lundstrom:2005za}%
  \BibitemOpen
  \bibfield  {author} {\bibinfo {author} {\bibfnamefont {E.}~\bibnamefont
  {Lundstrom}}, \bibinfo {author} {\bibfnamefont {G.}~\bibnamefont {Brodin}},
  \bibinfo {author} {\bibfnamefont {J.}~\bibnamefont {Lundin}}, \bibinfo
  {author} {\bibfnamefont {M.}~\bibnamefont {Marklund}}, \bibinfo {author}
  {\bibfnamefont {R.}~\bibnamefont {Bingham}}, \bibinfo {author} {\bibfnamefont
  {J.}~\bibnamefont {Collier}}, \bibinfo {author} {\bibfnamefont {J.~T.}\
  \bibnamefont {Mendonca}}, \ and\ \bibinfo {author} {\bibfnamefont
  {P.}~\bibnamefont {Norreys}},\ }\href {\doibase
  10.1103/PhysRevLett.96.083602} {\bibfield  {journal} {\bibinfo  {journal}
  {Phys. Rev. Lett.}\ }\textbf {\bibinfo {volume} {96}},\ \bibinfo {pages}
  {083602} (\bibinfo {year} {2005})},\ \Eprint {http://arxiv.org/abs/0510076}
  {arXiv:0510076 [hep-ph]} \BibitemShut {NoStop}%
\bibitem [{\citenamefont {Lundin}\ \emph {et~al.}(2006)\citenamefont {Lundin},
  \citenamefont {Marklund}, \citenamefont {Lundstr{\"{o}}m}, \citenamefont
  {Brodin}, \citenamefont {Collier}, \citenamefont {Bingham}, \citenamefont
  {Mendon{\c{c}}a},\ and\ \citenamefont {Norreys}}]{Lundin:2006wu}%
  \BibitemOpen
  \bibfield  {author} {\bibinfo {author} {\bibfnamefont {J.}~\bibnamefont
  {Lundin}}, \bibinfo {author} {\bibfnamefont {M.}~\bibnamefont {Marklund}},
  \bibinfo {author} {\bibfnamefont {E.}~\bibnamefont {Lundstr{\"{o}}m}},
  \bibinfo {author} {\bibfnamefont {G.}~\bibnamefont {Brodin}}, \bibinfo
  {author} {\bibfnamefont {J.}~\bibnamefont {Collier}}, \bibinfo {author}
  {\bibfnamefont {R.}~\bibnamefont {Bingham}}, \bibinfo {author} {\bibfnamefont
  {J.~T.}\ \bibnamefont {Mendon{\c{c}}a}}, \ and\ \bibinfo {author}
  {\bibfnamefont {P.}~\bibnamefont {Norreys}},\ }\href {\doibase
  10.1103/PhysRevA.74.043821} {\bibfield  {journal} {\bibinfo  {journal} {Phys.
  Rev. A}\ }\textbf {\bibinfo {volume} {74}},\ \bibinfo {pages} {043821}
  (\bibinfo {year} {2006})},\ \Eprint {http://arxiv.org/abs/0510076}
  {arXiv:0510076 [hep-ph]} \BibitemShut {NoStop}%
\bibitem [{\citenamefont {Tommasini}\ \emph {et~al.}(2009)\citenamefont
  {Tommasini}, \citenamefont {Ferrando}, \citenamefont {Michinel},\ and\
  \citenamefont {Seco}}]{Tommasini:2009nh}%
  \BibitemOpen
  \bibfield  {author} {\bibinfo {author} {\bibfnamefont {D.}~\bibnamefont
  {Tommasini}}, \bibinfo {author} {\bibfnamefont {A.}~\bibnamefont {Ferrando}},
  \bibinfo {author} {\bibfnamefont {H.}~\bibnamefont {Michinel}}, \ and\
  \bibinfo {author} {\bibfnamefont {M.}~\bibnamefont {Seco}},\ }\href {\doibase
  10.1088/1126-6708/2009/11/043} {\bibfield  {journal} {\bibinfo  {journal} {J.
  High Energy Phys.}\ }\textbf {\bibinfo {volume} {2009}},\ \bibinfo {pages}
  {043} (\bibinfo {year} {2009})},\ \Eprint {http://arxiv.org/abs/0909.4663}
  {arXiv:0909.4663} \BibitemShut {NoStop}%
\bibitem [{\citenamefont {Tommasini}\ and\ \citenamefont
  {Michinel}(2010)}]{Tommasini:2010fb}%
  \BibitemOpen
  \bibfield  {author} {\bibinfo {author} {\bibfnamefont {D.}~\bibnamefont
  {Tommasini}}\ and\ \bibinfo {author} {\bibfnamefont {H.}~\bibnamefont
  {Michinel}},\ }\href {\doibase 10.1103/PhysRevA.82.011803} {\bibfield
  {journal} {\bibinfo  {journal} {Phys. Rev. A}\ }\textbf {\bibinfo {volume}
  {82}},\ \bibinfo {pages} {011803} (\bibinfo {year} {2010})},\ \Eprint
  {http://arxiv.org/abs/1003.5932} {arXiv:1003.5932} \BibitemShut {NoStop}%
\bibitem [{\citenamefont {King}\ and\ \citenamefont
  {Keitel}(2012)}]{King:2012aw}%
  \BibitemOpen
  \bibfield  {author} {\bibinfo {author} {\bibfnamefont {B.}~\bibnamefont
  {King}}\ and\ \bibinfo {author} {\bibfnamefont {C.~H.}\ \bibnamefont
  {Keitel}},\ }\href {\doibase 10.1088/1367-2630/14/10/103002} {\bibfield
  {journal} {\bibinfo  {journal} {New J. Phys.}\ }\textbf {\bibinfo {volume}
  {14}},\ \bibinfo {pages} {103002} (\bibinfo {year} {2012})}\BibitemShut
  {NoStop}%
\bibitem [{\citenamefont {Gies}\ \emph {et~al.}(2018)\citenamefont {Gies},
  \citenamefont {Karbstein}, \citenamefont {Kohlf{\"{u}}rst},\ and\
  \citenamefont {Seegert}}]{Gies:2017ezf}%
  \BibitemOpen
  \bibfield  {author} {\bibinfo {author} {\bibfnamefont {H.}~\bibnamefont
  {Gies}}, \bibinfo {author} {\bibfnamefont {F.}~\bibnamefont {Karbstein}},
  \bibinfo {author} {\bibfnamefont {C.}~\bibnamefont {Kohlf{\"{u}}rst}}, \ and\
  \bibinfo {author} {\bibfnamefont {N.}~\bibnamefont {Seegert}},\ }\href
  {\doibase 10.1103/PhysRevD.97.076002} {\bibfield  {journal} {\bibinfo
  {journal} {Phys. Rev. D}\ }\textbf {\bibinfo {volume} {97}},\ \bibinfo
  {pages} {076002} (\bibinfo {year} {2018})},\ \Eprint
  {http://arxiv.org/abs/1712.06450} {arXiv:1712.06450} \BibitemShut {NoStop}%
\bibitem [{\citenamefont {King}\ \emph {et~al.}(2018)\citenamefont {King},
  \citenamefont {Hu},\ and\ \citenamefont {Shen}}]{King:2018wtn}%
  \BibitemOpen
  \bibfield  {author} {\bibinfo {author} {\bibfnamefont {B.}~\bibnamefont
  {King}}, \bibinfo {author} {\bibfnamefont {H.}~\bibnamefont {Hu}}, \ and\
  \bibinfo {author} {\bibfnamefont {B.}~\bibnamefont {Shen}},\ }\href {\doibase
  10.1103/PhysRevA.98.023817} {\  (\bibinfo {year} {2018}),\
  10.1103/PhysRevA.98.023817},\ \Eprint {http://arxiv.org/abs/1805.03688}
  {arXiv:1805.03688} \BibitemShut {NoStop}%
\bibitem [{\citenamefont {Gies}\ \emph {et~al.}(2013)\citenamefont {Gies},
  \citenamefont {Karbstein},\ and\ \citenamefont {Seegert}}]{Gies:2013yxa}%
  \BibitemOpen
  \bibfield  {author} {\bibinfo {author} {\bibfnamefont {H.}~\bibnamefont
  {Gies}}, \bibinfo {author} {\bibfnamefont {F.}~\bibnamefont {Karbstein}}, \
  and\ \bibinfo {author} {\bibfnamefont {N.}~\bibnamefont {Seegert}},\ }\href
  {\doibase 10.1088/1367-2630/15/8/083002} {\bibfield  {journal} {\bibinfo
  {journal} {New J. Phys.}\ }\textbf {\bibinfo {volume} {15}},\ \bibinfo
  {pages} {083002} (\bibinfo {year} {2013})},\ \Eprint
  {http://arxiv.org/abs/1305.2320} {arXiv:1305.2320} \BibitemShut {NoStop}%
\bibitem [{\citenamefont {Gies}\ \emph {et~al.}(2015)\citenamefont {Gies},
  \citenamefont {Karbstein},\ and\ \citenamefont {Seegert}}]{Gies:2014wsa}%
  \BibitemOpen
  \bibfield  {author} {\bibinfo {author} {\bibfnamefont {H.}~\bibnamefont
  {Gies}}, \bibinfo {author} {\bibfnamefont {F.}~\bibnamefont {Karbstein}}, \
  and\ \bibinfo {author} {\bibfnamefont {N.}~\bibnamefont {Seegert}},\ }\href
  {\doibase 10.1088/1367-2630/17/4/043060} {\bibfield  {journal} {\bibinfo
  {journal} {New J. Phys.}\ }\textbf {\bibinfo {volume} {17}},\ \bibinfo
  {pages} {043060} (\bibinfo {year} {2015})},\ \Eprint
  {http://arxiv.org/abs/1412.0951} {arXiv:1412.0951} \BibitemShut {NoStop}%
\bibitem [{\citenamefont {Yakovlev}(1967)}]{Yakovlev:1966}%
  \BibitemOpen
  \bibfield  {author} {\bibinfo {author} {\bibfnamefont {V.~P.}\ \bibnamefont
  {Yakovlev}},\ }\href {http://adsabs.harvard.edu/abs/1967JETP...24..411Y}
  {\bibfield  {journal} {\bibinfo  {journal} {Sov. Phys. JETP}\ }\textbf
  {\bibinfo {volume} {24}},\ \bibinfo {pages} {411} (\bibinfo {year}
  {1967})}\BibitemShut {NoStop}%
\bibitem [{\citenamefont {Fedotov}\ and\ \citenamefont
  {Narozhny}(2007)}]{Fedotov:2006ii}%
  \BibitemOpen
  \bibfield  {author} {\bibinfo {author} {\bibfnamefont {A.}~\bibnamefont
  {Fedotov}}\ and\ \bibinfo {author} {\bibfnamefont {N.}~\bibnamefont
  {Narozhny}},\ }\href {\doibase 10.1016/j.physleta.2006.09.085} {\bibfield
  {journal} {\bibinfo  {journal} {Phys. Lett. A}\ }\textbf {\bibinfo {volume}
  {362}},\ \bibinfo {pages} {1} (\bibinfo {year} {2007})},\ \Eprint
  {http://arxiv.org/abs/0604258} {arXiv:0604258 [hep-ph]} \BibitemShut
  {NoStop}%
\bibitem [{\citenamefont {{Di Piazza}}\ \emph
  {et~al.}(2008{\natexlab{a}})\citenamefont {{Di Piazza}}, \citenamefont
  {Hatsagortsyan},\ and\ \citenamefont {Keitel}}]{DiPiazza:2007prw}%
  \BibitemOpen
  \bibfield  {author} {\bibinfo {author} {\bibfnamefont {A.}~\bibnamefont {{Di
  Piazza}}}, \bibinfo {author} {\bibfnamefont {K.~Z.}\ \bibnamefont
  {Hatsagortsyan}}, \ and\ \bibinfo {author} {\bibfnamefont {C.~H.}\
  \bibnamefont {Keitel}},\ }\href {\doibase 10.1103/PhysRevLett.100.010403}
  {\bibfield  {journal} {\bibinfo  {journal} {Phys. Rev. Lett.}\ }\textbf
  {\bibinfo {volume} {100}},\ \bibinfo {pages} {010403} (\bibinfo {year}
  {2008}{\natexlab{a}})},\ \Eprint {http://arxiv.org/abs/0708.0475v1}
  {arXiv:0708.0475v1} \BibitemShut {NoStop}%
\bibitem [{\citenamefont {{Di Piazza}}\ \emph
  {et~al.}(2008{\natexlab{b}})\citenamefont {{Di Piazza}}, \citenamefont
  {Hatsagortsyan},\ and\ \citenamefont {Keitel}}]{DiPiazza:2009cq}%
  \BibitemOpen
  \bibfield  {author} {\bibinfo {author} {\bibfnamefont {A.}~\bibnamefont {{Di
  Piazza}}}, \bibinfo {author} {\bibfnamefont {K.~Z.}\ \bibnamefont
  {Hatsagortsyan}}, \ and\ \bibinfo {author} {\bibfnamefont {C.~H.}\
  \bibnamefont {Keitel}},\ }\href {\doibase 10.1103/PhysRevA.78.062109}
  {\bibfield  {journal} {\bibinfo  {journal} {Phys. Rev. A}\ }\textbf {\bibinfo
  {volume} {78}},\ \bibinfo {pages} {062109} (\bibinfo {year}
  {2008}{\natexlab{b}})},\ \Eprint {http://arxiv.org/abs/0906.5576v1}
  {arXiv:0906.5576v1} \BibitemShut {NoStop}%
\bibitem [{\citenamefont {King}\ \emph {et~al.}(2014)\citenamefont {King},
  \citenamefont {B{\"{o}}hl},\ and\ \citenamefont {Ruhl}}]{King:2014vha}%
  \BibitemOpen
  \bibfield  {author} {\bibinfo {author} {\bibfnamefont {B.}~\bibnamefont
  {King}}, \bibinfo {author} {\bibfnamefont {P.}~\bibnamefont {B{\"{o}}hl}}, \
  and\ \bibinfo {author} {\bibfnamefont {H.}~\bibnamefont {Ruhl}},\ }\href
  {\doibase 10.1103/PhysRevD.90.065018} {\bibfield  {journal} {\bibinfo
  {journal} {Phys. Rev. D}\ }\textbf {\bibinfo {volume} {90}},\ \bibinfo
  {pages} {065018} (\bibinfo {year} {2014})}\BibitemShut {NoStop}%
\bibitem [{\citenamefont {Gies}\ \emph {et~al.}(2014)\citenamefont {Gies},
  \citenamefont {Karbstein},\ and\ \citenamefont
  {Shaisultanov}}]{Gies:2014jia}%
  \BibitemOpen
  \bibfield  {author} {\bibinfo {author} {\bibfnamefont {H.}~\bibnamefont
  {Gies}}, \bibinfo {author} {\bibfnamefont {F.}~\bibnamefont {Karbstein}}, \
  and\ \bibinfo {author} {\bibfnamefont {R.}~\bibnamefont {Shaisultanov}},\
  }\href {\doibase 10.1103/PhysRevD.90.033007} {\bibfield  {journal} {\bibinfo
  {journal} {Phys. Rev. D}\ }\textbf {\bibinfo {volume} {90}},\ \bibinfo
  {pages} {033007} (\bibinfo {year} {2014})},\ \Eprint
  {http://arxiv.org/abs/1406.2972} {arXiv:1406.2972} \BibitemShut {NoStop}%
\bibitem [{\citenamefont {Gies}\ \emph {et~al.}(2016)\citenamefont {Gies},
  \citenamefont {Karbstein},\ and\ \citenamefont {Seegert}}]{Gies:2016czm}%
  \BibitemOpen
  \bibfield  {author} {\bibinfo {author} {\bibfnamefont {H.}~\bibnamefont
  {Gies}}, \bibinfo {author} {\bibfnamefont {F.}~\bibnamefont {Karbstein}}, \
  and\ \bibinfo {author} {\bibfnamefont {N.}~\bibnamefont {Seegert}},\ }\href
  {\doibase 10.1103/PhysRevD.93.085034} {\bibfield  {journal} {\bibinfo
  {journal} {Phys. Rev. D}\ }\textbf {\bibinfo {volume} {93}},\ \bibinfo
  {pages} {085034} (\bibinfo {year} {2016})},\ \Eprint
  {http://arxiv.org/abs/1603.00314} {arXiv:1603.00314} \BibitemShut {NoStop}%
\bibitem [{\citenamefont {Adler}\ \emph {et~al.}(1970)\citenamefont {Adler},
  \citenamefont {Bahcall}, \citenamefont {Callan},\ and\ \citenamefont
  {Rosenbluth}}]{Adler:1970gg}%
  \BibitemOpen
  \bibfield  {author} {\bibinfo {author} {\bibfnamefont {S.~L.}\ \bibnamefont
  {Adler}}, \bibinfo {author} {\bibfnamefont {J.~N.}\ \bibnamefont {Bahcall}},
  \bibinfo {author} {\bibfnamefont {C.~G.}\ \bibnamefont {Callan}}, \ and\
  \bibinfo {author} {\bibfnamefont {M.~N.}\ \bibnamefont {Rosenbluth}},\ }\href
  {\doibase doi.org/10.1103/PhysRevLett.71.247} {\bibfield  {journal} {\bibinfo
   {journal} {Phys. Rev. Lett.}\ }\textbf {\bibinfo {volume} {25}},\ \bibinfo
  {pages} {1061} (\bibinfo {year} {1970})}\BibitemShut {NoStop}%
\bibitem [{\citenamefont {Bialynicka-Birula}\ and\ \citenamefont
  {Bialynicki-Birula}(1970)}]{BialynickaBirula:1970vy}%
  \BibitemOpen
  \bibfield  {author} {\bibinfo {author} {\bibfnamefont {Z.}~\bibnamefont
  {Bialynicka-Birula}}\ and\ \bibinfo {author} {\bibfnamefont {I.}~\bibnamefont
  {Bialynicki-Birula}},\ }\href {\doibase 10.1103/PhysRevD.2.2341} {\bibfield
  {journal} {\bibinfo  {journal} {Phys. Rev. D}\ }\textbf {\bibinfo {volume}
  {2}},\ \bibinfo {pages} {2341} (\bibinfo {year} {1970})}\BibitemShut
  {NoStop}%
\bibitem [{\citenamefont {Adler}(1971)}]{Adler:1971wn}%
  \BibitemOpen
  \bibfield  {author} {\bibinfo {author} {\bibfnamefont {S.~L.}\ \bibnamefont
  {Adler}},\ }\href {\doibase 10.1016/0003-4916(71)90154-0} {\bibfield
  {journal} {\bibinfo  {journal} {Ann. Phys. (N. Y).}\ }\textbf {\bibinfo
  {volume} {67}},\ \bibinfo {pages} {599} (\bibinfo {year} {1971})},\ \Eprint
  {http://arxiv.org/abs/9604028} {arXiv:9604028 [hep-th]} \BibitemShut
  {NoStop}%
\bibitem [{\citenamefont {Papanyan}\ and\ \citenamefont
  {Ritus}(1972)}]{Papanyan:1971cv}%
  \BibitemOpen
  \bibfield  {author} {\bibinfo {author} {\bibfnamefont {V.~O.}\ \bibnamefont
  {Papanyan}}\ and\ \bibinfo {author} {\bibfnamefont {V.~I.}\ \bibnamefont
  {Ritus}},\ }\href {http://adsabs.harvard.edu/abs/1972JETP...34.1195P}
  {\bibfield  {journal} {\bibinfo  {journal} {Sov. Phys. JETP}\ }\textbf
  {\bibinfo {volume} {34}},\ \bibinfo {pages} {1195} (\bibinfo {year}
  {1972})}\BibitemShut {NoStop}%
\bibitem [{\citenamefont {{Di Piazza}}\ \emph {et~al.}(2007)\citenamefont {{Di
  Piazza}}, \citenamefont {Milstein},\ and\ \citenamefont
  {Keitel}}]{DiPiazza:2007yx}%
  \BibitemOpen
  \bibfield  {author} {\bibinfo {author} {\bibfnamefont {A.}~\bibnamefont {{Di
  Piazza}}}, \bibinfo {author} {\bibfnamefont {A.~I.}\ \bibnamefont
  {Milstein}}, \ and\ \bibinfo {author} {\bibfnamefont {C.~H.}\ \bibnamefont
  {Keitel}},\ }\href {\doibase 10.1103/PhysRevA.76.032103} {\bibfield
  {journal} {\bibinfo  {journal} {Phys. Rev. A}\ }\textbf {\bibinfo {volume}
  {76}},\ \bibinfo {pages} {032103} (\bibinfo {year} {2007})},\ \Eprint
  {http://arxiv.org/abs/0704.0695v2} {arXiv:0704.0695v2} \BibitemShut {NoStop}%
\bibitem [{\citenamefont {King}\ \emph
  {et~al.}(2010{\natexlab{a}})\citenamefont {King}, \citenamefont {{Di
  Piazza}},\ and\ \citenamefont {Keitel}}]{King:2013am}%
  \BibitemOpen
  \bibfield  {author} {\bibinfo {author} {\bibfnamefont {B.}~\bibnamefont
  {King}}, \bibinfo {author} {\bibfnamefont {A.}~\bibnamefont {{Di Piazza}}}, \
  and\ \bibinfo {author} {\bibfnamefont {C.~H.}\ \bibnamefont {Keitel}},\
  }\href {\doibase 10.1038/nphoton.2009.261} {\bibfield  {journal} {\bibinfo
  {journal} {Nat. Photonics}\ }\textbf {\bibinfo {volume} {4}},\ \bibinfo
  {pages} {92} (\bibinfo {year} {2010}{\natexlab{a}})},\ \Eprint
  {http://arxiv.org/abs/1301.7038} {arXiv:1301.7038} \BibitemShut {NoStop}%
\bibitem [{\citenamefont {King}\ \emph
  {et~al.}(2010{\natexlab{b}})\citenamefont {King}, \citenamefont {{Di
  Piazza}},\ and\ \citenamefont {Keitel}}]{King:2013zz}%
  \BibitemOpen
  \bibfield  {author} {\bibinfo {author} {\bibfnamefont {B.}~\bibnamefont
  {King}}, \bibinfo {author} {\bibfnamefont {A.}~\bibnamefont {{Di Piazza}}}, \
  and\ \bibinfo {author} {\bibfnamefont {C.~H.}\ \bibnamefont {Keitel}},\
  }\href {\doibase 10.1103/PhysRevA.82.032114} {\bibfield  {journal} {\bibinfo
  {journal} {Phys. Rev. A}\ }\textbf {\bibinfo {volume} {82}},\ \bibinfo
  {pages} {032114} (\bibinfo {year} {2010}{\natexlab{b}})}\BibitemShut
  {NoStop}%
\bibitem [{\citenamefont {Kryuchkyan}\ and\ \citenamefont
  {Hatsagortsyan}(2011)}]{Hatsagortsyan:2011}%
  \BibitemOpen
  \bibfield  {author} {\bibinfo {author} {\bibfnamefont {G.~Y.}\ \bibnamefont
  {Kryuchkyan}}\ and\ \bibinfo {author} {\bibfnamefont {K.~Z.}\ \bibnamefont
  {Hatsagortsyan}},\ }\href {\doibase 10.1103/PhysRevLett.107.053604}
  {\bibfield  {journal} {\bibinfo  {journal} {Phys. Rev. Lett.}\ }\textbf
  {\bibinfo {volume} {107}},\ \bibinfo {pages} {053604} (\bibinfo {year}
  {2011})},\ \Eprint {http://arxiv.org/abs/1102.4013} {arXiv:1102.4013}
  \BibitemShut {NoStop}%
\bibitem [{\citenamefont {{Della Valle}}\ \emph {et~al.}(2016)\citenamefont
  {{Della Valle}}, \citenamefont {Ejlli}, \citenamefont {Gastaldi},
  \citenamefont {Messineo}, \citenamefont {Milotti}, \citenamefont {Pengo},
  \citenamefont {Ruoso},\ and\ \citenamefont {Zavattini}}]{DellaValle:2015xxa}%
  \BibitemOpen
  \bibfield  {author} {\bibinfo {author} {\bibfnamefont {F.}~\bibnamefont
  {{Della Valle}}}, \bibinfo {author} {\bibfnamefont {A.}~\bibnamefont
  {Ejlli}}, \bibinfo {author} {\bibfnamefont {U.}~\bibnamefont {Gastaldi}},
  \bibinfo {author} {\bibfnamefont {G.}~\bibnamefont {Messineo}}, \bibinfo
  {author} {\bibfnamefont {E.}~\bibnamefont {Milotti}}, \bibinfo {author}
  {\bibfnamefont {R.}~\bibnamefont {Pengo}}, \bibinfo {author} {\bibfnamefont
  {G.}~\bibnamefont {Ruoso}}, \ and\ \bibinfo {author} {\bibfnamefont
  {G.}~\bibnamefont {Zavattini}},\ }\href {\doibase
  10.1140/epjc/s10052-015-3869-8} {\bibfield  {journal} {\bibinfo  {journal}
  {Eur. Phys. J. C}\ }\textbf {\bibinfo {volume} {76}},\ \bibinfo {pages} {24}
  (\bibinfo {year} {2016})},\ \Eprint {http://arxiv.org/abs/1510.08052}
  {arXiv:1510.08052} \BibitemShut {NoStop}%
\bibitem [{\citenamefont {Cadene}\ \emph {et~al.}(2014)\citenamefont {Cadene},
  \citenamefont {Fouche}, \citenamefont {Battesti},\ and\ \citenamefont
  {Rizzo}}]{Cadene:2013bva}%
  \BibitemOpen
  \bibfield  {author} {\bibinfo {author} {\bibfnamefont {A.}~\bibnamefont
  {Cadene}}, \bibinfo {author} {\bibfnamefont {M.}~\bibnamefont {Fouche}},
  \bibinfo {author} {\bibfnamefont {R.}~\bibnamefont {Battesti}}, \ and\
  \bibinfo {author} {\bibfnamefont {C.}~\bibnamefont {Rizzo}},\ }in\ \href
  {\doibase 10.1109/CPEM.2014.6898345} {\emph {\bibinfo {booktitle} {29th Conf.
  Precis. Electromagn. Meas. (CPEM 2014)}}}\ (\bibinfo  {publisher} {IEEE},\
  \bibinfo {year} {2014})\ pp.\ \bibinfo {pages} {234--235},\ \Eprint
  {http://arxiv.org/abs/1109.4792} {arXiv:1109.4792} \BibitemShut {NoStop}%
\bibitem [{\citenamefont {Fan}\ \emph {et~al.}(2017)\citenamefont {Fan},
  \citenamefont {Kamioka}, \citenamefont {Inada}, \citenamefont {Yamazaki},
  \citenamefont {Namba}, \citenamefont {Asai}, \citenamefont {Omachi},
  \citenamefont {Yoshioka}, \citenamefont {Kuwata-Gonokami}, \citenamefont
  {Matsuo}, \citenamefont {Kawaguchi}, \citenamefont {Kindo},\ and\
  \citenamefont {Nojiri}}]{Fan:2017fnd}%
  \BibitemOpen
  \bibfield  {author} {\bibinfo {author} {\bibfnamefont {X.}~\bibnamefont
  {Fan}}, \bibinfo {author} {\bibfnamefont {S.}~\bibnamefont {Kamioka}},
  \bibinfo {author} {\bibfnamefont {T.}~\bibnamefont {Inada}}, \bibinfo
  {author} {\bibfnamefont {T.}~\bibnamefont {Yamazaki}}, \bibinfo {author}
  {\bibfnamefont {T.}~\bibnamefont {Namba}}, \bibinfo {author} {\bibfnamefont
  {S.}~\bibnamefont {Asai}}, \bibinfo {author} {\bibfnamefont {J.}~\bibnamefont
  {Omachi}}, \bibinfo {author} {\bibfnamefont {K.}~\bibnamefont {Yoshioka}},
  \bibinfo {author} {\bibfnamefont {M.}~\bibnamefont {Kuwata-Gonokami}},
  \bibinfo {author} {\bibfnamefont {A.}~\bibnamefont {Matsuo}}, \bibinfo
  {author} {\bibfnamefont {K.}~\bibnamefont {Kawaguchi}}, \bibinfo {author}
  {\bibfnamefont {K.}~\bibnamefont {Kindo}}, \ and\ \bibinfo {author}
  {\bibfnamefont {H.}~\bibnamefont {Nojiri}},\ }\href {\doibase
  10.1140/epjd/e2017-80290-7} {\bibfield  {journal} {\bibinfo  {journal} {Eur.
  Phys. J. D}\ }\textbf {\bibinfo {volume} {71}},\ \bibinfo {pages} {308}
  (\bibinfo {year} {2017})},\ \Eprint {http://arxiv.org/abs/1705.00495}
  {arXiv:1705.00495} \BibitemShut {NoStop}%
\bibitem [{\citenamefont {Mignani}\ \emph {et~al.}(2017)\citenamefont
  {Mignani}, \citenamefont {Testa}, \citenamefont {{Gonz{\'{a}}lez Caniulef}},
  \citenamefont {Taverna}, \citenamefont {Turolla}, \citenamefont {Zane},\ and\
  \citenamefont {Wu}}]{Mignani:2016fwz}%
  \BibitemOpen
  \bibfield  {author} {\bibinfo {author} {\bibfnamefont {R.~P.}\ \bibnamefont
  {Mignani}}, \bibinfo {author} {\bibfnamefont {V.}~\bibnamefont {Testa}},
  \bibinfo {author} {\bibfnamefont {D.}~\bibnamefont {{Gonz{\'{a}}lez
  Caniulef}}}, \bibinfo {author} {\bibfnamefont {R.}~\bibnamefont {Taverna}},
  \bibinfo {author} {\bibfnamefont {R.}~\bibnamefont {Turolla}}, \bibinfo
  {author} {\bibfnamefont {S.}~\bibnamefont {Zane}}, \ and\ \bibinfo {author}
  {\bibfnamefont {K.}~\bibnamefont {Wu}},\ }\href {\doibase
  10.1093/mnras/stw2798} {\bibfield  {journal} {\bibinfo  {journal} {Mon. Not.
  R. Astron. Soc.}\ }\textbf {\bibinfo {volume} {465}},\ \bibinfo {pages} {492}
  (\bibinfo {year} {2017})},\ \Eprint {http://arxiv.org/abs/1610.08323}
  {arXiv:1610.08323} \BibitemShut {NoStop}%
\bibitem [{\citenamefont {Capparelli}\ \emph {et~al.}(2017)\citenamefont
  {Capparelli}, \citenamefont {Damiano}, \citenamefont {Maiani},\ and\
  \citenamefont {Polosa}}]{Capparelli:2017mlv}%
  \BibitemOpen
  \bibfield  {author} {\bibinfo {author} {\bibfnamefont {L.~M.}\ \bibnamefont
  {Capparelli}}, \bibinfo {author} {\bibfnamefont {A.}~\bibnamefont {Damiano}},
  \bibinfo {author} {\bibfnamefont {L.}~\bibnamefont {Maiani}}, \ and\ \bibinfo
  {author} {\bibfnamefont {A.~D.}\ \bibnamefont {Polosa}},\ }\href {\doibase
  10.1140/epjc/s10052-017-5342-3} {\bibfield  {journal} {\bibinfo  {journal}
  {Eur. Phys. J. C}\ }\textbf {\bibinfo {volume} {77}},\ \bibinfo {pages} {754}
  (\bibinfo {year} {2017})},\ \Eprint {http://arxiv.org/abs/1705.01540}
  {arXiv:1705.01540} \BibitemShut {NoStop}%
\bibitem [{\citenamefont {Turolla}\ \emph {et~al.}(2017)\citenamefont
  {Turolla}, \citenamefont {Zane}, \citenamefont {Taverna}, \citenamefont
  {Caniulef}, \citenamefont {Mignani}, \citenamefont {Testa},\ and\
  \citenamefont {Wu}}]{Turolla:2017tqt}%
  \BibitemOpen
  \bibfield  {author} {\bibinfo {author} {\bibfnamefont {R.}~\bibnamefont
  {Turolla}}, \bibinfo {author} {\bibfnamefont {S.}~\bibnamefont {Zane}},
  \bibinfo {author} {\bibfnamefont {R.}~\bibnamefont {Taverna}}, \bibinfo
  {author} {\bibfnamefont {D.~G.}\ \bibnamefont {Caniulef}}, \bibinfo {author}
  {\bibfnamefont {R.~P.}\ \bibnamefont {Mignani}}, \bibinfo {author}
  {\bibfnamefont {V.}~\bibnamefont {Testa}}, \ and\ \bibinfo {author}
  {\bibfnamefont {K.}~\bibnamefont {Wu}},\ }\href
  {http://arxiv.org/abs/1706.02505} {\ ,\ \bibinfo {pages} {1} (\bibinfo {year}
  {2017})},\ \Eprint {http://arxiv.org/abs/1706.02505} {arXiv:1706.02505}
  \BibitemShut {NoStop}%
\bibitem [{\citenamefont {Karbstein}\ and\ \citenamefont
  {Shaisultanov}(2015{\natexlab{a}})}]{Karbstein:2014fva}%
  \BibitemOpen
  \bibfield  {author} {\bibinfo {author} {\bibfnamefont {F.}~\bibnamefont
  {Karbstein}}\ and\ \bibinfo {author} {\bibfnamefont {R.}~\bibnamefont
  {Shaisultanov}},\ }\href {\doibase 10.1103/PhysRevD.91.113002} {\bibfield
  {journal} {\bibinfo  {journal} {Phys. Rev. D}\ }\textbf {\bibinfo {volume}
  {91}},\ \bibinfo {pages} {113002} (\bibinfo {year} {2015}{\natexlab{a}})},\
  \Eprint {http://arxiv.org/abs/1412.6050} {arXiv:1412.6050} \BibitemShut
  {NoStop}%
\bibitem [{\citenamefont {Karbstein}(2015)}]{Karbstein:2015qwa}%
  \BibitemOpen
  \bibfield  {author} {\bibinfo {author} {\bibfnamefont {F.}~\bibnamefont
  {Karbstein}}\ }(\bibinfo {year} {2015})\ \Eprint
  {http://arxiv.org/abs/1510.03178} {arXiv:1510.03178} \BibitemShut {NoStop}%
\bibitem [{\citenamefont {Gies}\ \emph {et~al.}(2017)\citenamefont {Gies},
  \citenamefont {Karbstein},\ and\ \citenamefont
  {Kohlf{\"{u}}rst}}]{Gies:2017ygp}%
  \BibitemOpen
  \bibfield  {author} {\bibinfo {author} {\bibfnamefont {H.}~\bibnamefont
  {Gies}}, \bibinfo {author} {\bibfnamefont {F.}~\bibnamefont {Karbstein}}, \
  and\ \bibinfo {author} {\bibfnamefont {C.}~\bibnamefont {Kohlf{\"{u}}rst}},\
  }\href {\doibase 10.1103/PhysRevD.97.036022} {\bibfield  {journal} {\bibinfo
  {journal} {Phys. Rev. D}\ }\textbf {\bibinfo {volume} {97}},\ \bibinfo
  {pages} {036022} (\bibinfo {year} {2017})},\ \Eprint
  {http://arxiv.org/abs/1712.03232} {arXiv:1712.03232} \BibitemShut {NoStop}%
\bibitem [{\citenamefont {Böhl}\ \emph {et~al.}(2015)\citenamefont {Böhl},
  \citenamefont {King},\ and\ \citenamefont {Ruhl}}]{Bohl:2015uba}%
  \BibitemOpen
  \bibfield  {author} {\bibinfo {author} {\bibfnamefont {P.}~\bibnamefont
  {Böhl}}, \bibinfo {author} {\bibfnamefont {B.}~\bibnamefont {King}}, \ and\
  \bibinfo {author} {\bibfnamefont {H.}~\bibnamefont {Ruhl}},\ }\href {\doibase
  10.1103/PhysRevA.92.032115} {\bibfield  {journal} {\bibinfo  {journal} {Phys.
  Rev.}\ }\textbf {\bibinfo {volume} {A92}},\ \bibinfo {pages} {032115}
  (\bibinfo {year} {2015})},\ \Eprint {http://arxiv.org/abs/1503.05192}
  {arXiv:1503.05192 [physics.plasm-ph]} \BibitemShut {NoStop}%
\bibitem [{\citenamefont {Carneiro}\ \emph {et~al.}(2016)\citenamefont
  {Carneiro}, \citenamefont {Grismayer}, \citenamefont {Fonseca},\ and\
  \citenamefont {Silva}}]{Domenech:2016xx}%
  \BibitemOpen
  \bibfield  {author} {\bibinfo {author} {\bibfnamefont {P.}~\bibnamefont
  {Carneiro}}, \bibinfo {author} {\bibfnamefont {T.}~\bibnamefont {Grismayer}},
  \bibinfo {author} {\bibfnamefont {R.}~\bibnamefont {Fonseca}}, \ and\
  \bibinfo {author} {\bibfnamefont {L.}~\bibnamefont {Silva}},\ }\href@noop {}
  {\  (\bibinfo {year} {2016})},\ \Eprint {http://arxiv.org/abs/1607.04224}
  {arXiv:1607.04224 [physics.plasm-ph]} \BibitemShut {NoStop}%
\bibitem [{\citenamefont {Pons~Domenech}\ and\ \citenamefont
  {Ruhl}(2016)}]{Carneiro:2016qus}%
  \BibitemOpen
  \bibfield  {author} {\bibinfo {author} {\bibfnamefont {A.}~\bibnamefont
  {Pons~Domenech}}\ and\ \bibinfo {author} {\bibfnamefont {H.}~\bibnamefont
  {Ruhl}},\ }\href@noop {} {\  (\bibinfo {year} {2016})},\ \Eprint
  {http://arxiv.org/abs/1607.00253} {arXiv:1607.00253 [physics.comp-ph]}
  \BibitemShut {NoStop}%
\bibitem [{\citenamefont {Blinne}\ \emph {et~al.}(2018)\citenamefont {Blinne},
  \citenamefont {Kuschel}, \citenamefont {Tietze},\ and\ \citenamefont
  {Zepf}}]{Blinne:2018}%
  \BibitemOpen
  \bibfield  {author} {\bibinfo {author} {\bibfnamefont {A.}~\bibnamefont
  {Blinne}}, \bibinfo {author} {\bibfnamefont {S.}~\bibnamefont {Kuschel}},
  \bibinfo {author} {\bibfnamefont {S.}~\bibnamefont {Tietze}}, \ and\ \bibinfo
  {author} {\bibfnamefont {M.}~\bibnamefont {Zepf}},\ }\href
  {http://arxiv.org/abs/1801.04812} {\bibfield  {journal} {\bibinfo  {journal}
  {Arxiv e-Print}\ } (\bibinfo {year} {2018})},\ \Eprint
  {http://arxiv.org/abs/1801.04812} {arXiv:1801.04812} \BibitemShut {NoStop}%
\bibitem [{\citenamefont {Gal'tsov}\ and\ \citenamefont
  {Nikitina}(1983)}]{Galtsov:1982}%
  \BibitemOpen
  \bibfield  {author} {\bibinfo {author} {\bibfnamefont {D.~V.}\ \bibnamefont
  {Gal'tsov}}\ and\ \bibinfo {author} {\bibfnamefont {N.~S.}\ \bibnamefont
  {Nikitina}},\ }\href@noop {} {\bibfield  {journal} {\bibinfo  {journal} {Zh.
  Eksp. Teor. Fiz.}\ }\textbf {\bibinfo {volume} {84}},\ \bibinfo {pages}
  {1217} (\bibinfo {year} {1983})}\BibitemShut {NoStop}%
\bibitem [{\citenamefont {Karbstein}\ and\ \citenamefont
  {Shaisultanov}(2015{\natexlab{b}})}]{Karbstein:2015cpa}%
  \BibitemOpen
  \bibfield  {author} {\bibinfo {author} {\bibfnamefont {F.}~\bibnamefont
  {Karbstein}}\ and\ \bibinfo {author} {\bibfnamefont {R.}~\bibnamefont
  {Shaisultanov}},\ }\href {\doibase 10.1103/PhysRevD.91.085027} {\bibfield
  {journal} {\bibinfo  {journal} {Phys. Rev. D}\ }\textbf {\bibinfo {volume}
  {91}},\ \bibinfo {pages} {085027} (\bibinfo {year} {2015}{\natexlab{b}})},\
  \Eprint {http://arxiv.org/abs/1503.00532} {arXiv:1503.00532} \BibitemShut
  {NoStop}%
\bibitem [{\citenamefont {Briscese}(2018)}]{Briscese:2017htx}%
  \BibitemOpen
  \bibfield  {author} {\bibinfo {author} {\bibfnamefont {F.}~\bibnamefont
  {Briscese}},\ }\href {\doibase 10.1103/PhysRevA.97.033803,
  10.1103/PhysRevA.97.033803.} {\bibfield  {journal} {\bibinfo  {journal}
  {Phys. Rev.}\ }\textbf {\bibinfo {volume} {A97}},\ \bibinfo {pages} {033803}
  (\bibinfo {year} {2018})},\ \Eprint {http://arxiv.org/abs/1710.07703}
  {arXiv:1710.07703 [physics.optics]} \BibitemShut {NoStop}%
\bibitem [{\citenamefont {Briscese}(2017)}]{Briscese:2017wuh}%
  \BibitemOpen
  \bibfield  {author} {\bibinfo {author} {\bibfnamefont {F.}~\bibnamefont
  {Briscese}},\ }\href {\doibase 10.1103/PhysRevA.96.053801} {\bibfield
  {journal} {\bibinfo  {journal} {Phys. Rev.}\ }\textbf {\bibinfo {volume}
  {A96}},\ \bibinfo {pages} {053801} (\bibinfo {year} {2017})},\ \Eprint
  {http://arxiv.org/abs/1710.03338} {arXiv:1710.03338 [physics.optics]}
  \BibitemShut {NoStop}%
\bibitem [{\citenamefont {Davis}(1979)}]{Davis:1979zz}%
  \BibitemOpen
  \bibfield  {author} {\bibinfo {author} {\bibfnamefont {L.~W.}\ \bibnamefont
  {Davis}},\ }\href {\doibase 10.1103/PhysRevA.19.1177} {\bibfield  {journal}
  {\bibinfo  {journal} {Phys. Rev. A}\ }\textbf {\bibinfo {volume} {19}},\
  \bibinfo {pages} {1177} (\bibinfo {year} {1979})}\BibitemShut {NoStop}%
\bibitem [{\citenamefont {Barton}\ and\ \citenamefont
  {Alexander}(1989)}]{Barton:1989}%
  \BibitemOpen
  \bibfield  {author} {\bibinfo {author} {\bibfnamefont {J.~P.}\ \bibnamefont
  {Barton}}\ and\ \bibinfo {author} {\bibfnamefont {D.~R.}\ \bibnamefont
  {Alexander}},\ }\href {\doibase 10.1063/1.344207} {\bibfield  {journal}
  {\bibinfo  {journal} {J. Appl. Phys.}\ }\textbf {\bibinfo {volume} {66}},\
  \bibinfo {pages} {2800} (\bibinfo {year} {1989})}\BibitemShut {NoStop}%
\bibitem [{\citenamefont {Salamin}\ \emph {et~al.}(2002)\citenamefont
  {Salamin}, \citenamefont {Mocken},\ and\ \citenamefont
  {Keitel}}]{Salamin:2002dd}%
  \BibitemOpen
  \bibfield  {author} {\bibinfo {author} {\bibfnamefont {Y.~I.}\ \bibnamefont
  {Salamin}}, \bibinfo {author} {\bibfnamefont {G.~R.}\ \bibnamefont {Mocken}},
  \ and\ \bibinfo {author} {\bibfnamefont {C.~H.}\ \bibnamefont {Keitel}},\
  }\href {\doibase 10.1103/PhysRevSTAB.5.101301} {\bibfield  {journal}
  {\bibinfo  {journal} {Phys. Rev. Spec. Top. - Accel. Beams}\ }\textbf
  {\bibinfo {volume} {5}},\ \bibinfo {pages} {101301} (\bibinfo {year}
  {2002})}\BibitemShut {NoStop}%
\bibitem [{\citenamefont {Salamin}\ \emph {et~al.}(2006)\citenamefont
  {Salamin}, \citenamefont {Hu}, \citenamefont {Hatsagortsyan},\ and\
  \citenamefont {Keitel}}]{Salamin:2006ff}%
  \BibitemOpen
  \bibfield  {author} {\bibinfo {author} {\bibfnamefont {Y.~I.}\ \bibnamefont
  {Salamin}}, \bibinfo {author} {\bibfnamefont {S.~X.}\ \bibnamefont {Hu}},
  \bibinfo {author} {\bibfnamefont {K.~Z.}\ \bibnamefont {Hatsagortsyan}}, \
  and\ \bibinfo {author} {\bibfnamefont {C.~H.}\ \bibnamefont {Keitel}},\
  }\href {\doibase 10.1016/j.physrep.2006.01.002} {\bibfield  {journal}
  {\bibinfo  {journal} {Phys. Rep.}\ }\textbf {\bibinfo {volume} {427}},\
  \bibinfo {pages} {41} (\bibinfo {year} {2006})}\BibitemShut {NoStop}%
\bibitem [{\citenamefont {Salamin}(2007)}]{Salamin:2006}%
  \BibitemOpen
  \bibfield  {author} {\bibinfo {author} {\bibfnamefont {Y.}~\bibnamefont
  {Salamin}},\ }\href {\doibase 10.1007/s00340-006-2442-4} {\bibfield
  {journal} {\bibinfo  {journal} {Appl. Phys. B}\ }\textbf {\bibinfo {volume}
  {86}},\ \bibinfo {pages} {319} (\bibinfo {year} {2007})}\BibitemShut
  {NoStop}%
\bibitem [{\citenamefont {Waters}\ and\ \citenamefont
  {King}(2018)}]{Waters:2017tgl}%
  \BibitemOpen
  \bibfield  {author} {\bibinfo {author} {\bibfnamefont {W.~J.}\ \bibnamefont
  {Waters}}\ and\ \bibinfo {author} {\bibfnamefont {B.}~\bibnamefont {King}},\
  }\href {\doibase 10.1088/1555-6611/aa94dc} {\bibfield  {journal} {\bibinfo
  {journal} {Laser Phys.}\ }\textbf {\bibinfo {volume} {28}},\ \bibinfo {pages}
  {015003} (\bibinfo {year} {2018})},\ \Eprint
  {http://arxiv.org/abs/1705.08554} {arXiv:1705.08554} \BibitemShut {NoStop}%
\bibitem [{\citenamefont {Monden}\ and\ \citenamefont
  {Kodama}(2011)}]{Monden:2011}%
  \BibitemOpen
  \bibfield  {author} {\bibinfo {author} {\bibfnamefont {Y.}~\bibnamefont
  {Monden}}\ and\ \bibinfo {author} {\bibfnamefont {R.}~\bibnamefont
  {Kodama}},\ }\href {\doibase 10.1103/PhysRevLett.107.073602} {\bibfield
  {journal} {\bibinfo  {journal} {Phys. Rev. Lett.}\ }\textbf {\bibinfo
  {volume} {107}},\ \bibinfo {pages} {073602} (\bibinfo {year}
  {2011})}\BibitemShut {NoStop}%
\bibitem [{\citenamefont {Karbstein}\ and\ \citenamefont
  {Sundqvist}(2018)}]{Karbstein:2018}%
  \BibitemOpen
  \bibfield  {author} {\bibinfo {author} {\bibfnamefont {F.}~\bibnamefont
  {Karbstein}}\ and\ \bibinfo {author} {\bibfnamefont {C.}~\bibnamefont
  {Sundqvist}},\ }\href@noop {} {\bibfield  {journal} {\bibinfo  {journal} {in
  preparation}\ } (\bibinfo {year} {2018})}\BibitemShut {NoStop}%
\bibitem [{\citenamefont {Monden}\ and\ \citenamefont
  {Kodama}(2012)}]{Monden:2012}%
  \BibitemOpen
  \bibfield  {author} {\bibinfo {author} {\bibfnamefont {Y.}~\bibnamefont
  {Monden}}\ and\ \bibinfo {author} {\bibfnamefont {R.}~\bibnamefont
  {Kodama}},\ }\href {\doibase 10.1103/PhysRevA.86.033810} {\bibfield
  {journal} {\bibinfo  {journal} {Phys. Rev. A}\ }\textbf {\bibinfo {volume}
  {86}},\ \bibinfo {pages} {033810} (\bibinfo {year} {2012})}\BibitemShut
  {NoStop}%
\bibitem [{CIL(2018)}]{CILEX}%
  \BibitemOpen
  \href@noop {} {}\bibinfo {howpublished} {CILEX, \url{http://cilexsaclay.fr/}}
  (\bibinfo {year} {2018})\BibitemShut {NoStop}%
\bibitem [{CoR(2018)}]{CoReLS}%
  \BibitemOpen
  \href@noop {} {}\bibinfo {howpublished} {CILEX,
  \url{http://corels.ibs.re.kr/}} (\bibinfo {year} {2018})\BibitemShut
  {NoStop}%
\bibitem [{ELI(2018)}]{ELI}%
  \BibitemOpen
  \href@noop {} {}\bibinfo {howpublished} {ELI, \url{https://eli-laser.eu/}}
  (\bibinfo {year} {2018})\BibitemShut {NoStop}%
\bibitem [{\citenamefont {Zhu}\ \emph {et~al.}(2016)\citenamefont {Zhu},
  \citenamefont {Xie}, \citenamefont {Yang}, \citenamefont {Kang},
  \citenamefont {Zhu}, \citenamefont {Guo}, \citenamefont {Zhu}, \citenamefont
  {Gao}, \citenamefont {Liu}, \citenamefont {Fan}, \citenamefont {Liu},
  \citenamefont {Oyang}, \citenamefont {Wei},\ and\ \citenamefont
  {Wang}}]{SG-II}%
  \BibitemOpen
  \bibfield  {author} {\bibinfo {author} {\bibfnamefont {J.}~\bibnamefont
  {Zhu}}, \bibinfo {author} {\bibfnamefont {X.}~\bibnamefont {Xie}}, \bibinfo
  {author} {\bibfnamefont {Q.}~\bibnamefont {Yang}}, \bibinfo {author}
  {\bibfnamefont {J.}~\bibnamefont {Kang}}, \bibinfo {author} {\bibfnamefont
  {H.}~\bibnamefont {Zhu}}, \bibinfo {author} {\bibfnamefont {A.}~\bibnamefont
  {Guo}}, \bibinfo {author} {\bibfnamefont {P.}~\bibnamefont {Zhu}}, \bibinfo
  {author} {\bibfnamefont {Q.}~\bibnamefont {Gao}}, \bibinfo {author}
  {\bibfnamefont {Z.}~\bibnamefont {Liu}}, \bibinfo {author} {\bibfnamefont
  {Q.}~\bibnamefont {Fan}}, \bibinfo {author} {\bibfnamefont {D.}~\bibnamefont
  {Liu}}, \bibinfo {author} {\bibfnamefont {X.}~\bibnamefont {Oyang}}, \bibinfo
  {author} {\bibfnamefont {H.}~\bibnamefont {Wei}}, \ and\ \bibinfo {author}
  {\bibfnamefont {X.}~\bibnamefont {Wang}},\ }in\ \href {\doibase
  10.1364/CLEO_SI.2016.SM1M.7} {\emph {\bibinfo {booktitle} {Conf. Lasers
  Electro-Optics}}}\ (\bibinfo  {publisher} {OSA},\ \bibinfo {year} {2016})\
  p.\ \bibinfo {pages} {SM1M.7}\BibitemShut {NoStop}%
\bibitem [{\citenamefont {Gusynin}\ and\ \citenamefont
  {Shovkovy}(1996)}]{Gusynin:1995bc}%
  \BibitemOpen
  \bibfield  {author} {\bibinfo {author} {\bibfnamefont {V.~P.}\ \bibnamefont
  {Gusynin}}\ and\ \bibinfo {author} {\bibfnamefont {I.~A.}\ \bibnamefont
  {Shovkovy}},\ }\href {\doibase 10.1139/p96-044} {\bibfield  {journal}
  {\bibinfo  {journal} {Can. J. Phys.}\ }\textbf {\bibinfo {volume} {74}},\
  \bibinfo {pages} {282} (\bibinfo {year} {1996})},\ \Eprint
  {http://arxiv.org/abs/9509383} {arXiv:9509383 [hep-ph]} \BibitemShut
  {NoStop}%
\bibitem [{\citenamefont {Gusynin}\ and\ \citenamefont
  {Shovkovy}(1999)}]{Gusynin:1998bt}%
  \BibitemOpen
  \bibfield  {author} {\bibinfo {author} {\bibfnamefont {V.~P.}\ \bibnamefont
  {Gusynin}}\ and\ \bibinfo {author} {\bibfnamefont {I.~A.}\ \bibnamefont
  {Shovkovy}},\ }\href {\doibase 10.1063/1.533037} {\bibfield  {journal}
  {\bibinfo  {journal} {J. Math. Phys.}\ }\textbf {\bibinfo {volume} {40}},\
  \bibinfo {pages} {5406} (\bibinfo {year} {1999})},\ \Eprint
  {http://arxiv.org/abs/9804143} {arXiv:9804143 [hep-th]} \BibitemShut
  {NoStop}%
\bibitem [{\citenamefont {Ritus}(1975)}]{Ritus:1975}%
  \BibitemOpen
  \bibfield  {author} {\bibinfo {author} {\bibfnamefont {V.~I.}\ \bibnamefont
  {Ritus}},\ }\href@noop {} {\bibfield  {journal} {\bibinfo  {journal} {Sov.
  Phys. JETP}\ }\textbf {\bibinfo {volume} {42}},\ \bibinfo {pages} {774}
  (\bibinfo {year} {1975})}\BibitemShut {NoStop}%
\bibitem [{\citenamefont {Gies}\ and\ \citenamefont
  {Karbstein}(2017)}]{Gies:2016yaa}%
  \BibitemOpen
  \bibfield  {author} {\bibinfo {author} {\bibfnamefont {H.}~\bibnamefont
  {Gies}}\ and\ \bibinfo {author} {\bibfnamefont {F.}~\bibnamefont
  {Karbstein}},\ }\href {\doibase 10.1007/JHEP03(2017)108} {\bibfield
  {journal} {\bibinfo  {journal} {J. High Energy Phys.}\ }\textbf {\bibinfo
  {volume} {2017}},\ \bibinfo {pages} {108} (\bibinfo {year} {2017})},\ \Eprint
  {http://arxiv.org/abs/1612.07251v2} {arXiv:1612.07251v2} \BibitemShut
  {NoStop}%
\end{thebibliography}%

\end{document}